\documentclass[preprint, superscriptaddress, amsmath,amssymb, aps,floatfix]{revtex4-2}
\usepackage{graphicx}
\usepackage[caption=false]{subfig}
\usepackage{dcolumn}
\usepackage{bm}
\usepackage{xcolor}
\usepackage{multirow}
\usepackage{float}
\usepackage{soul}
\usepackage{hyperref}
\usepackage[mathlines]{lineno}

\begin{document}


\title{Scaling signatures of near-critical dynamics in large-scale human resting-state electrophysiology}

\author{Irem Topal}
\affiliation{Department of Biomedical Sciences, University of Padova, Padova, Italy}%

\author{Anna Poggialini}
\affiliation{Department of Biomedical Sciences, University of Padova, Padova, Italy}%

\author{Marco Dal Maschio}
\affiliation{Department of Biomedical Sciences, University of Padova, Padova, Italy}%

\author{Daniele De Martino}
\affiliation{Biofisika Institute (CSIC,UPV-EHU) and Ikerbasque Foundation, Bilbao 48013, Spain}

\author{Oren Shriki}
\affiliation{Department of Industrial Engineering and Management, School of Brain Sciences and Cognition, Ben-Gurion University of the Negev, Beer-Sheva, Israel}%

\author{Fabrizio Lombardi}
\email{fabrizio.lombardi\_at\_unipd.it}
\affiliation{Department of Biomedical Sciences, University of Padova, Padova, Italy}%

\date{\today}


\begin{abstract}

From 1/f noise to neuronal avalanches, evidence of scaling in brain activity has been increasingly linked to tuning to or near criticality. The concept of scaling is intimately related to the renormalization group (RG), in essence providing coarse-grained, simplified descriptions that generalize to classes of diverse physical systems. Following the RG idea, scaling laws have been reported in populations of spiking neurons at microscopic scales. Whether similar scaling principles govern large-scale neural activity in the human brain and how they relate to underlying neural physiology remains unresolved. Here, we analyze large-scale electrophysiological recordings (MEG) of human resting-state brain activity and apply a RG-inspired coarse-graining approach to track collective neural dynamics across spatial scales. We find that multiple observables exhibit robust scale-invariant behavior under coarse-graining: activity variance and correlations grow according to power laws, covariance eigenspectra follow a characteristic scaling relation, and neuronal avalanche statistics and scaling relationships are conserved. Using an analytically tractable neural network model with parameters inferred from data, we show that the observed scaling signatures arise close to, but slightly below, criticality, and that the scaling exponents can be modulated by network inhibitory drives. These findings demonstrate that RG-inspired scaling analysis can uncover signatures of critical dynamics in non-invasive human electrophysiology and suggest a principled route toward estimating excitation-inhibition balance from large-scale brain recordings.

\end{abstract}



\maketitle

\section{\label{sec:intro}Introduction}

Criticality has emerged in recent years as an effective framework for quantifying the internal state of the brain \cite{fl_2014_crit_brain,Zimmern_2020,shew25neuron,algom2025concrit}. The idea that the brain may operate near a phase transition dates back to the early  1950's  and originates from the analogy between collective behaviors in neural populations and (inanimate) physical systems---in particular, ferromagnets---at criticality  \cite{cragg1954ising}, a special point of the parameter space separating an ordered from a disordered phase. Intuitively, one would link the emergence of collective, cooperative behaviors to the presence of numerous direct interactions among the many  components of a system. However, stunning collective properties emerge at criticality even in systems with local, short range interactions \cite{stan:pt}, above all scale invariance (or simply scaling) and long-range power-law correlations. 

So, what are we expected to observe, practically, if a population of neurons  operated at criticality? The recent advancement in neural recording techniques has allowed  the simultaneous tracking of an ever increasing number of neurons, prompting a renovated interest in brain criticality. Observations of neuronal avalanches across scales, systems,  and species have provided important evidence of power-law, scale-invariant features in collective neural activity \cite{plenz:pl03,plenz:awa,taglia_2012_point,shriki13,fontanele_2019_prl,Scarpetta_2023, Burrows3259}. \textcolor{black}{Similarly,  long-range spatio-temporal correlations  have been  reported in brain activity,  in particular in humans and  rats \cite{linken01,THURNER_2003_scaling_fmri,linken05,Expert_selfsim_2010,fraiman12,palva13,meisel2017lrtc,fl2020_lrtc,fl2020jneurosci,wang2019plos,Huo_LC_2024}.} All of these  could be considered hints of criticality that arise from tracking spatio-temporal, collective neural dynamics \cite{levina,menesse2022homeo,fosque_quasi_crit_2022}. On the other hand, focusing on strictly  static or equilibrium (thermodynamic) aspects, evidence of tuning to criticality has come from maximum entropy models of neuronal populations \cite{tkavcik2015thermodynamics,mora2015dynamical,simoes2025maximumentropymodels}. In this context, signatures of  criticality are given by extrema in thermodynamic quantities of the data-inferred model, such as maxima in the susceptibility and in the specific heat \cite{stan:pt,tkavcik2014searching}.  

Those hallmarks of criticality  boil down  to one key  fact, i.e. the dynamics are independent of the microscopic details.  This  implies that, at criticality, most degrees of freedom are irrelevant to obtain an essential understanding (or model) of the system, and opens the way for simplified descriptions in terms of coarse-grained variables integrating out ``irrelevant details''. This idea is formalized in the renormalization group (RG) of equilibrium statistical physics \cite{Cardy_1996}. Intuitively, one initially deals with many variables---these could be binary neurons---which define the microscopic scale of the system under scrutiny. Imagine now performing a ``local'' coarse–graining by averaging over small neighborhoods in space or, more generally, by replacing a group of local units with a single representative unit, which needs to be chosen appropriately \cite{kadanoff_box}. As one iterates such a process, the joint distribution of coarse–grained variables evolves  and, in most cases, becomes simpler, as fast fluctuations are averaged out. Under the assumption of the RG, the resulting coarse-grained, macroscopic description should be more general than the underlying  microscopic structures and form a universality class. Within a universality class,  the  measurable properties of microscopically diverse systems are described 
by common scaling functions and power laws, with  a set of critical exponents that 
solely depend on general features such as symmetries and dimensionality \cite{stan:pt}. A major obstacle in applying RG ideas to neural data  lies in the fact that, unlike most physical systems, the underlying interaction structure  is highly complex---including the highly intricate and extended morphology of individual neurons---and remains largely unknown.

A simplification in the spirit of the RG has recently been  proposed for populations of real neurons in the hippocampus \cite{meshulam2019}. In this approach, spatial coarse-graining of local variables is replaced by a correlation-based coarse-graining of individual neurons. This choice is motivated by the extended, non-local  nature of neurons, which also make long-range synaptic connections. Since in systems with local interactions microscopic variables are most strongly correlated with nearest neighbors, correlation is used in this approach as a proxy for neighborhood. To guide intuition, suppose that we start with a population of $N$ neurons. We quantify the correlations between all possible couples of neurons, and then ``couple'' each neuron with its most correlated neuron by summing their respective activity at each time interval. In doing so, we create $N/2$ new time series, which we refer to generically as variables. The process can be iterated until we are left with a single variable. Conceptually, this procedure is identical to the so called block-renormalization, or spatial averaging \cite{kadanoff_box}.  
In  \cite{meshulam2019}, it was shown that under iterated coarse-graining,  the activity of a network of hippocampal neurons approached a fixed non-Gaussian distribution, and several quantities, e.g. correlation time, showed a non-trivial power–law scaling as a function of the  coarse-graining level. These properties signal that the system is near a non-trivial fixed point of the RG, i.e., criticality \cite{meshulam2019}

Similar results have been found in  population of neurons from different regions of the mouse cortex \cite{morales_quasiuniversal_2023,cambrainha_criticality_2024} and, to some extent, in Blood Oxygenation Level Dependent (BOLD) signals \cite{ponce-alvarez_critical_2023}.   
However, such evidence  has not been confirmed in large-scale  electrophysiological recordings, and its link with distance to criticality (DTC) and basic network physiology remains unclear. Here, we aim to close this gap by investigating the  magnetoencephalography (MEG) of the awake resting-state under coarse-graining, and verify  that common fingerprints of brain criticality, e.g. neuronal avalanches, are invariant under the same coarse-graining procedure---a key prediction of the RG. Our analysis shows  robust scaling behaviors in resting-state MEG, with exponents close to those measured in populations of spiking neurons and stability of avalanche dynamics across coarse-grained scales. By means of an analytically tractable model, we are able to locate the emergence of such behaviors within a slightly subcritical region of the parameter space, and \textcolor{black}{show that scaling exponents can be modulated by  network inhibitory drives.} 

This is, to our knowledge, the first demonstration that RG-like scaling and coarse-graining invariance of avalanche dynamics can be robustly observed in large-scale non-invasive human electrophysiology.  M/EEG signals  result (mostly)  from   ionic currents to which large populations of neurons contribute \cite{pfurt99,stam2005nonlin}, and can  be considered in turn as an effective  coarse-grained measure of the underlying neural  activity. Thus,  MEG scaling laws in line with those reported for populations of spiking neurons represent an  effective bridge between spatial  scales, and our modeling results  suggest a  non-invasive approach to estimate the E/I balance  from RG-derived scaling exponents.

\section{Results}

\subsection{Resting-state MEG obeys scaling laws indicative of criticality}

We analyze  MEG during 4 minutes of  eyes-closed resting-state in 100 healthy subjects (Methods \ref{data}). We normalize MEG signals to zero mean and unit standard deviation (SD) and identify positive and negative  extremes of signal fluctuations greater than a threshold $e = \pm3$SD (Methods \ref{data}).   We refer to them as extreme events or simply events, and in our analysis they will be the equivalent of spikes in populations of neurons \cite{meshulam2019}. This preprocessing  maps  continuous signals  into discrete time series of number of events  and ensures that coarse-grainig is performed on the same time series that are used to define neuronal avalanches in the MEG \cite{shriki13,fl2020_lrtc}.

\begin{figure*}[ht!]
    \centering
    \includegraphics[width=0.95\linewidth]{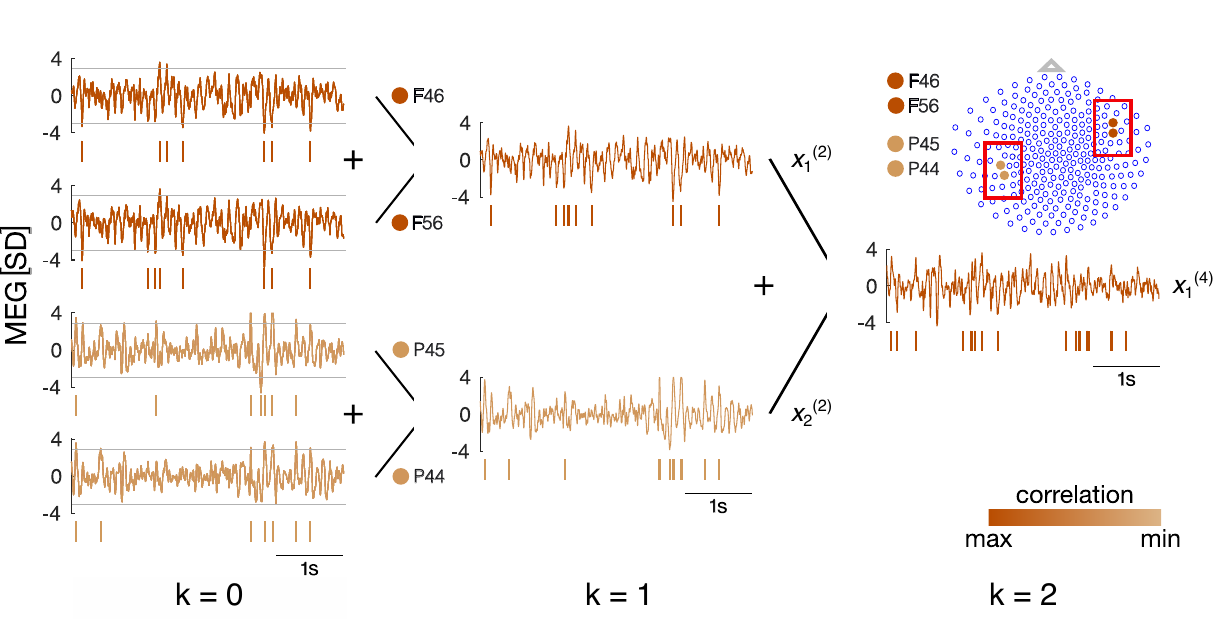}
    \caption{\scriptsize{\textbf{Phenomenological Renormalization Group analysis of MEG data.} 
    Schematic presentation of the coarse-graining procedure for four MEG sensors. Sensor signals (z-normalized, amplitude in unit of standard deviation (SD); see Methods \ref{data}) are initially binarized by marking as $1$'s the extremes of positive and negative  excursions beyond a threshold $e = \pm 3$ SD (horizontal gray lines)  and as $0$'s all other points (\textbf{left}). Resulting binary time series (colored bars) are shown at the bottom of each signal, where vertical lines represent tips of above-threshold excursions, which we refer to as extreme events. The case $k = 0$ corresponds to the original MEG signals (left). 
    Signals are sorted in couples, from the most to the least correlated. The first couple is made by the sensors F46 and F56 (dark red dots), the second by the sensors P45 and P44 (light red dots). Location of these sensors  on the scalp is shown on the right upper corner.
    At each step $k$ of the coarse-graining, pairwise correlations between sensor time series are calculated. Each sensor is paired with its most correlated sensor,  and their activity  is summed and results in a new time-series (\textbf{middle}, $k = 1$). On top of each  time series of events we show the signal resulting from summing together the original sensor (or variable) signals, which is the continuous, z-normalized coarse-grained variable. 
    The resulting summed activity defines the \emph{coarse-grained variables}. At each successive iteration of the procedure, we compute the correlations between coarse-grained variables and group the most  correlated pairs by summing their activity. This process is repeated iteratively until only one variable is left (\textbf{right}, $k = 2$).}
    }
    \label{Fig_1}
\end{figure*}

We illustrate the coarse-graining procedure in Fig.~\ref{Fig_1}. For pedagogical purposes,  we start with  $N=4$ MEG signals from sensors in the frontal and parietal areas (F46, F56, P44 and  P45, Fig.~\ref{Fig_1}, left). For each signal, we show  the corresponding extreme events at the bottom (colored bars). The initial step of the procedure is labeled $k = 0$ (original, non coarse-grained signals). Then,  the sequence of events from each sensor is summed with the sequence of events of its most correlated sensor, leading to $N_k = N/K = N/2^k$ new  variables in step $k = 1$ (Fig.~\ref{Fig_1}, middle).  
The new  variables are then normalized in such a way that the average number of events in non-empty time bins is one, as in the initial step $k=0$ (Methods \ref{app_prg}).  The procedure continues until $N_k = 1$, i.e.,  when only a single variable is left (Fig.~\ref{Fig_1}, right). Throughout the paper, $k$ will indicate the coarse-graining step, $K = 2^k$ the number of signals iteratively summed  up to step $k$,  and $N_k = N/K$ the number of variables at each step $k$. Hence, the variables in  step $k$ are clusters of size $K$  and are labeled as $x_i ^{(K)}$ ($i=1, \dots, N_k$).

\begin{figure*}[t!]
    \centering
    \includegraphics[width=\linewidth]{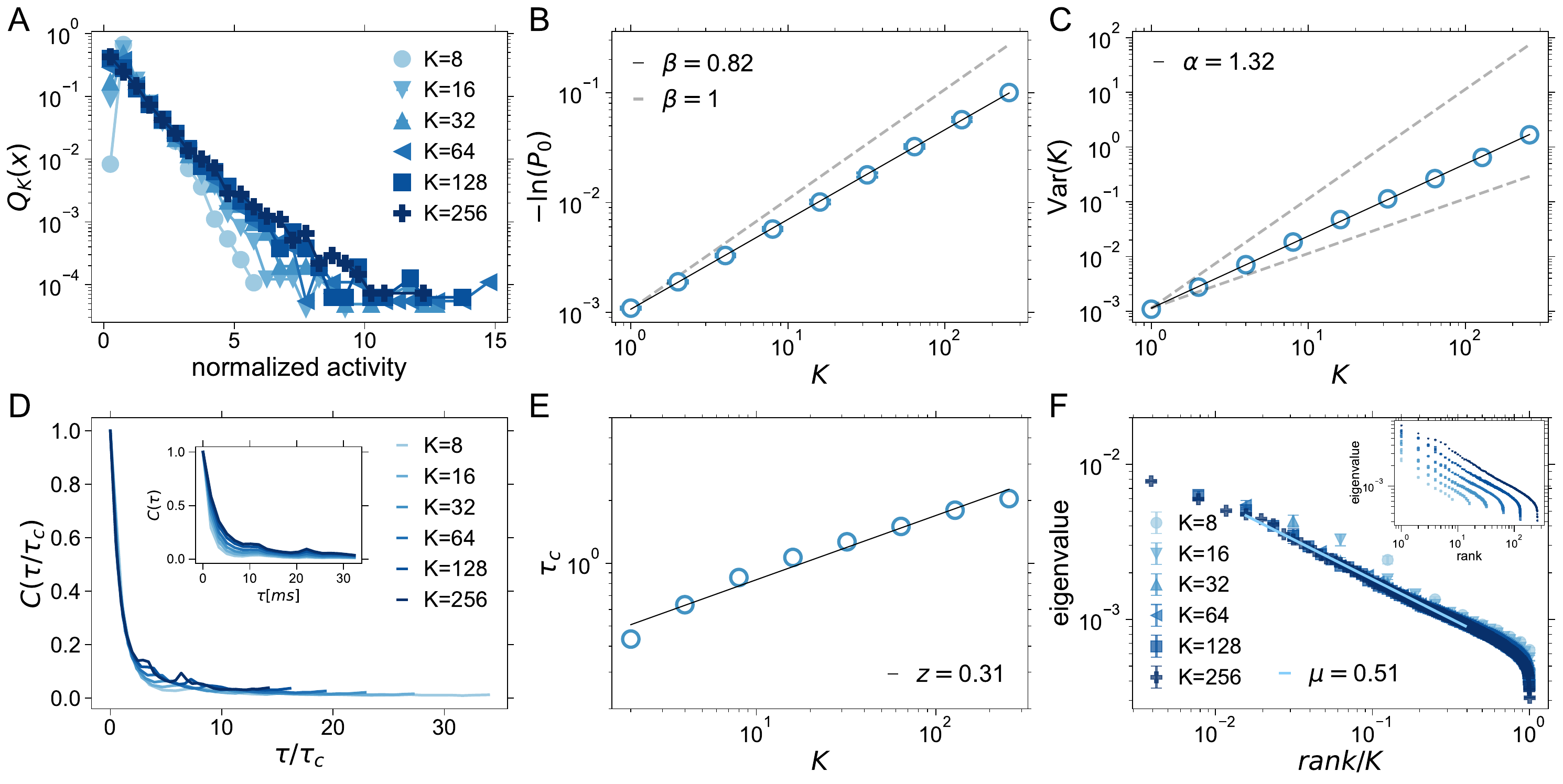}
    \caption{\scriptsize{{\textbf{Scaling in resting-state human MEG.} 
    \textbf{A.} Probability distribution of normalized non-zero activity for the cluster sizes $K \geq 8$. Darker blue = Increasing cluster size. Only data in the range $(0,15]$ are shown. \textbf{B.} The log-probability of silence in coarse-grained variables ($ln P_0$) as a function of cluster size ($K$). The solid black line represents a linear least-squares fit of $-ln P_0 = bK^\beta$ in log-log scale, where $\beta = 0.82$ ($R^2=0.99$, $p < 10^{-5}$).
    The gray dashed line indicates the prediction for independent variables. For each $K$, SEM is calculated across coarse-grained variables. Error bars are always smaller than the marker size.  \textbf{C.} Variance (Var) of the non-normalized coarse-grained variables  as a function of the cluster size, $K$. The solid black line corresponds to the linear least-square fit,  $log \ Var(K) = \alpha \cdot logK + b$, where $\alpha = 1.32$ ($R^2=0.99$, $p < 10^{-5}$).
    gray dashed lines indicate reference growth rates: bottom, linear scaling ($\alpha=1$) characteristic of independent  variables; top, quadratic scaling ($\alpha=2$) expected for fully correlated variables. \textbf{D.} Average correlation functions for different cluster sizes, $K$, collapse on a single curve after rescaling time, $\tau$, by the auto-correlation time $\tau_c$. Inset: Original autocorrelation for different $K$.
    \textbf{E.} The autocorrelation time, $\tau_c$, as a function of $K$, grows as $\tau_c \propto K^z$. The solid black line corresponds to the linear least-square fit,  $log \ \tau_c(K) = z \cdot logK + b$, where $z = 0.31$ ($R^2=0.97$, $p < 10^{-5}$). 
    \textbf{F.} Eigenvalues of covariance matrices within clusters  of different   sizes. Larger clusters correspond to darker colors. The solid line is a linear least squares fit to $log \ \lambda = log \ b (rank/K)^{-\mu}$ performed for $K=128$. The fitting range is $[2/128, 50/128]$. Inset: $\lambda$ as a function of non-rescaled ranks. The analysis is performed using $e = \pm 3$SD and $\delta = 1T_s = 1.67$ ms, where $T_s$ is the sampling time (Methods \ref{data}).}}}
    \label{fig_prg_data}
\end{figure*}

To implement this procedure on  MEG recordings, at each step $k$, we calculate the correlation matrix $C_{ij} ^k$ among the $N_k$  variables and uniquely couple them according to its entries, as illustrated in Fig.~\ref{Fig_1} and detailed in Methods \ref{app_prg}. At each step $k$, we first analyze two basic quantities:  the distribution of activity, $Q_K (x)$, and the probability of silence $P_0 (K)$ across the $N_k$ variables (Methods \ref{app_prg}). We find that the distribution of activity rapidly approaches a non-Gaussian, exponential-like fixed form for increasing values of the cluster size $K$ (Fig.~\ref{fig_prg_data}A). This is in agreement with previous observations in populations of spiking neurons \cite{meshulam2019, morales_quasiuniversal_2023}, and indicates that brain activity is close to a non-Gaussian fixed point of the renormalization group, namely the statistics of the activity is independent of the scale of observation (coarse-graining). We note that for weakly dependent or independent variables coarse-graining would lead to a fixed Gaussian distribution.
Similarly,  for a collection of independent variables the probability of silence, $P_0 (K)$,   decays exponentially with $K$ \cite{meshulam2019}. In contrast, we find that in resting-state activity the probability of silence decays as a stretched exponential, $P_0(K) \propto e^{-aK^\beta}$, with $\beta = 0.82 \pm 0.02$ (mean $\pm$ SD; Fig. \ref{fig_prg_data}B, individual subject; Fig. S1B, average over $n = 100$ subjects), in close agreement with previous reports for populations of spiking neurons \cite{meshulam2019,cambrainha_criticality_2024,morales_quasiuniversal_2023}. These behaviors are consistent across subjects (see Supplementary Information, Fig.~S1, for the average over $n = 100$ subjects) and very robust with respect to binarization threshold and temporal binning (Figs.~S2-S4).

To understand whether this fixed point of resting-state brain activity is consistent with criticality, we next iteratively analyzed cluster dynamics as a function of $K$, the cluster size.  Since our coarse-graining procedure simply consists in iteratively summing an increasing number of signals, the mean activity scales linearly with $K$ (see Methods \ref{app_prg}). For independent variables, we also expect  the variance to scale linearly with $K$, since the covariance between  variables is zero. In the opposite scenario of a collection of fully correlated  variables, the variance scales instead as $Var (K) \propto K^2$ (Methods \ref{app_prg}). We find that resting-state activity falls between these two limiting cases (Fig.~\ref{fig_prg_data}C). In fact, we observe that the variance of the non-normalized cluster activity scales as a power-law of $K$, $Var(K) \propto K^{\alpha}$, with $\alpha = 1.32 \pm 0.03$ (mean $\pm$ SD; n = 100). This intermediate value is close to previous observations in populations of neurons \cite{meshulam2019,morales_quasiuniversal_2023,cambrainha_criticality_2024}, and indicates a self-similar structure of the underlying correlations associated to a non-trivial growth of the activity variance. 
The scaling of activity variance is robust and weakly dependent on the binarization threshold and temporal binning used in the analysis, as demonstrated by the analysis across all subjects  (Figs.~S1) and in a range of thresholds and time bins  (Figs.~S3-S4).

If we are near criticality, then we expect that the correlations scale  with the cluster size $K$, as we include more and more sensors, i.e. an  increasing portion of the system. Specifically, this implies that the autocorrelation time of the variables, $\tau_c$, increases with $K$ as a power-law,  i.e. $\tau_c (K) \propto K^z$. 
To test whether this scaling holds in resting-state MEG, we first evaluated the average autocorrelation for each cluster of size $K$, $C_K(\tau)$ (Methods \ref{app_prg}). We observe that the autocorrelation depends on $K$ (Fig.~\ref{fig_prg_data}D, inset), and its decay slows as $K$ increases. However, when we rescale the time by the autocorrelation time, $\tau_c$, calculated at each $K$, all different curves collapse onto the same functional form (Fig.~\ref{fig_prg_data}D). Importantly, we find that $\tau_c$ depends on $K$ according to a power-law with exponent $z = 0.33 \pm 0.03$ (mean $\pm$ SD; Fig.~\ref{fig_prg_data}E, individual subject).  The scaling of $\tau_c (K)$ is robust across subjects (see Supplementary Information, Fig.~S1),   and is  stable with respect to temporal binning and  binarization thresholds (Figs.~S3-S4).
We note that $\tau_c$ tends to decrease with increasing threshold values, in particular for $K > 64$. This behavior must be ascribed to the increasing sparsity of time series induced by higher thresholds, which  likely weakens  the correlations present in the original data. 

Next, we look inside the clusters of size $K$ and analyze their covariance matrix (Methods). If  correlations are self-similar, as our analysis suggests,   then 
the eigenvalues of the covariance matrix and their respective ranks should follow a power-law relationship,  independently of $K$ (Methods \ref{app_prg}). This  requirement is condensed  in Eq. \ref{eq:lambda}, stating that the eigenspectrum of the covariance matrix, i.e. $\lambda(rank)$, has the same form when ranks are rescaled by the cluster size $K$. As a consequence, plotting $\lambda$ as a function of $rank/K$ should result in a collapse of all eigenspectra for different $K$ on the same power-law with exponent $\mu$. This prediction is robustly satisfied by resting-state MEG. In Fig.~\ref{fig_prg_data}F, we show the eigenspectra of the covariance matrix for different $K$ in an individual subject. They all collapse on a single power-law, $\lambda \propto (rank/K)^{-\mu}$, with $\mu = 0.51$. 
The scaling of $\lambda (rank)$  shows  little variability between subjects  (Fig.~S1F; \textcolor{black}{$\mu = 0.46 \pm 0.04$}; mean $\pm$ SD; $n = 100$) and in a range of thresholds (Fig.~S3) and bin sizes (Fig.~S4).

\begin{figure}[t!]
    \centering
    \includegraphics[width=0.5\linewidth]{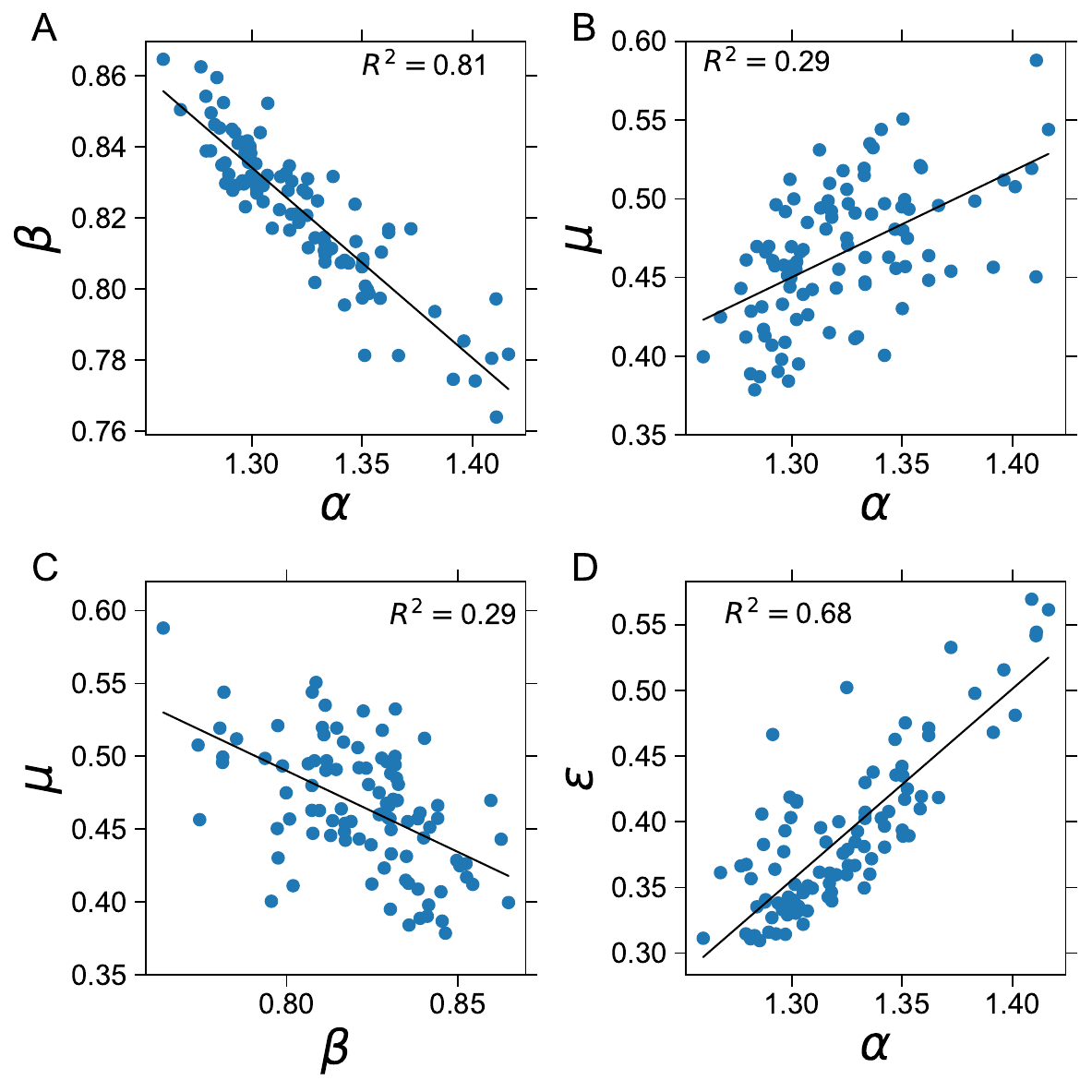}
    \caption{\scriptsize{\textbf{Relationships between scaling exponents in the MEG of the resting state.}
    \textbf{A.} The scaling exponent of the cluster variance is inversely proportional to the exponent $\beta$ that characterizes the probability of silence $P_0 (K)$, i.e. $\beta = -0.54\alpha + 1.53$ (linear least square fit: $R^2 = 0.81$, $p < 10^{-5}$; $n = 98$. Two subjects were excluded based on Cook's distance \cite{Cook_1977}).  \textbf{B.} The exponent $\mu$ characterizing the eigenvalue spectrum of the covariance matrix increases linearly with $\alpha$: $\mu = 0.67\alpha - 0.42$ (linear least square fit: $R^2 = 0.3$, $p < 10^{-5}$; $n=98$). 
       \textbf{C.} As expected from A and  B, the exponent $\mu$ decreases for increasing values of $\beta$. The relationship between $\mu$ and $\beta$ is well described by a linear equation, $\mu = -1.1\beta + 1.38$ (linear least square fit: $R^2 = 0.3$, $p < 10^{-5}$; $n=98$).  \textbf{D.} The scaling exponent of the largest eigenvalue of the covariance matrix, $\epsilon$, increases for increasing values of $\alpha$, as shown by the linear relationship $\epsilon = 1.46\alpha - 1.54$ (linear least square fit: $R^2 = 0.68$, $p < 10^{-5}$; $n = 98$). 
     }}
     \vspace{-0.0cm}
     \label{fig_exp_relation}
\end{figure}

We then asked whether there is any empirical relationship between  scaling exponents characterizing the MEG of the resting-state. 
For example, for independent variables we know that the exponent $\beta$, controlling the scaling of $P_0$ with $K$, must tend to one, and so the exponent $\alpha$ for the variance of cluster activity. When correlations  increase, instead,   $\beta$ is expected to decrease and $\alpha$ to increase. Thus, we can anticipate that $\beta$ and $\alpha$ should be anti-correlated. Plotting $\alpha$ versus $\beta$ for all subjects (Fig.~\ref{fig_exp_relation}A), we find that $\beta$  decreases linearly with $\alpha$, i.e. $\beta = -0.54\cdot\alpha + 1.53$. 
Additionally, we observe that the value of the exponent $\alpha$  correlates with the scaling exponent of the eigenspectrum of the covariance matrix (Fig.~\ref{fig_exp_relation}B): $\mu = 0.61\cdot\alpha -0.39$.  
This relationship indicates that $\mu \to 0$ (no scaling) for uncorrelated variables,  and increases linearly with $\alpha$ over the range of non-trivial values measured in resting-state activity (Fig.~\ref{fig_exp_relation}B). Due to the relationship between $\alpha$ and $\beta$, the exponent $\mu$ must in turn  be correlated with $\beta$, as we indeed find  (Fig.~\ref{fig_exp_relation}C; $\mu = -1.0\cdot\beta + 1.24$).
Next, we consider the largest eigenvalue, $\lambda_1$, of the covariance matrix and measure how it scales with $K$. We  observe that $\lambda_1 (K) \sim K^\epsilon$, and that the exponent $\epsilon$  is related to  the exponent $\alpha$ as $\epsilon \approx \alpha -1$,  
as shown in \cite{castro2024interdependent} (Fig.~\ref{fig_exp_relation}D). Specifically, we find  $\epsilon = 1.46\alpha -1.54$.  
However, outliers analysis shows   that the linear regression between $\alpha$ and $\epsilon$ is strongly influenced by certain subjects \cite{Cook_1977}. When removing those subjects, we obtain $\epsilon = 1.09 \alpha -1.07$ ($R^2 = 0.53$, $p < 10^{-5}$; $n = 91$; errors on the fit parameters estimated via bootstrap: $0.04$ and $0.05$, respectively), very close to theoretical prediction \cite{castro2024interdependent}.


\bigskip

\begin{figure*}[t!]
    \centering
    \includegraphics[width=\linewidth]{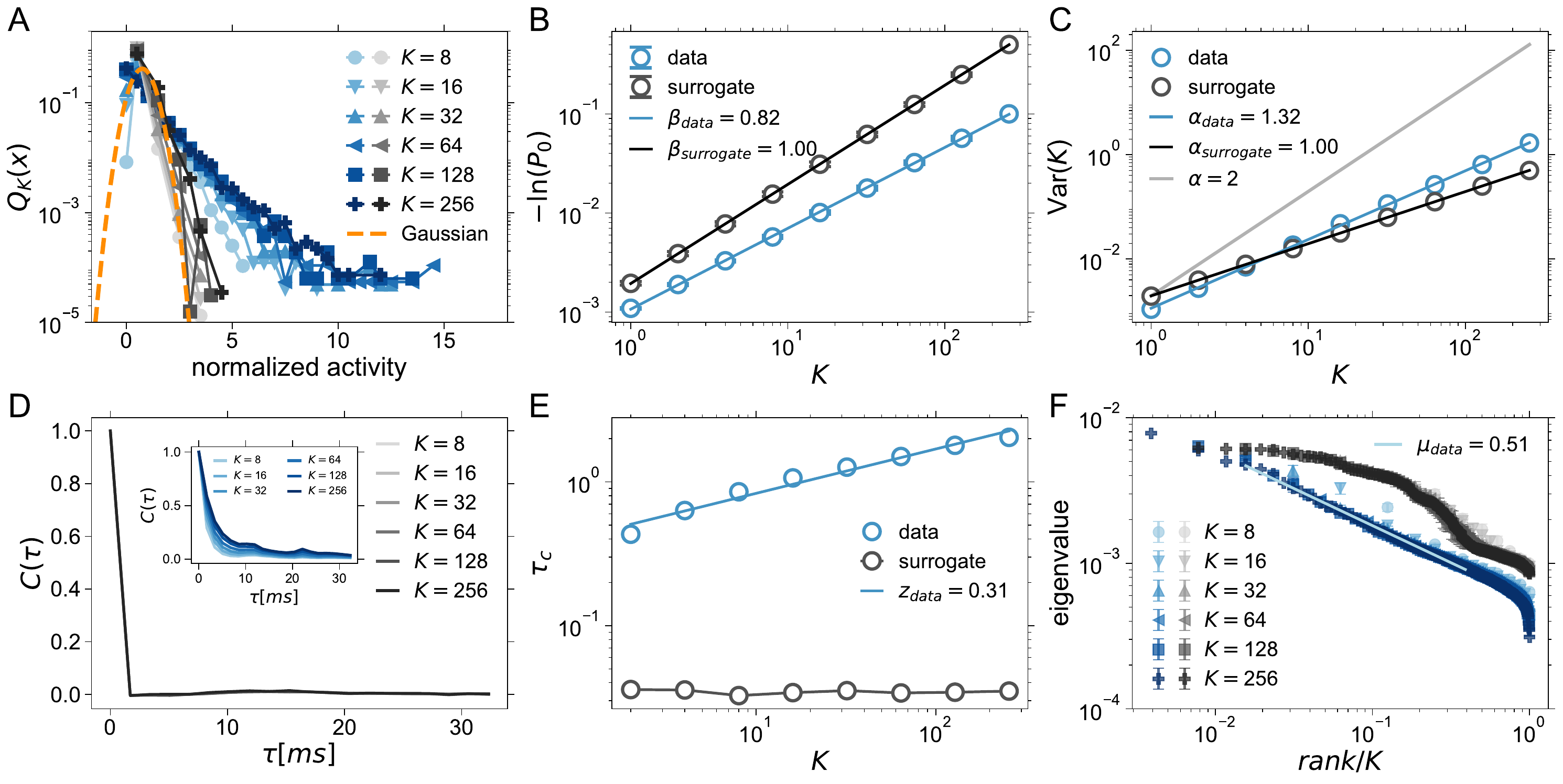}
    \caption{\scriptsize{{\textbf{Absence of scaling in phase randomized MEG signals.}  \textbf{A}. Probability distributions for normalized non-zero activity ($K \geq 8$) approach a Gaussian in phase-randomized data (gray scale). In blue: original data. Darker color = Increasing cluster size.  Only data in the range $(0,15]$ are shown. \textbf{B}. Probability of silence in coarse-grained variables ($-ln \ P_0$) as a function of the cluster size,  $K$. Probability of silence decays as  $P_0(K) = e^{-bK^\beta}$ with $\beta = 1$ in case of phase randomized signals (solid black line), as for independent variables. For the original data, $\beta_{data} = 0.82$ (blue solid line).  
     \textbf{C.} Variance (Var) of the coarse-grained variables  as a function of the cluster size, $K$. The solid  lines correspond to  linear least-square fits,  $log \ Var(K) = \alpha \cdot logK + b$. For phase randomized data $\alpha = 1$ (solid black line), while for original data $\alpha = 1.32$. Gray thick line: $Var(K) \sim K^2$.  \textbf{D}. Average correlation functions for different cluster sizes, $K$, of the phase randomized data. Autocorrelation is independent of $K$ and shows a very fast decay to zero. Inset: Autocorrelation for the original data.
    \textbf{E}. 
    The autocorrelation time, $\tau_c$, is independent of $K$ for the phase randomized data (black circles). For the original data   $\tau_c \propto K^z$. Solid blue line: linear least-square fit,  $log \ \tau_c(K) = z \cdot logK + b$, where $z = 0.31$. 
    \textbf{F.} Eigenvalues of covariance matrices within clusters for different  cluster sizes. For randomized data (gray scale), scaling is disrupted.  Larger clusters correspond to darker colors. Original data are shown in blue. Solid blue line: least squares fit to $\log  \lambda = \log  b (rank/K)^{-\mu}$ performed for $K = 128$. The fitting range is $2/128 - 50/128$. Results are shown for an individual subject. Error bars represent SEM and are calculated across coarse-grained variables for the corresponding $K$. Error bars are always smaller than the marker size.}}
    }
    \label{fig:surrogate}
\end{figure*}

 Taken together, the non-Gaussian form of the distributions of activity and the scaling behaviors observed  in all analyzed quantities indicate that MEG resting-state is at (or near) a non-Gaussian fixed point of the   renormalization group transformation applied.  
 To support this evidence, we performed three different surrogate tests. First, we performed the analysis on  MEG signals with randomized phases (Methods \ref{surrogate}). Phase randomization preserves autocorrelation of individual signals while destroying cross-correlations among signals. Analysis of phase-randomized data (Fig.~\ref{fig:surrogate}; see Fig.~S6 for the average over n=100 subjects) shows a (nearly) Gaussian distribution of activity---activity that is also strongly suppressed---, trivial scaling for the probability of silence ($\beta = 1.0003 \pm 0.0001$; mean $\pm$ SEM; $n = 100$) and for the variance ($\alpha = 0.9995 \pm 0.0001$; mean $\pm$ SEM; $n = 100$), and no scaling of the correlation time  $\tau_c$ ($z \approx 0$), with fast decaying autocorrelation independent of $K$ (Fig.~\ref{fig:surrogate}D-E). As for the eigenspectra, we observe absence of clear scaling compared with original data (Fig.~\ref{fig:surrogate}F), with the exception of a very short regime for high rank eigenvalues. Furthermore, the relationship between  $\alpha$ ($\beta$)  and $\epsilon$ is not satisfied (see SI, Fig.~S5). 

 Second, we scrumble signal amplitudes in such a way to destroy the autocorrelation of individual signals while preserving their cross-correlations (trace randomization; see Methods \ref{surrogate}). In this case,  all analyzed quantities behave as in the original data (observed differences are not significant: $p>0.9$, $n = 100$), except for the autocorrelation of variables, which becomes weaker and does not scale with $K$ (Fig.~S7). This indicates that autocorrelation of individual signals influences the scaling of $\tau_c$, while all other quantities  depend mainly on the collective correlation structure.

   Third, we keep the signals unchanged, but pair variables randomly over all steps of the procedure (random coarse-graining; Methods \ref{surrogate}). In this case, we observe both deviations from scaling and, in particular, from data collapse in comparison to original data (Fig.~S8). The distributions of cluster activity tend to stay non-Gaussian, but they are not consistent over coarse-graining steps and seem not to converge to a fixed functional form, although we cannot exclude it for $K \to \infty$---because of the limited number of sensors. Deviations from scaling occur at large cluster sizes, $K > 64$, in the probability of silence ($-lnP_0$) and in the  variance (Fig.~S8B, C). Concerning the eigenspectra of the covariance matrices, we observe clear deviations from the data collapse compared to the original, non-random coarse-graining procedure, although the eigenspectra  still follow (approximate)  power-law behaviors (Fig.~S8F). This implies a $K$-dependent exponent  $\mu$. 

\bigskip

Surrogate tests indicate that, altogether, robust scaling behaviors and data collapses in resting-state MEG result from the intrinsic structure of the underlying correlations. Random coarse-graining might not fully destroy the scaling features of the original data because  correlations among  MEG sensors are  non-negligible over large areas, i.e.  they are non-local and decay rather slowly with distance \cite{shriki13}. Summing the activity of most correlated sensors to generate coarse-grained variables is equivalent to grouping  nearest neighbors in systems with local interactions (Fig.~\ref{Fig_1}), as in the block renormalization technique \cite{kadanoff_box}. If correlations were strictly local, random coarse-graining would produce the same effects of phase randomization (Fig.~\ref{fig:surrogate}). However, since in the MEG sensor array correlations are rather long-range, the effect of random pairing becomes apparent only when the full structure of the correlations kicks in: (i) in the co-activation of sensors, implied in $Q_K(x)$, (ii) in the scaling of cluster activity at large cluster sizes $K$, (iii) in the scaling of the eigenspectrum of the covariance matrix, which depends on $K$ and breaks down at low ranks (long wavelength).

\bigskip

Before turning our attention to dynamics of neuronal avalanches under coarse-graining, we discuss the effect of the variable normalization and signal discretization choice on the non-Gaussian  distributions and scaling behaviors of resting-state brain activity. We have shown that the results presented in Fig.~\ref{fig_prg_data} are robust in a wide range of  thresholds (Fig.~S2-S3). We only considered  threshold $e> 2.7$ SD, where the distribution of signal amplitudes  deviates from a Gaussian \cite{shriki13} (see Methods \ref{data}), and took both positive ($x > +e$) and negative extremes ($x < -e$). First, we notice that results are stable when only positive or negative extremes are considered (SI, Table I).  If we lower the binarization threshold  to progressively include an increasing number of signal points (and noise), we observe that the distribution of activity tends to a Gaussian for increasing $K$ (Fig.~S9). However, scaling is preserved in the probability of silence, activity variance, and (partially) in the eigenspectra, but not in the autocorrelation time (Fig.~S9). 
In the limit of continuous signals (variables), we observe that the distribution of activity deviates from a Gaussian and shows long tails (positive and negative) before coarse-graining and for a number of coarse-graining steps (Fig.~S10)---nearly independently of $K$---and converges to its Gaussian core for large $K$. At the same time, we observe no scaling in the silence probability and in the autocorrelation time, while the variance shows marked deviations from scaling for $K \geq 32$  
and the eigenspectrum of the covariance matrix shows limited scaling (also dependent on $K$)  (Fig.~S10).


An alternative to selecting only extremes of the above-threshold excursions is represented by taking all points belonging to them. Also in this case, we found robust, fixed non-Gaussian distribution of activity and non-trivial scaling in all analyzed quantities (Fig.~S11), although with different exponents (see SI, Table II-III for a comparison between the two methods). Importantly, the relations between scaling exponents are similar to those found when considering  extreme events only (compare Fig.~S12 with Fig.~\ref{fig_exp_relation}). Moreover, in this case scaling behaviors tend to be well preserved for low discretization thresholds ($e = 0.5$SD), i.e., when more Gaussian noise is included (Fig.~S13). \\
Finally, we note that  repeating the analysis on non-normalized variables---i.e. without normalizing variables at each step $k$ as explained in Methods \ref{app_prg}---gives comparable results (Fig.~S14).


\subsection{Scaling of neuronal avalanches is invariant under coarse-graining}

The coarse-graining analysis suggests that resting-state brain activity is close to a fixed point of the dynamics characterized by scale invariance, i.e. criticality. Prior  evidence of brain tuning to criticality in resting MEG was provided  by power-law distributions of neuronal  avalanches and long-range temporal correlations \cite{shriki13,fl2020_lrtc,fosque_quasi_crit_2022}. Here we ask  how neuronal avalanche dynamics evolve under the coarse-graining procedure studied above, and show that they are invariant. 

\begin{figure}[t!]
    \centering
    \includegraphics[width=0.45\linewidth]{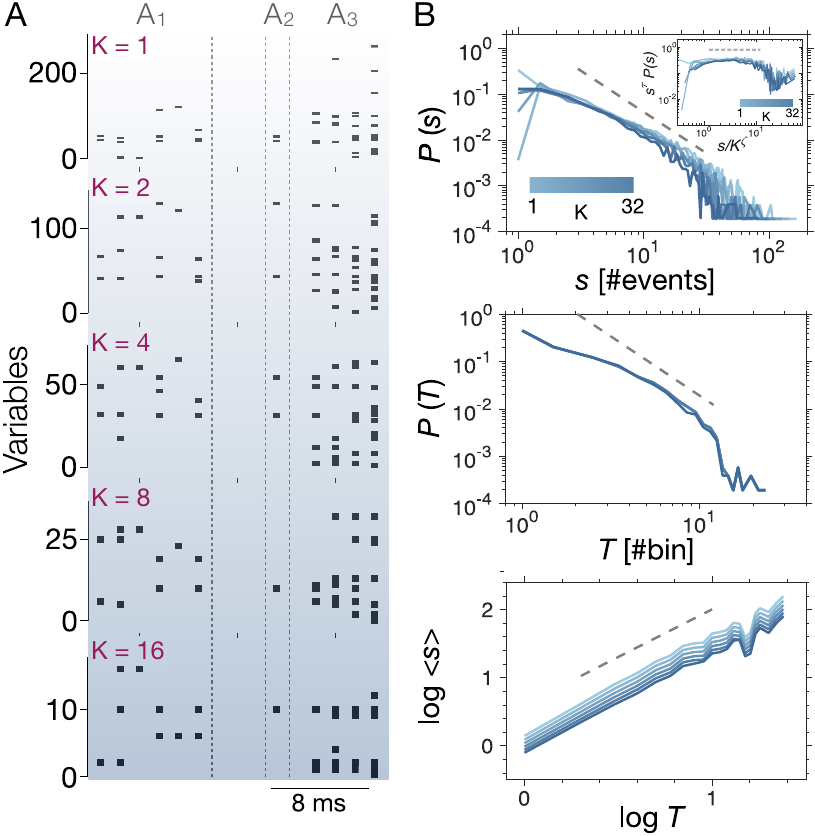}
    \caption{\scriptsize{\textcolor{black}{\textbf{Invariant avalanche scaling  under coarse-graining.} \textbf{A.} Evolution of neuronal avalanches across coarse-graining steps (left). 
    At each step, a reduced set of coarse-grained variables are obtained, from the initial 273 sensors ($K = 1$, top) to 17 variables for $K = 16$ (bottom).  Neuronal avalanches are defined as consecutive sequences of events across sensors an time bins (delimited by dashed lines), bounded by periods with no events. Avalanche size $s$ is defined as the number events (blue dots) across all sensors during the lifetime of the avalanche; duration $T$ is the number of consecutive time bins belonging to the avalanche times the bin size. \textbf{B.} The distribution of avalanche sizes, $P(S)$, and avalanche durations, $P(T)$, do not change over consecutive coarse-graining steps (top and middle). Inset: the plot of $s^\tau_s P(s)$ versus $s/K^\zeta$ shows that the  
    distributions $P(s)$ for  different cluster sizes, $K$,  approximately collapse onto a single curve  for $\tau = 1.3$ and $\zeta = 0.2$. The exponent $\tau$ obtained from the data collapse is slightly smaller than the value obtained from  maximum likelihood estimates over the entire range of avalanche sizes. 
    The scaling relation between $s$ and $T$ is also invariant under coarse-graining (bottom). The analysis is performed on an individual subject (same as Fig.~\ref{fig_prg_data}) using $e = \pm 3$SD and $\delta = 1T_s = 1.67$ ms, with $T_s$ the sampling time (Methods \ref{data}). Results are consistent across subjects}}
    }
    \label{fig:avalanche_data}
\end{figure}

In Fig.~\ref{fig:avalanche_data}A, we show the evolution of neuronal avalanches over four coarse-graining steps, from $K = 2$ to $K = 16$; the upper row, $K = 1$, corresponds to the raster plot with avalanches obtained from the original MEG signals. Neuronal avalanches are labeled as $A_i$ and defined as  sequences  of time bins continuously populated with at least one extreme event (blue dots) in any of the MEG sensors or variables (Methods \ref{avalanches}). As we proceed with  coarse-graining, we notice that the number of variables decreases, as well as the number of extreme events (blue dots), whereas the time windows occupied by the avalanches (i.e. the durations) remain unchanged (Fig.~\ref{fig:avalanche_data}A). To quantify the dynamics of avalanches across coarse-graining steps, we plot the distributions of avalanche sizes and durations as a function of $K$ (Fig.~\ref{fig:avalanche_data}B). We observe that the largest avalanche size decreases as  $K$ increases (Fig.~\ref{fig:avalanche_data}B, top). However, the scaling regime of the distribution (approximately two decades) remains nearly unchanged  and is followed by an exponential cutoff whose onset in turn depends on $K$ .  This suggests that the distribution of avalanche sizes for different $K$  follows the  scaling form $P(s) = s^{-\tau}f(s/K^\zeta)$, where $\tau_s$ is the power-law exponent and $f(s/K^\zeta)$  a scaling function, the exponent $\zeta$ capturing the dependence of the exponential cutoff on $K$. By plotting $s^\tau_s P(s)$ versus $s/K^\zeta$, we found that distributions of avalanche sizes for different cluster sizes, $K$, collapse onto a single curve, confirming that the finite-size scaling ansatz, $P(s) = s^{-\tau}f(s/K^\zeta)$, is  valid \textcolor{black}{(Fig.~\ref{fig:avalanche_data}B; $\zeta = 0.2$)}. This indicates that the scaling regime of $P(s)$ is stable across coarse-graining steps. 

Similarly, the scaling behavior of  the duration distribution is independent of $K$  (Fig.~\ref{fig:avalanche_data}B, middle and Fig.~S15). Furthermore, coarse-graining  preserves the scaling relation between avalanche sizes and durations, $\langle s \rangle \propto T^\gamma$, with  the exponent $\gamma$ remaining nearly unchanged over coarse-graining steps (Fig.~\ref{fig:avalanche_data}B, bottom and Fig.~S15). 
Because the number of variables progressively decreases as  $K$ increases, the cutoff of the power-law regime shifts towards smaller $s$. In contrast, durations are almost unaffected by coarse-graining, i.e. the  temporal localization  of avalanches is nearly fixed. As a consequence, $\langle s \rangle (T)$ is vertically shifted, but its scaling behavior remains stable (Fig.~\ref{fig:avalanche_data}B, bottom). Maximum likelihood estimates of $\tau_s$ showed a slight increase with $K$ (Fig.~S15).
On the other hand,  $\gamma$ and the avalanche duration exponent, $\alpha_t$, remained nearly unchanged (Fig.~S15).


\subsection{\textcolor{black}{Inhibitory neural feedback modulates scaling exponents near criticality}}

To understand the emergence of scaling in resting-state brain activity in relation to neural dynamics and criticality,  we simulated resting-state activity using a recently proposed interpretable neural network model~\cite{FL-2023}. This model  has a genuine non-equilibrium critical point and reproduces both local, individual sensor dynamics---including neural oscillations---, and collective scale-free neuronal avalanches. Importantly, its analytical tractability  enables us to make direct contact with MEG data  and infer its two free parameters (Methods \ref{model}).

The model consists of a population of $N$ interacting neurons whose dynamics is self-regulated by a time-varying feedback  that depends on the ongoing population activity level. Neurons are all excitatory, and are simulated as binary units  $s_i = \pm 1$ ($i = 1,2,...,N$, with $N=273 \cdot 10^3$, unless specified differently)  that are active when $s_i = + 1$ or inactive when $s_i = -1$. Here we consider the simplest, fully  homogeneous scenario in which neurons interact with each other through  synapses of equal strength  $J_{ij} = J = 1$ (see Methods \ref{model} for details).  The ongoing network  activity is defined as $m(t) = \frac{1}{N}\sum_{i=1} ^N s_i (t)$,  
and each neuron experiences a uniform negative feedback $h$ that depends on $m$ as $\dot{h} = -cm$, with $c$ a positive constant. The negative feedback can be identified with a combination of  inhibitory drives from the neuronal  population that  affect  individual neurons with a delay given by the characteristic time $1/c$.  
Neurons $s_i$ are stochastically activated according to Glauber dynamics, where the new state of the neuron $s_i$ is drawn from the marginal Boltzmann-Gibbs distribution $P(s_i) \propto \exp(\beta_T \tilde{h}_is_i)$, with $\tilde{h}_i = \sum_{j\neq i} J_{ij} s_j + h$, where $\beta_T$ is a parameter reminiscent of the inverse temperature for an Ising model. 
For our simulations, we  consider an all-to-all network connectivity, i.e. $J_{ij} = 1, \forall{i,j}$ and $i \neq j$. Network behavior is determined by the two parameters $c$ and  $\beta_T$, which control feedback strength and proximity to criticality, respectively. For $c>0$, the model is driven out of equilibrium and has a critical point at $ \beta_T =\beta_c =1 $ ~\cite{ddemartino_jphys_2019,FL-2023}. Slightly below  the critical point, the model exhibits both  neuronal avalanches and oscillations, characteristics of the MEG signals \cite{FL-2023}.  

To estimate the parameters $\beta_T$ and $c$, we fit the analytical form of the model autocorrelation to the  MEG signal autocorrelation (Methods \ref{model}). Parameters inferred across all sensors and subjects suggested  baseline values of $\beta_T = 0.99$ and $c = 0.01$ \cite{FL-2023}. 
 Then, we make contact with the data by parceling our simulated network into $M = 273$ equally-sized disjoint subsystems of $n_{\rm sub} = N/M = 1000$ neurons each, and consider each subsystem activity $m_\mu$, $\mu=1,\dots,M$, as the equivalent of a single MEG sensor signal. 
All quantities for the model then follow the same definition as for the data, allowing us to perform direct side-by-side comparisons with data (Methods \ref{model}). \textcolor{black}{Threshold $e$ and time bin $\delta$ used for model analyses were determined by approximately matching the network activity rate, $\langle A \rangle = \sum_{t=1} ^T \sum_{s=1} ^N x_s(t)/T$, to the empirical activity rate (Methods).} 
Extensive robustness analysis confirms that  model behaviors  are stable in a range of thresholds (Fig.~S16, S17), and  we detail them  one by one in the following.

\begin{figure*}[t!]
    \centering
    \includegraphics[width=1\linewidth] {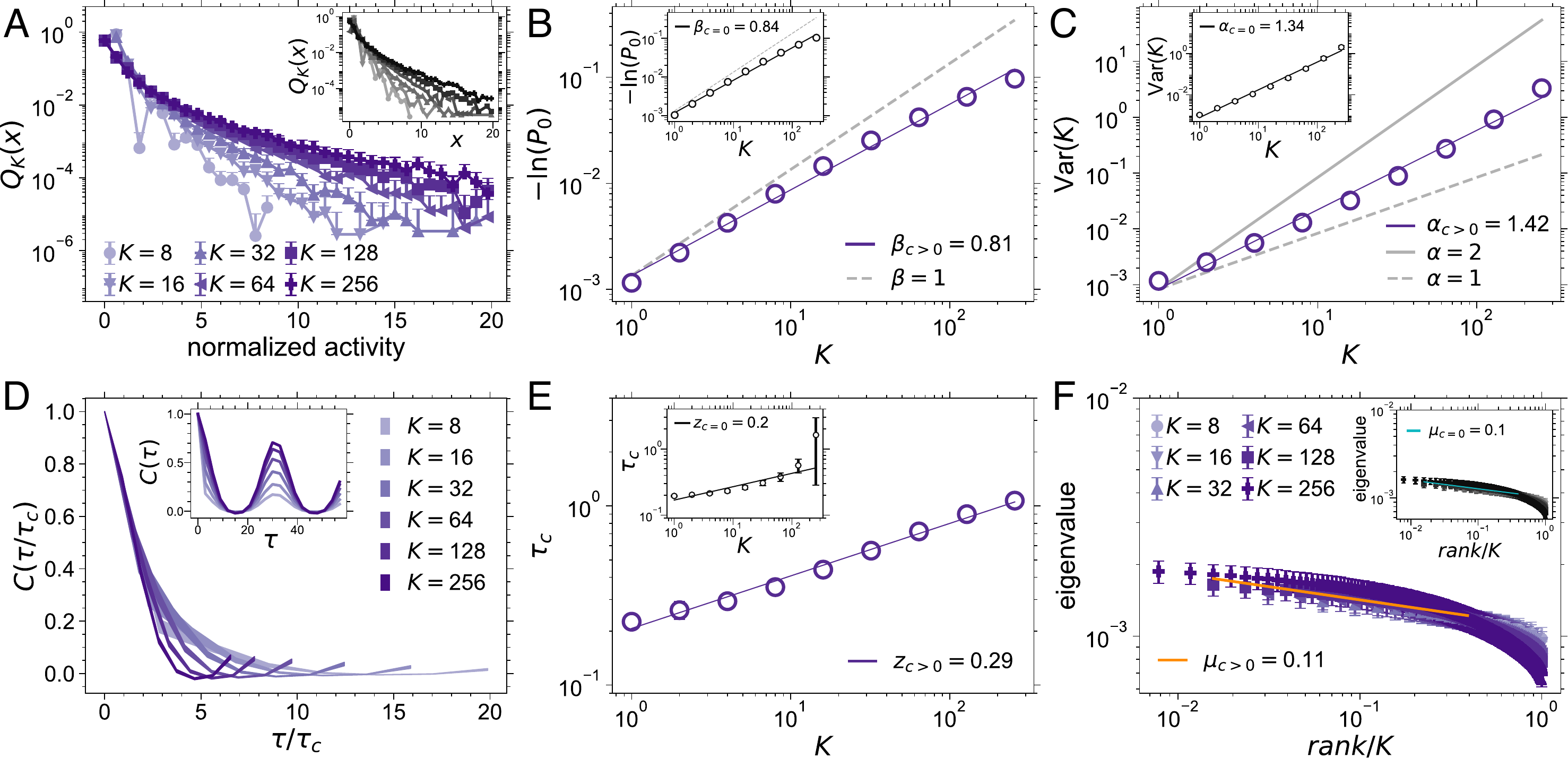}
    \caption{\scriptsize{\textcolor{black}{\textbf{Scaling behaviors emerge near criticality and are modulated by the neural inhibitory feedback.}  \textbf{A.} Probability distributions of normalized non-zero activity for cluster sizes $K \geq 8$ in the model with feedback ($c = 0.01$). Inset: $Q_K(x)$ in the model  without feedback ($c=0$). Only data in the range $(0,20]$ are shown. \textbf{B.} Log-probability of silence in coarse-grained variables ($-\ln P_0$) as a function of cluster size ($K$).  Solid line: linear least-squares fits of $log (-\ln P_0) = \beta_{model} \cdot logK + b$ ($\beta_{c > 0} =   0.809 \pm 0.014$; mean $\pm$ SD). Gray dashed line:  prediction for independent variables. Inset: Log-probability of silence for  $c=0$ ($\beta_{c = 0} = 0.840 \pm 0.009$).  \textbf{C.} Variance of non-normalized coarse-grained variables as a function of $K$.  Solid  line: linear least-squares fit, $log \ Var(K) = \alpha \cdot logK + b$ ($\alpha_{model} = 1.417 \pm 0.032 $). Gray dashed and thick lines indicate reference growth rates: $\alpha = 1$ for independent variables and $\alpha = 2$ for fully correlated variables.  Inset: variance in the model with $c=0$ ($\alpha_{c = 0} = 1.340 \pm 0.019$). \textbf{D.} Rescaled average autocorrelation functions for different cluster sizes ($K$) in the model with $c > 0$. 
    Inset: Autocorrelation functions before rescaling. \textbf{E.} The autocorrelation time, $\tau_c$, increases with the cluster size as $\tau_c \propto K^{z}$, with $z_{c>0} = 0.290 \pm 0.110$.  Solid line: linear least-squares fit, $\log \tau_c(K) = z \cdot \log K + b$. Inset: the exponent for the model with $c=0$ is significantly smaller ($z_{c=0} = 0.190 \pm 0.040$). \textbf{F.} Eigenvalues of covariance matrices for different   $K$ in  the model with $c > 0$ and $c = 0$ (inset). Solid  lines: least squares fit to $log \ \lambda = -\mu \cdot log \ (rank/K) + b$ performed for $K=128$ ($\mu_{c>0} = 0.111 \pm 0.007$; $\mu_{c=0} = 0.098 \pm 0.003$). Fitting range: $[2/128, 50/128]$. Simulations ($n=10$): fully connected network of $273000$ neurons with $\beta_T=0.9999$;  $N_s = 273$ subsystems  each including $1000$ neurons. Error bars and shaded areas represent SD, and where not shown are smaller than the symbol size. All differences between the model with $c=0.01$ and $c=0$ are significant ($p < 10^{-3}$). 
    The analysis is performed using $e = \pm 3.2$SD ($c>0$) and $e = 3.15$SD ($c=0$)   with $\delta = 3dt$.
    }}
    }
    \label{fig_prg_model}
\end{figure*}

First, the distribution of activity, $Q_K (x)$, tends to a non-Gaussian fixed form for increasing cluster size $K$ (Fig. \ref{fig_prg_model}A), and the probability of silence scales non-trivially with $K$, i.e. $P_0(K) = e^{-aK^{\beta}}$, with $\beta \approx 0.8$ in agreement with the empirical value (Fig. \ref{fig_prg_model}B). Similarly, and in line with the relation between exponents found in empirical data,  the activity variance grows as a power-law with $K$, with an exponent $\alpha \approx 1.4$, close to the empirical value (Fig. \ref{fig_prg_model}C). 
As in  the data, also in the model we observe that temporal correlations scale with the cluster size (Fig. \ref{fig_prg_model}D), and the auto-correlation time, $\tau_c$,  increases  as a power-law, $\tau_c \propto K^z$, with  $z \approx 0.3$ (Fig. \ref{fig_prg_model}E). Turning to the eigenspectrum of the covariance matrix, the model shows a scaling regime with an exponent $\mu \approx 0.1$, which is approximately independent of the cluster size $K$, in agreement with the prediction $\lambda \propto (rank/K)^{-\mu}$ (Fig. \ref{fig_prg_model}F). We note that  $\alpha$  is  close but slightly larger than the  corresponding empirical value ($1.417 \pm 0.032$ vs $1.32 \pm 0.03$), while $\mu$ is significantly smaller ($\mu$: $0.111 \pm 0.007$ vs $0.41 \pm 0.03$). \textcolor{black}{Despite these differences, the exponents $\beta$ and $\alpha$ and the largest eigenvalue, $\epsilon$, obey scaling relationships similar to those observed in the data (Fig.~S18) and predicted in \cite{castro2024interdependent}. Specifically, we  find $\beta = -0.46\alpha + 1.46$ and $\epsilon = 1.07\alpha - 1.05$ (Fig.~S18).}
Scaling behaviors show little  variability for  bin sizes $2dt \leq \delta \leq 5dt$---mostly around the largest cluster size---,  where $dt$ is the simulation time and corresponds to a  system sweep of Monte Carlo updates (Fig.~S17). \textcolor{black}{Importantly, we observe that  scaling laws of neuronal avalanches in the model are also invariant under coarse-graining (Fig.~S19). In particular, the distribution of avalanche sizes depends on $K$ as $P(s) = s^{-\tau}f(s/K^\zeta)$, with $\zeta = 0.2$, as found in empirical data (Fig.~S19).}

\textcolor{black}{To investigate the role of the negative feedback on scaling behaviors, we first turn it off by  setting $c = 0$, while keeping the network very close to criticality. Absence of  negative feedback effectively makes the network fully excitatory.}  
\textcolor{black}{In this scenario, we observe a mild yet significant modulation of all exponents, in particular $\beta$, $\alpha$ and $z$ (Fig.~\ref{fig_prg_model}). Specifically,  $\beta$ increases from $0.809 \pm 0.014$ for $c>0$ to $0.840 \pm 0.009$ for $c=0$,  while $\alpha$ decreases from $1.417 \pm 0.032$ to $1.340 \pm 0.019$ (mean $\pm$ SD; $p < 10^{-4}$).  
The dynamical exponent, $z$, decreases from $0.29 \pm 0.11$ in the model with feedback   to $0.19 \pm 0.04$ in absence of feedback ($c=0$).}   
\textcolor{black}{To further assess whether the observed differences were actually due to absence of feedback, we performed a permutation test (Methods), which confirmed that all exponents obtained for $c=0$  were significantly different from those measured in the network with feedback, $c>0$ ($p < 10^{-3}$).} 

Overall, this indicates that  feedback plays a sizeable role in reproducing the scaling behaviors of  resting-state brain activity in humans. In its simplest configuration, the feedback represents an inhibitory drive that   influences  neurons on a certain time scale controlled by $c$ (Methods \ref{model}). Absence of feedback mimics a scenario without  inhibition, where neural dynamics is solely controlled by excitatory neuronal interactions ($J_{ij}$). The opposite scenario, i.e. neural activity driven only by feedback (and noise), is obtained by setting $J_{ij} = 0$, e.g. non-interacting neurons. In this case, the distribution of activity approaches a Gaussian, and all other quantities show  trivial, i.e. consistent with independent variables, or no scaling (Fig.~S20). \textcolor{black}{Importantly, we observe  a similar tendency towards scaling exponents consistent with a weakly correlated scenario for $J_{ij} \neq 0$  in the limit $c \gg 0.01$ (Fig.~S21, S22),  
despite the proximity to the critical point. Such a  de-correlating effect of large negative feedback is in line with effects of increased inhibition \textit{in vivo}, e.g. scaling of power spectrum. 
}

\bigskip

\begin{figure}[t]
    \centering
    \includegraphics[width=0.4\linewidth]{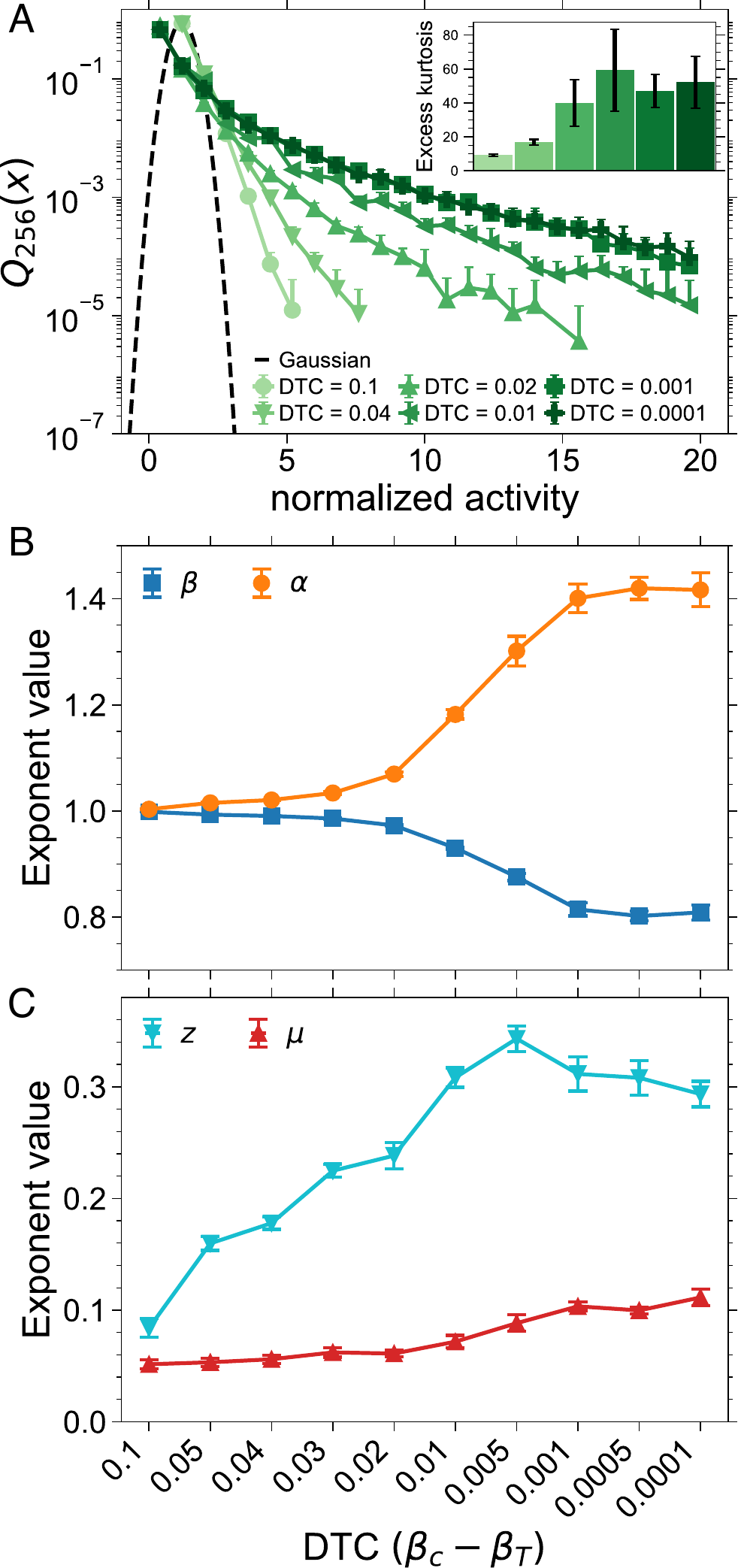}
    \caption{\scriptsize{\textcolor{black}{\textbf{Distribution of activity and scaling exponents as a function of the distance to criticality.} \textbf{A.} Distribution of activity for  $K = 256$  and  increasing values of distance to criticality, DTC $= \beta_c - \beta_T$. The distribution is stable (non-Gaussian) in a narrow range of DTC $\leq 0.02$, and rapidly tend to a Gaussian for DTC $> 0.02$. A similar behavior is observed for smaller cluster size, $K$ (Fig.~S23). Inset: kurtosis of the distributions $Q(x)$ as a function of DTC. Only data in the range $(0,20]$ are shown. \textbf{B.} The exponents controlling the scaling of silence and non-normalized activity variance, $\beta$ and $\alpha$, rapidly approach 1 for increasing DTC (top). The exponent $z$ and $\mu$ both decrease for increasing DTC, and rapidly drop for DTC $> 0.01$ (bottom). Simulations ($n=10$): fully connected network of $273000$ neurons with $c = 0.01$;  $N_s = 273$ subsystems  each including $1000$ neurons. On each data point, error bars represent standard deviation.}}}
    \label{fig_exp_beta}
\end{figure}

The inferred model is very  close to its critical point. A longstanding  question about the emergence of collective behaviors in  brain activity  concerns the dynamical regime in which they can all coexist \cite{fl2023_natbrief,disanto_2018_pnas}. Is this a localized or an extended region of the parameter's space? If, as our empirical results suggest, this dynamical regime corresponds to criticality, how close to it does one need to be in order to observe  scaling in cluster activity? And how do exponents depend on the distance to criticality? To address this question, we investigate the behavior of the  model for increasing distance to criticality (DTC), defined as DTC $ = \beta_c -\beta_T$. We find that both the distribution of activity  and the scaling of all analyzed quantities  depend on DTC, and are not-trivial in a (rather) small DTC range (Fig.~\ref{fig_exp_beta}). In Fig.~\ref{fig_exp_beta}A, we show the distribution of activity in clusters of size $K = 256$ for a range of DTC (see Fig.~S23 for other cluster sizes). We observe that the distribution of activity remains stable and non-Gaussian in a narrow range of DTC comprised between $0.0001$ and $0.02$. The distribution becomes increasingly sensitive to DTC and rapidly approaches a Gaussian behavior for DTC $> 0.04$ (Fig.~\ref{fig_exp_beta}A). \textcolor{black}{We note that the kurtosis of $Q_K(x)$ is much larger than the Gaussian kurtosis, and fluctuates around fifty for DTC $< 0.01$, while rapidly decreases for DTC $> 0.01$ (Fig.~\ref{fig_exp_beta}).}
Similarly, the exponents $\beta$ and $\alpha$ rapidly approach one as DTC increases, and are equal to one for DTC $= 0.1$  (Fig.~\ref{fig_exp_beta}). In parallel, the autocorrelation of the variables  drops markedly  for DTC $> 0.01$ and becomes nearly independent of the cluster $K$ for  DTC $> 0.1$  (Fig.~\ref{fig_exp_beta}). Furthermore,   the eigenspectrum of the covariance matrix shows power-law scaling on a narrow range of DTC, with $\mu \approx 0.1$, approaching zero (flat eigenspectrum) for DTC $ \geq 0.1$ (Fig.~\ref{fig_exp_beta}B). These results suggest that one needs to stay rather close to criticality to observe the non-trivial scaling behaviors observed in resting-state MEG.

\bigskip

Before summarizing our findings, a word of caution is needed to avoid misinterpretation of the results and approach we have taken here. 
 First, in equilibrium critical phenomena, critical exponents have precise values and are well defined at the critical point of the phase transition, $\beta_T = \beta_c$ \cite{stan:pt}. 
 Deviations from scaling and theoretical estimate of  exponents are typically associated with finite size effects. 
 In the brain, instead, we face a rather different scenario. First, we are out of equilibrium. Second, we still do not know exactly whether there is a ``unique'' critical point or a critical region \cite{munoz_griffiths_2013} and, if so, what its nature is and what exponents we should expect to measure---including potential crossover between exponent values that may occur out of equilibrium \cite{Tauber_2014}.
 Biological considerations and empirical evidence suggest a dynamical region within which the brain operates across physiological and functional states, and thus not necessarily presence of scaling behaviors only in a specific point and complete absence everywhere else.  Last but importantly, we took a phenomenological approach to renormalization of neural data \cite{meshulam2019}. From the modeling perspective---an out-of-equilibrium one---, it is then  reasonable to ask  whether and how the quantities we track  respond to proximity to criticality. 
 Once interesting (non-trivial) scaling features are identified near criticality, the role of biology (which may map in some way into RG symmetry and dimensionality arguments) in determining exponents becomes of interest, and we interrogated the model about that.
 
 \textcolor{black}{Second, the adaptive Ising model on a fully-connected network with uniform feedback recapitulated  most of the MEG scaling behaviors, with the important exception of the exponent $\mu$, which we found to be significantly smaller than the empirical value at all DTC. The motivation for working with this simplified structure came from (i) the analytical tractability, which allows to identify a specific parameter region directly from the data, and (ii) from evidence of a wide spectrum of  collective behaviors observed in MEG data \cite{FL-2023}. However, this also implies a simplified structure of the sensor covariance matrix, with a large  eigenvalue (not shown) and a nearly flat ($\mu \approx 0.1$) eigenvalue spectrum (Fig.~\ref{fig_prg_model}). On the other hand, the (current) lack of analytical results makes it  difficult to precisely assess the distance to criticality beyond the all-to-all structure. As a consequence, in a heterogeneous or even 2D nearest-neighbours scenario, we would not be able to reliably decouple effects on scaling behaviors due to proximity to the critical point---we do not know the exact value of $\beta_c$, which may  also depend on the feedback strength---from  those related to the network structure. Our  current understanding suggests that a certain level of network heterogeneity would significantly improve the agreement with the empirical exponent $\mu$, but would not have a crucial effects on other exponents for this class of adaptive Ising models (see Fig.~S24 for results on a small-world network and Fig.~S25 for 2D lattice with nearest-neighbors interactions), assuming same DTC. 
 }



\section{Discussion}

We have shown that resting-state brain activity in humans is close to a (non-trivial) fixed point of a phenomenological renormalization group (PRG). Activity in coarse-grained clusters approaches a fixed non-Gaussian distribution as their sizes  increase, while the probability of silence $P_0$ (no activity) decreases sub-exponentially. 
 The  variance and autocorrelation of cluster activity  both scale as a power-law of the cluster size, indicating a scale invariant structure of correlations in resting-state brain activity. This was further  confirmed by the size-independent scaling of the eigenspectra of the cluster covariance matrices. In addition, we demonstrated that  the scaling features of neuronal avalanches are invariant under the adopted coarse-graining scheme. 
 Our neural network model, with its two parameters inferred from data, suggests that empirical scaling exponents depend on the balance of excitation and inhibition in the network, paving the way for  robust assessment  of this key quantity in neuroscience research.

\subsection{Scaling in collective neural activity: from populations of neurons to fields} Evidence of a  non-Gaussian fixed point in collective neural dynamics has recently been reported for populations of neurons in the hippocampus \cite{meshulam2019,predrag_sf_2026}, and in different regions of the mouse cortex \cite{morales_quasiuniversal_2023,cambrainha_criticality_2024},  with population activity showing consistent scaling features and signatures of non-Gaussianity. However, such reports were limited to the mouse  brain and to relatively small spatial scales. \textcolor{black}{Complementarily, scaling in spatial and temporal correlations has been demonstrated for populations of neurons in the rat cortex \cite{camargo_PRE_2023}}.   

In humans, access to neural population activity is still limited, and most of neuroscience research is   based on non-invasive electrophysiological recordings, such as electro- and magneto-encephalography. Both measures can be seen as an effective coarse-graining of neural activity, i.e. fields to which large populations of neurons contribute \cite{pfurt99,stam2005nonlin}. Our analysis shows that these fields, although originating from filtered and  highly non-linear superposition of neural/synaptic activity \cite{stam2005nonlin,deste10}, are collectively described  by scaling laws that are remarkably similar---and in some  cases comparable---to those observed in small populations of neurons.  Despite the fact that we could not directly compare these findings  with activity of spiking neurons in the human brain,   our analysis indicates that collective  behaviors of small- and large-scale neural activity are consistent and point to a non-trivial fixed point with common features. Recent analysis of BOLD fMRI in humans \cite{ponce-alvarez_critical_2023,castro2024interdependent} reported scaling exponents close to ours for the probability of silence ($\beta$) and for the activity variance ($\alpha$),  and a  relation between them similar to the one we measured in the MEG \cite{castro2024interdependent}. 
We also found  robust scaling in the eigenspectra of the covariance matrices, with an exponent $\mu$ that correlates with the scaling exponents $\beta$ and $\alpha$. However, a close comparison with fMRI is difficult due, for instance, to the specific choice in the discretization threshold \cite{ponce-alvarez_critical_2023,castro2024interdependent}.

That the MEG signals  may carry over the same scaling  properties of (local) populations of spiking neurons  and, to some extent, even the same scaling exponents measured in a different species, is rather remarkable. However, our analysis indicates that simultaneous recording of spiking activity and fields, e.g. local field potentials, is needed to verify whether exponents are preserved across scales (and recorded signals) and species.  In our approach, we consider only the extremes of all signal fluctuations above a certain threshold, and find that scaling behaviors are robust and persist from very low to very high values of the threshold, with an almost-complete  decimation of the signals. While this could indicate that extreme events carry most  information about underlying neural activity, we still observed scaling in all  quantities  when including in the analysis all the points belonging to the above-threshold excursions.  However, all exponents changed significantly. In particular, the exponents $\beta$ and $z$ decreased, while  $\alpha$ and $\mu$ increased (SI, Table II-III). 
Then, in the limit of continuous signals (Fig.~S10), i.e. $e \to 0$,  we found that, for increasing $K$, the cluster  activity converges to the Gaussian core of the original signal amplitude distribution, and all other analyzed quantities show no scaling, except for the activity variance that grows as $\approx K^2$ over a range of $K$ values (Fig.~S10).

Overall, this indeed suggests that the relevant information about underlying neural dynamics is carried by high amplitude fluctuations. In contrast, small amplitude fluctuations within  the Gaussian core of the signal amplitude distribution are mostly `noise'. To a large extent, coarse-graining seems to effectively `integrate out' this noise, scaling behaviors being very robust even when a consistent part of that noise is included in the analysis (see Figs.~S9, S13). However, scaling breaks down when all small fluctuations are included in the analysis (Fig.~S10).

\subsection{Coarse-graining and avalanche criticality}

The fields captured by MEG sensors across the scalp may have a significant  overlap, and thus MEG signals may be highly correlated, even over relatively long distances \cite{shriki13}. Locally in particular, fields captured by nearest-neighbour sensors tend to have an important overlap. For this reason, in a correlation-based coarse-graining of the activity most of the sensors that are coupled, if not all, are likely to be  spatially close. Thus, one can expect that the $K$-clusters nucleate and grow in a localized area of the scalp until they are merged (i.e., their respective activity is summed) with other clusters that nucleated over a different area to form the few large clusters of the last steps of the PRG. If this intuition is correct, at least to some extent,  then a coarse-graining strategy based on spatial proximity  may provide similar results.  

In \cite{shriki13}, Shriki et al. proposed  a coarse-graining of MEG sensor activity based on spatial proximity to study the stability of neuronal avalanche dynamics. They showed that, under such coarse-graining, the scaling behavior of avalanches was preserved. This is in line with our results on avalanche dynamics across coarse-graining steps (Fig.~\ref{fig:avalanche_data}) and suggests that, for MEG data, the two procedures may be equivalent (at least to some extent).

 The relation between scaling in coarse-grained brain activity and  typical measures of brain criticality (or near-criticality) based on avalanche dynamics remain to be studied, and is not clear whether they provide a complementary picture of brain criticality. Our analysis shows that the difference  $\big |\frac{\alpha_t -1}{\tau_s - 1} - \gamma \big |$  tends to zero with increasing cluster sizes (Fig.~S15), indicating that  the scaling relation  $\frac{\alpha_t -1}{\tau_s - 1} = \gamma $ is satisfied. This suggests in turn that the system moves towards criticality \cite{fontanele_2019_prl}. 
 Further, numerical simulations of our network  model, although limited to a simplified scenario, suggest that avalanche criticality and scaling behaviors in resting-state activity  coexist in a narrow subcritical region  \cite{FL-2023}, as also suggested in \cite{Calvo_2026}. However, analysis of dynamic quantities such as neuronal avalanches may require rescaling of time to properly assess scaling features (and their invariance) near criticality, as in dynamical RG \cite{Tauber_2014}. 

\subsection{E/I balance and scaling behavior in the resting-state} The balance between excitation and inhibition is widely considered  key  for brain function, and has been linked to brain tuning to criticality \cite{shew09,simoes2025maximumentropymodels,silvia_2013_bio,ponce_alv_iScience_26}. Yet, how to reliably assess the E/I ratio from non-invasive, large scale electrophysiological recordings remains an open challenge. Computational models indicate that the E/I ratio is connected  with the  exponent of the power-law decay of the spectral power \cite{fl2017chaos,gao_ei_2017}. Our numerical results suggest that the E/I ratio may also modulate the scaling behaviors of resting-state activity under coarse-graining. The   adaptive Ising model, in which  negative feedback mimics the effect of an inhibitory population on the network \cite{FL-2023}, showed significant modulations in scaling exponents compared to the model without feedback, in particular $\alpha$ and $z$. 
\textcolor{black}{For  strong feedback in particular, scaling exponents were driven toward their trivial values and the distribution of activity toward a Gaussian behavior (Figs.~S21, S22).}
\textcolor{black}{Although this is a simplification of the true E/I balance involving distinct excitatory and inhibitory populations, it suggests that alterations of the E/I balance could be assessed by measuring one of the RG scaling exponents}, which would generally be  easier and more robust than  estimating the scaling exponent of  (often noisy) power spectra, where the scaling regime  may not be unambiguously   identifiable and crossover between different power-law regimes may exist \cite{he_2014_scale_free}.

\textcolor{black}{We note that in the absence of feedback, the network is at equilibrium and corresponds to a ferromagnetic Ising model that operates  very close to (but below) the mean-field critical point $\beta_c = 1$. In this regime, the magnetization fluctuations are  non-Gaussian due to close proximity to the transition, and correlations between the subsystems that mimic MEG sensors are not negligible. As a consequence, the distribution $Q(x)$ remains non-Gaussian, as in the off equilibrium case. In line with that, the activity variance and the probability of silence scale non-trivially with $K$. On the contrary, when moving sufficiently away from the critical point, $Q(x)$ approaches a Gaussian and all other  quantities scale trivially, both with and without feedback. This is line with results on other off-equilibrium models, which show that  (sufficiently) away from the critical state the distribution of activity becomes Gaussian   and all scaling exponents approach trivial values \cite{nicoletti_scaling_2020,Nascimento_2026}---as we also observed in the off-equilibrium case ($c > 0$). However, within a 10\% increase of the control parameter, which would correspond to  $\beta_T = 0.9$ in our model, the analysis in \cite{nicoletti_scaling_2020} provided non-trivial scaling in most quantities, but a  Gaussian distribution of activity (in the momentum space); in our model at $\beta_T = 0.9$ instead, we found that most quantities do approach (and are consistent with) a trivial scaling behavior.}

Recently, scaling behaviors in neural activity have been suggested to emerge due to a slowly varying external drive \cite{morrell_latent_2021,Mariani_2022}, as could be interpreted the  time-varying global feedback and/or the  all-to-all connectivity. However, we note that (i) the negative feedback is endogenous rather than external; (ii) non-homogeneous feedbacks are possible and do not disrupt scaling behaviors; (iii) we find similar scaling behaviors   in 2D networks with nearest-neighbors interactions (Fig.~S25). Furthermore, in the non-interacting, feedback-driven case, we find trivial scaling in activity variance and  $P_0$, and no scaling in the other quantities  (Fig.~S20).

\subsection{Network dynamics, topology, and phenomenological renormalization}

Although the model satisfactorily recapitulates most of the empirical  results, we openly acknowledge two quantitative shortfalls: (i) the distribution of cluster activity (non-Gaussian) deviates from the empirical distribution, showing higher probabilities of large cluster activity;  
(ii) the scaling exponent $\mu$ of the eigenspectrum is significantly lower than the empirical value. Both these deviations from empirical data may depend  on the over-simplified spatial structure of the model. We considered the simplest realization of the adaptive Ising class: uniform,  all-to-all excitatory (ferromagnetic) connectivity and inhibitory drive (negative feedback). In this configuration, we are able to fit the two model parameters from data by matching analytical and empirical auto-correlations, and thus  locate empirical data in the phase diagram of the model \cite{FL-2023}. \textcolor{black}{However, the underlying network  is far from  realistic. Furthermore, the comparison of collective behaviors between model and MEG data  involves building  subsystems  of a certain size (number of neurons) to mimic the signals captured by MEG sensors, and fixing two additional parameters specific of the MEG data analysis, i.e., temporal bin and discretization threshold \cite{fl2020_lrtc,FL-2023}. These subsystems were disjoint and equally sized, creating an effective regular lattice with  $M$ units over which we performed  coarse-graining of network activity. This  setting may affect the correlation structure and, in turn, the eigenspectrum of the covariance matrix and the collective activity over large clusters $K$, and also limit our assessment of the influence of network connectivity on scaling exponents.}  Despite these valid points of concern,  this simple model shows  remarkable predictive power on the scaling  behaviors of the resting human brain, while simultaneously reproducing  avalanches and oscillations near criticality \cite{FL-2023}. 

Future work should systematically investigate the role of network connectivity (if any) in scaling behaviors and  exponents, including more realistic topologies,  inhibitory synapses, and heterogeneous synaptic strength. \textcolor{black}{Our small-world simulations show scaling behaviors similar to those observed in the fully-connected network (mean-field), but a better (qualitative)  agreement of $\mu$ and $Q(x)$ with empirical data   (Fig.~S24). This is in line with recent analysis in non-equilibrium models \cite{nicoletti_scaling_2020} and   
in different areas of the mouse cortex \cite{morales_quasiuniversal_2023}, both suggesting a weak dependence on network topologies or long-range connections. Yet, a precise analysis of the effect of network heterogeneity and long-range connections in the adaptive Ising class   would require knowing precisely the critical temperature and its dependence on the feedback strength, $c$---both currently not available. Despite all this, it seems plausible, as other models also suggest \cite{Hu_spectrum_cov_2022}, that these exponents depend on the qualitative nature of the connectivity and thus on network excitation-inhibition tuning, as observed for the scaling exponents of activity correlations, power spectra, and neuronal avalanches \cite{palva13,fl2016,fl2017chaos,shew09,oren18}.} \textcolor{black}{From a more theoretical perspective, the model is suitable for a formal RG analysis that would provide a set of scaling exponents and scaling relationships, and allow to investigate the connection with the phenomenological approach to renormalization followed in this work.} 


\begin{acknowledgments}
FL acknowledges support  from the European Union's Horizon research and innovation program under the Marie Sklodowska-Curie Grant Agreement No. 101066790. FL and IT acknowledge support from the program TAlent in ReSearch@University of Padua -- STARS@UNIPD  (project  BRAINCIP---Brain criticality and information processing).
\end{acknowledgments}

\section*{Data and code availability}

Data analyzed in this study were collected at the MEG facility of the NIH for a previously published study \cite{shriki13}. Data belongs to NIH and are available from O.S. (shrikio@bgu.ac.il) on reasonable request. Codes used in the current study are publicly available on GitHub.  Data analysis:   \href{https://github.com/lffrrnt/CoarseGraining}{https://github.com/lffrrnt/CoarseGraining}. \\ 
Model: \href{https://github.com/demartid/stat_mod_ada_nn}{https://github.com/demartid/stat\textunderscore mod\textunderscore ada\textunderscore nn}.

\section{Methods}

\subsection{Data acquisition and pre-processing}
\label{data}
Resting-state brain activity was recorded from 100 healthy  participants in the MEG core facility at the NIMH (Bethesda, MD, USA) for a duration of 4 min (eyes closed). All experiments were carried out in accordance with NIH guidelines for human subjects. The sampling rate was 600 Hz, and the data were band-pass filtered between 1 and 150 Hz. Power-line interferences were removed using a 60 Hz notch filter designed in Matlab (Mathworks). The sensor array consisted of 275 axial first-order gradiometers. Two dysfunctional sensors were removed, leaving 273 sensors in the analysis. Analysis was performed directly on the axial gradiometer waveforms. The data analyzed here were selected from a set of MEG recordings for a previously published study \cite{shriki13}, where further details can be found.

For each sensor,  positive and negative excursions beyond a threshold $e = \pm 3$SD were identified.  
In each excursion beyond the threshold, a single event was identified  at the most extreme value (maximum for positive excursions and minimum for negative excursions). For each signal, this resulted  in a time series of number of events per time bin, including zeros, indicating absence of events in the bin.    Comparison of the signal distribution with  the best  Gaussian  fit indicates that the two distributions start to deviate from one another around $\pm 2.7$SD \cite{shriki13}. Thus, thresholds smaller than $\pm 2.7$SD would lead to the detection of many events related to noise in addition to real events whereas much larger thresholds would miss many of the real events. To avoid noise-related events while preserving most of the relevant events, in this study  the threshold $e$ was set at $\pm 3.0$SD. The raster of identified events was binned at a temporal resolution $\delta = 1.67$, the sampling frequency, and  used for the PRG analysis. Stability of results with respect to  thresholds, bin sizes, and definition of events is analyzed in Figs.~S2-S4.

\subsection{Phenomenological Renormalization Group}
\label{app_prg}

\subsubsection{Coarse-graining of brain activity}

To work in the same data format that is used for the  analysis of neuronal avalanches in MEG data, we performed coarse-graining on the binarized MEG signals, i.e. on the extreme events that are used to identify avalanches \cite{shriki13,fl2020_lrtc}. At each step $k$ of the PRG, which includes $N_k$ variables $x_i$ ($i = 1, 2, ..., N_k$), the entries of correlation coefficient matrix $c_{ij}$ are calculated as
\begin{equation}
    c_{ij} = \frac{\langle \delta x_i \delta x_j \rangle} { [\langle \delta x_i^2 \rangle \langle \delta x_j^2 \rangle]^{1/2} },
\end{equation}
where $\delta x_i = x - \langle x_i \rangle $. Then the most  correlated pair $(i,j)$ is identified by finding the largest non-diagonal element in the matrix,  and the variables $x_i$ and $x_j$ are summed to  define the new coarse-grained variable 
\begin{equation}
    x_i^{(K)} = Z_i^{(K)} (x_i ^{(K/2)} + x_j ^{(K/2)}),
\end{equation}
where $Z_i^{(K)} = 1/\langle x_i ^{(K)} \rangle_ {x_i \neq 0}$ is a normalization constant and $K = 2^k$ indicates the  cluster size at the step $k$, that is the number of original sensor signals that contribute to the variables $x_i ^{(K)}$ at the step $k$ of the procedure. The notation $\langle \rangle_ {x_i \neq 0}$ indicates average over non-zero values.  
Normalization is such that the average number of extreme events in non-empty time bins is equal to one across all steps of the coarse-graining procedure.
 Each variable can only participate in one pair. 
 
 Thus,  the pair $(i,j)$ is removed  and the next most  correlated pair is identified until the original $N_k$ variables have been grouped into $N_k/2$ pairs. This procedure is iterated until $N_k = 1$, namely only a single variable is left. At each step $k$ of the procedure, coarse-grained variables are the result of an iterative clustering of $K = 2^k$ original sensors signals.

 \subsubsection{Distribution of activity and probability of silence in coarse-grained variables} The distribution of individual coarse-grained variables, $P_K(x)$, can be split into two contributions. The first one, $Q_K (x)$, is the probability of finding $x \neq 0$  extreme events in non-empty time bins, i.e. time bin in which one detects at least one extreme event in the time series of coarse-grained variables  resulting from clustering $K$ MEG signals at the step $k$ of the coarse-graining procedure. The second, $P_0 (K)$, is the probability of finding a time bin with no extreme events (silence). The behavior of these two contributions is analyzed as a function of $K$ (Fig.~\ref{fig_prg_data}). 

\subsubsection{Mean activity and activity variance}
The mean activity of coarse-grained variables, $x_i ^{(K)}$, is 
\begin{equation}
    \langle x^{(K)} \rangle = \frac{1}{N_k} \sum_{i = 1} ^{N_k} \langle  x_i ^{(K)}\rangle = K \langle x^{(1)}\rangle,
\end{equation}
and grows linearly with the mean of the original signals,  $\langle x^{(1)}\rangle$. The variance of the non-normalized activity for cluster of size $K$ is defined as 
\begin{equation}
    Var (K) = \frac{1}{N_k} \sum_{i = 1} ^{N_k} \Bigl[ \langle  (x_i ^{(K)})^2\rangle - \langle x_i ^{(K)}\rangle^2 \Bigr].
\end{equation}
For independent variables the variance grows linearly with the cluster size $K$, i.e. $Var(K) \propto K$. In the opposite case of fully correlated variables, the variance would grow quadratically with $K$, i.e. $Var(K) \propto K^2$, because of the maximally positive covariance between all variables. Scaling of the variance with $K$ as $Var(K) \propto K^\alpha$ with an intermediate value of the exponent alpha is indicative of a self-similar structure in the correlations, expected at criticality,   and  thus of a non-trivial fixed point in the renormalization  group. Variance is calculated on non-normalized variables.

\subsubsection{Autocorrelation $C(\tau)$ and correlation time $\tau_c$}

The autocorrelation of activity in clusters of size $K$ is defined as  
\begin{equation}
    C^{(K)}_i (\tau) = \frac{\langle \delta x_i^{(K)}(t_0) \delta x_i^{(K)}(t_0 + \tau) \rangle} {Var(x_i^{(K)})}.
\end{equation}
The correlation time, $\tau_c$, is defined as the characteristic time of the exponential fit, $y (t) = A \cdot e^{-t/\tau_c}$, to $C_K(\tau)$, the average autocorrelation for the variables \{$x_i ^ {(K)}$\}. Under the hypothesis of dynamic scaling occurring near criticality, correlations scale with the cluster size $K$, and grow as a power-law of $K$, $\tau_c(K) \propto K^z$, with $z > 0$.  
The exponential fit was performed in the range $t \in [0,\tau_{max}]$. For empirical data, we set $\tau_{max} = 5$, where $C(\tau)$  approaches zero for most $K$. Increasing $\tau_{max}$ does not significantly affect $\tau_c$ and the scaling exponent $z$. For numerical data ($c>0$) in Fig.~\ref{fig_prg_model}, a double exponential fit of the form $C(\tau) = Ae^{-\tau/\tau_c} + Be^{-\tau/\tau_2}cos(\omega t)$, with $\tau_{max} \in [25,29]$, and $\tau_c$ corresponds to the faster time scale. 
In this case, $\tau_{max} \in [5,10]$ provided a good data collapse on the initial decay of $C(\tau)$ over different $K$, and a stable scaling exponent $z$. For  $K$ and  $\tau_{max} > 10$, due to increasing oscillatory trends, the initial exponential decay is followed by a slower decay of the autocorrelation, which can be described by an additional exponential term that modulates a sinusoidal function. In all cases, $\tau_c$ refers to the time scale of  the initial decay of the autocorrelation. In Figs.~\ref{fig_prg_model}, S24, and S25, a double exponential fit was used, and $\tau_c$ corresponds to the faster time scale. 


\subsubsection{Eigenspectrum of the covariance matrix}

For each cluster size $K$, the eigenvalues of the corresponding covariance matrix are calculated as follows. First, 
the original sensors that contributed to each cluster of size $K$ are retrieved. Then, the corresponding covariance matrix is calculated from the original binary activity (extreme events) of the sensors and the eigenspectrum obtained for each individual cluster. The representative eigenspectrum  is the average over the $N_k$ eigenspectra for the clusters of size $K$. The eigenvalues, $\lambda$, of the covariance matrix are ordered from the highest to the lowest. At a non-trivial fixed  point of the renormalization group, as the one associated to criticality, one expects $\lambda (rank) \propto 1/rank^\mu$ \cite{meshulam2019,Bradde_2017}. The higher eigenvalues correspond to large wavelength (and low rank) and should `feel' the system size, whereas smaller eigenvalues correspond to short wavelength (higher rank) and should not depend on the system size. As a consequence, for different cluster size $K$, one should have
\begin{equation}
    \lambda  \propto \Big(\frac{rank}{K}\Big)^{-\mu}.
    \label{eq:lambda}
\end{equation}
The Eq.\ref{eq:lambda} implies: (i) the scaling of $\lambda$ is independent of $K$; (ii) when the rank is rescaled by $K$ one observes a data collapse, i.e. all curves $\lambda (rank/K)$ fall onto a single one (see Fig.~\ref{fig_prg_data}F). The exponent $\mu$ in Eq.~\ref{eq:lambda} was estimated in data and model  using the same number of time bins, which provided  the same ratio $T/N$ between the number of time points ($T$) and sensors ($N$).

\subsubsection{Neuronal avalanches}
\label{avalanches}

For $K = 1$ (no coarse-graining, original signals), an avalanche is defined as a continuous sequence of time bins, $\delta$, in which there is at least an event on any sensor, ending with at least a time bin with no events. The size of an avalanche, $s$, is given by the number of events in the avalanche; its duration, $T$, is given by the number of time bins belonging to the avalanche times $\delta$.  For further details see \cite{shriki13,fl2020_lrtc}. Avalanche analysis was performed on the original signals and in the first $4$ PRG steps ($K = 2, 4, 8, 16$), which retain a sufficient number of variables ($N_2=136$, $N_4=68$, $N_8=34 $, $N_{16}=17$) for  avalanche statistics.  For the coarse-grained variables ($K \geq 2$), an event is defined as any non-zero activity in the time series. Then, avalanches follow the same definition as for non-coarse-grained signals ($K = 1$). 


\subsection{Surrogate data}
\label{surrogate}
\subsubsection{Surrogate MEG signals}
\paragraph*{Phase randomization.} Random phase shuffling of the original continuous MEG signals is performed for each sensor independently. A Fourier transform of each sensor signal is performed, the corresponding phases are randomized while amplitudes are preserved. The surrogate signals are then obtained by performing an inverse Fourier transform. The random phase shuffling destroys phase synchronization across cortical sites while preserving the linear properties of the original signals, such as power spectral density and two-point correlations \cite{theiler92}.

\paragraph*{Trace randomization.} Trace shuffling  is performed by applying a unique random permutation of the time points to  all sensor signals. In this way, one destroys signal autocorrelations but preserves  cross-correlations between MEG sensors. Together with phase randomization, this surrogate test  allows one   to disentangle the contribution of the sensor cross-correlation structure and local temporal dynamics to the observed collective behaviors.

\subsubsection{Random coarse graining.} Sensors are randomly paired  without considering their actual cross-correlations. This test is performed for a single subject with $100$ independent realizations of the random sensor pairing procedure. The full PRG analysis is repeated for each realization, and results are averaged across the $100$ realizations. This test allows one to evaluate to what extent  the observed results rely specifically on the actual sensor-to-sensor correlation hierarchy.  
\textcolor{black}{\subsection{Permutation test}
For each exponent in Fig.~\ref{fig_prg_model}, the 20 estimates were pooled and randomly reassigned to two groups of 10, and the two-sample $t$-statistic was recalculated over 10000 random permutations. The observed $t$-statistics were compared with the resulting null distributions, yielding permutation $p$-values of $1.0\times10^{-4}$ for $\beta$, $2.0\times10^{-4}$ for $\alpha$, $2.0\times10^{-4}$ for $\mu$, and $1.0\times10^{-4}$ for $z$.}

\subsection{Further details on the Adaptive Ising model}
\label{model}
The model is composed of a collection of $N$ spins $s_i = \pm 1$ ($i = 1,2,...,N$) that  interact with each other with a coupling strength $J_{ij}$. In our analysis, the  $N$ spins  represent excitatory neurons that are active when $s_i = + 1$ or inactive when $s_i = -1$, and $J_{ij} > 0$. Furthermore, we consider the fully  homogeneous scenario, with neurons interacting  with each other through  synapses of equal strength $J_{ij} = J = 1$.   The $s_i$ are stochastically activated according to the Glauber dynamics, where the state of a neuron is drawn from the marginal Boltzmann-Gibbs distribution 
\begin{equation}
P(s_i) \propto \exp(\beta \tilde{h}_is_i) \quad \tilde{h}_i = \sum_j J_{ij} s_j + h_i
\end{equation}
Spins  experience an external field $h$, a negative feedback that depends on network activity according  to the following equation, 
\begin{equation}
\dot{h}_i = -c \frac{1}{|\mathcal{N}_i|}\sum_{j\in \mathcal{N}_i}  ^{|N_i|} s_j,
\label{eq:feed}
\end{equation}
where $c$ is a constant that controls the feedback strength, and the sum runs over a neighborhood of the neuron $i$ specified by $\mathcal{N}_i$; index $j$ enumerates over all the elements of this neighborhood. Depending on the choice of $\mathcal{N}_i$, the feedback may depend on the activity of the neuron $i$ itself (self-feedback), its nearest neighbors, or the entire network---the case we considered in the main text. In a more realistic setting including both excitatory ($J_{ij} > 0$) and inhibitory neurons ($J_{ij} < 0$), one could then take into account the different structural and functional properties of excitatory and inhibitory neurons by considering different interaction and feedback properties \cite{bonifazi2009gabae}.
In our simulations, one  time step, $dt$,  corresponds to one system sweep---i.e. $N$ spin flips---of Monte Carlo updates, and Eq~(\ref{eq:feed}) is integrated using  $\Delta t = 1/N$. Note that this choice of timescales for deterministic vs stochastic dynamic is important, as it interpolates between the quasi-equilibrium regime where spins fully equilibrate with respect to the field $h$, and the regime where the field is updated by feedback after each spin-flip and so spins can constantly remain out of equilibrium.    $\Delta t$ is generally much smaller than the characteristic time of the adaptive feedback that is controlled by the parameter $c$. Model data analysis followed the same procedure illustrated for empirical data.
\textcolor{black}{\paragraph*{Small-world network.} The small-world adaptive Ising model (Fig.~S24)  was constructed from a square lattice with $N = 5120 \times 5120 = 26.2 \times 10^6$ spins, nearest-neighbour ferromagnetic couplings ($J = 1$, coordination number $z = 4$), and periodic boundary conditions. Each of the $2N$  bonds was  rewired with probability $p=0.279$. One endpoint was replaced by a uniformly chosen random site, excluding self-loops and duplicate bonds. This keeps the number of bonds and the mean coordination number $z = 4$ fixed while adding long-range links. The  model without feedback has an equilibrium phase transition whose critical temperature dependence on $p$ is known numerically \cite{herrero_SW_ising}. Numerical simulations indicate that the location of the critical temperature for the corresponding non-equilibrium phase transition  does not strongly change for the (small) feedback strength relevant for this work, as also confirmed previously for both the mean field \cite{FL-2023} and 2D model  \cite{sinelshchikov_feedback_2d}.}

\bibliography{bib_file_all,prg}

\clearpage

\onecolumngrid
\appendix

\centering{\Large {\bf Supplementary Information}}

\vspace{4cm}

\renewcommand{\figurename}{Fig. S\hspace{-0.11cm}} 

\setcounter{figure}{0}

\clearpage

\section*{Supplementary Figures}



\begin{figure*}[ht]
    \centering
    \includegraphics[width=\linewidth]{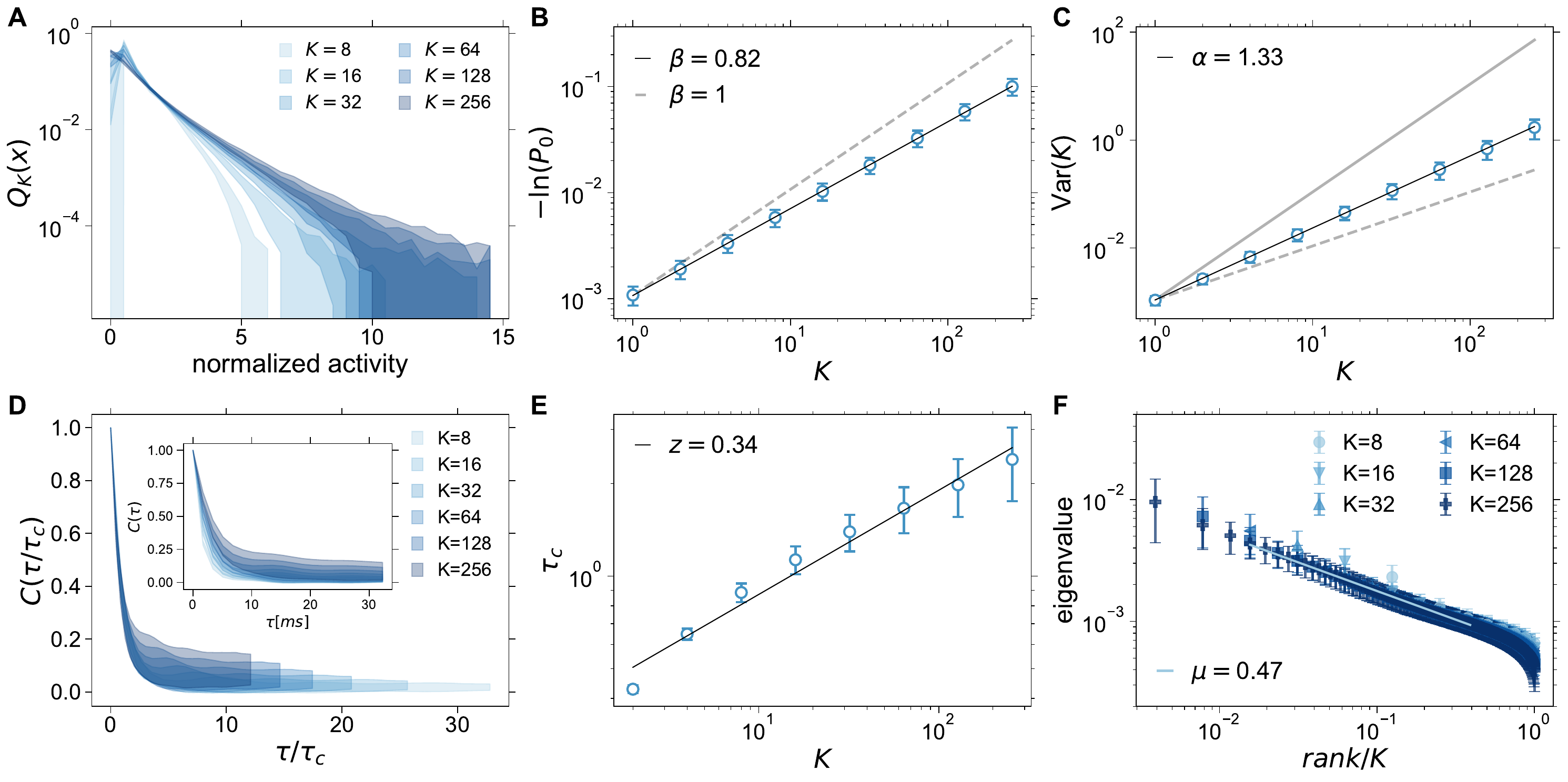}
    \caption{\footnotesize{\textbf{PRG analysis of resting state brain activity shows little variability across subjects.}
    \textbf{A.} Average probability distribution for normalized nonzero activity for the cluster sizes $K \geq 8$. Darker blue = Increasing cluster size. Only data in the range $(0,15]$ are shown. \textbf{B.} Average log-probability of silence in coarse-grained variables ($ln P_0$) as a function of cluster size, $K$ (n = 100). The solid black line represents a linear least-squares fit of $log(-ln P_0) = log\cdot K^\beta$ + b, where $\beta = 0.82 \pm 0.002$ (fit $\pm$ error on the fit) 
    The gray dashed line indicates prediction for independent variables. \textbf{C.} Average variance (Var) of the non-normalized coarse-grained variables across n = $100$ subjects as a function of the cluster size, $K$. The solid black line corresponds to the linear least-square fit,  $log \ Var(K) = \alpha \cdot logK + b$, where $\alpha = 1.33 \pm 0.005$ (fit $\pm$ error on the fit). The gray dashed lines indicate reference growth rates: linear scaling ($\alpha=1$) characteristic of independent  variables and quadratic scaling ($\alpha=2$) expected for fully correlated variables. \textbf{D.} Average correlation functions (n = $100$) for different cluster sizes, $K$, collapse on a single curve after rescaling time, $\tau$, by the correlation time $\tau_c$. Inset: Original autocorrelation for different $K$.  
    \textbf{E.} The autocorrelation time, $\tau_c$, as a function of $K$, grows as $\tau_c \propto K^z$. The solid black line corresponds to the linear least-square fit,  $log \ \tau_c(K) = z \cdot logK + b$, where $z = 0.34 \pm 0.02$ (fit $\pm$ error on the fit).  
    \textbf{F.} Eigenvalues of cluster covariance matrices  for different  cluster sizes. Results are averaged over $n = 100$ subjects. Larger clusters correspond to darker colors. The solid line is a least squares fit to $log \ \lambda = log \ b (rank/K)^{-\mu}$ performed for $K=128$, \textcolor{black}{, yielding $\mu = 0.47 \pm 0.004$ (fit $\pm$ error on the fit)}. The fitting range is $2/128 - 50/128$. All results are averaged over  $n = 100$ subjects. Error bars and shaded areas represent standard deviation (SD).}}
\end{figure*}

\clearpage

\begin{figure*}[ht]
    \centering
    \includegraphics[width=\linewidth]{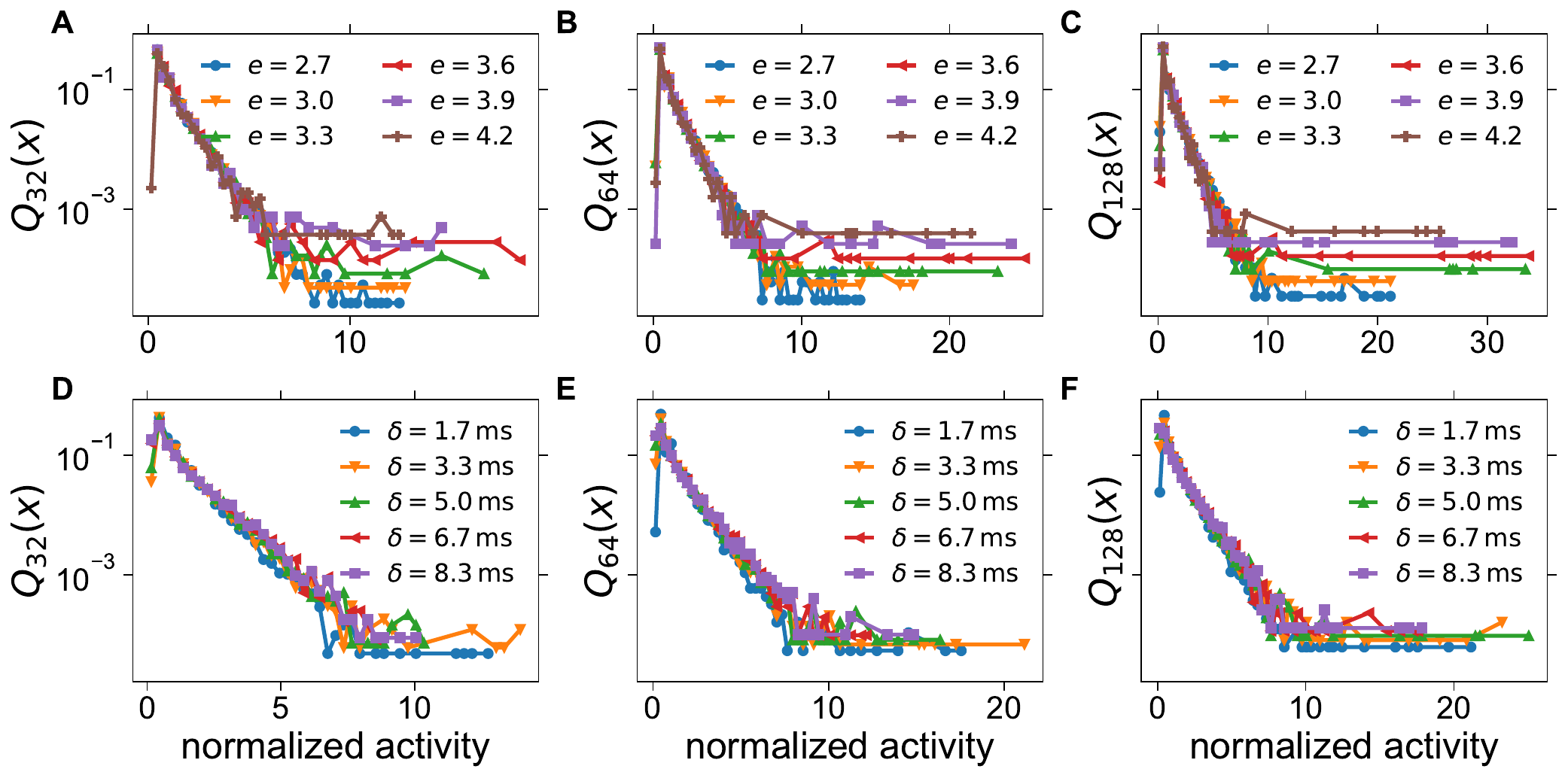}
    \caption{\textbf{Distribution of normalized cluster activity is independent of the binarization threshold, $e$, and bin size $\delta$.} \textbf{Top row}.  Distribution of normalized cluster activity for $K = 32$, $K = 64$, and $K = 128$ for different values of the binarization threshold $e$. \textbf{Bottom row}.  Distribution of normalized cluster activity for $K = 32$, $K = 64$, and $K = 128$ for different bin sizes $\delta$.}
\end{figure*}

\clearpage

\begin{figure*}[ht]
    \centering
    \includegraphics[width=0.8\linewidth]{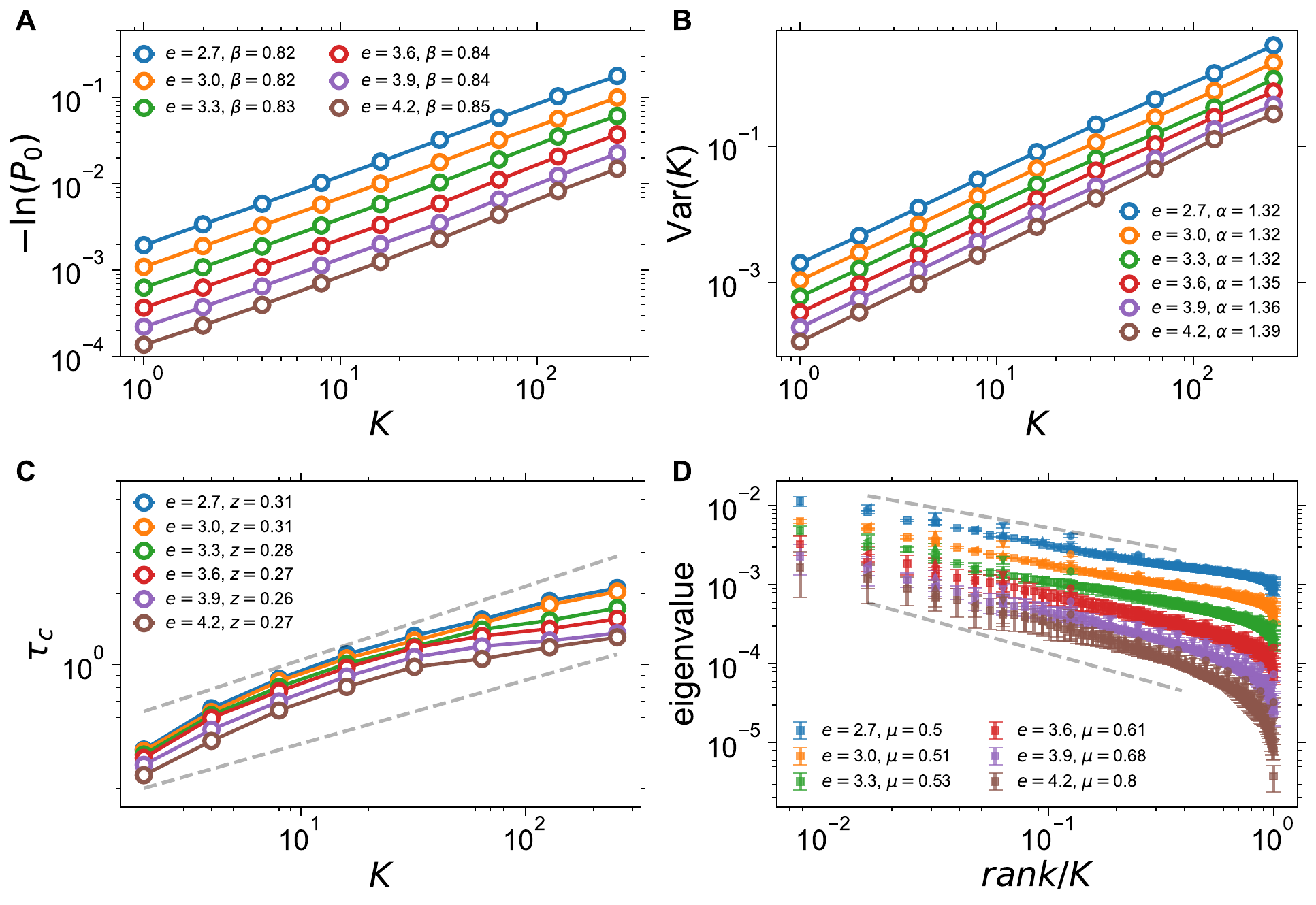}
    \caption{\textbf{Robustness of scaling behaviors with respect to data binarization threshold.} \textbf{A.} Distributions of silence robustly follow the the functional form $P_0 (K) = e^{-aK^\beta}$ for a range of binarization threshold $2.7 \leq e \leq 4.2$ SD. The exponent $\beta$ is stable, and shows small variability, $0.82 \leq \beta \leq 0.85$, increasing for higher  thresholds. This is expected since more and more points are removed from the timeseries as one increases the threshold. \textbf{B.} Scaling of the variance $Var(K)$ with $K$ is robust and shows little dependence on binarization threshold $e$. \textbf{C.} Autocorrelation time, $\tau_c$, as a function of $K$. Deviations from scaling are observed for higher thresholds $e$ at large cluster sizes, $K > 64$. This must be ascribed to increasing sparsity of time series (or increasing sub-sampling), which weakens original correlations. 
    \textbf{D.} Dependence of eigenvalue spectra on the threshold $e$. Scaling and exponent are robust for a range of $e$ values. Deviations are observed for $e \geq 3.6$. We notice that, for a given cluster size, variability across clusters  increases for increasing thresholds $e$. All quantities are calculated for an individual subject with bin size $\delta = 1.67$ ms, the sampling time. Results are consistent across subjects.}
\end{figure*}

\begin{figure*}[ht]
    \centering
    \includegraphics[width=0.8\linewidth]{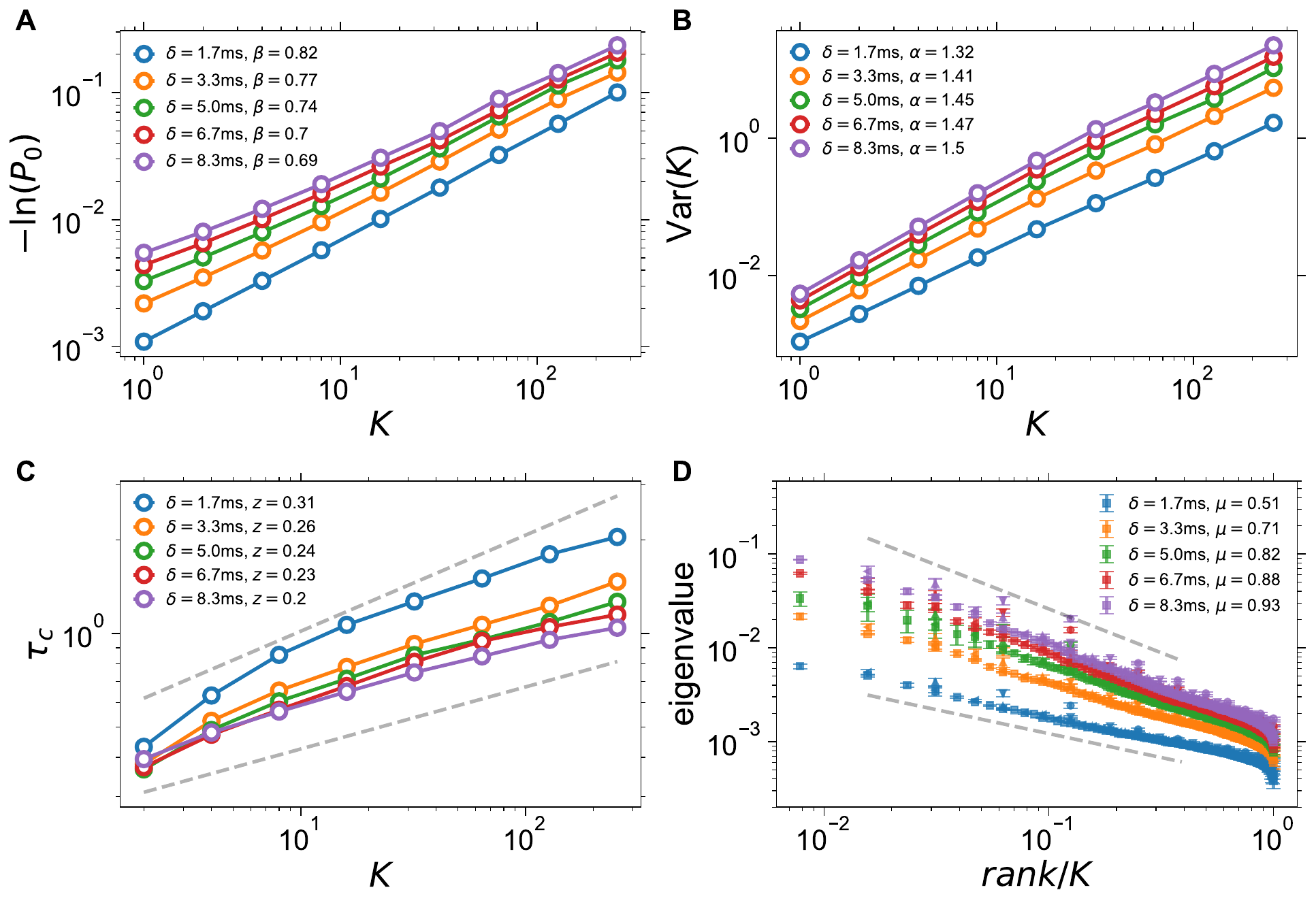}
    \caption{\textbf{Dependence of scaling behaviors on data binning.} \textbf{A.} Distributions of silence follow the functional form $P_0 (K) = e^{-aK^\beta}$ in the range of time bin sizes from $1T_s$ to $5T_s$, where $T_s = 1/600$s is the sampling time. The exponent $\beta$ tends to decrease for increasing bin sizes, and goes from $0.82$ at $\delta = 1T_s$ to $\beta = 0.69$ at $\delta = 5T_s$. Time binning mostly affects behavior at $K \leq 8 $. \textbf{B.} The scaling exponent of the variance, $\alpha$, increases with the bin size $\delta$. Although rather limited, this increase indicates that binning increases correlations among variables.  \textbf{C.} Autocorrelation time, $\tau_c$, as a function of $K$. For $\delta = 1T_s$, $\tau_c$ is consistently larger than for larger bin sizes.      Asymptotic scaling (large $K$) is well preserved  across bin sizes. Small changes in the scaling exponent $z$ are mostly due to effects of binning on autocorrelation at small $K$.
    As for $P_0$, time binning mostly affects $\tau_c$ scaling  at small cluster sizes $K$. \textbf{D.} The scaling and exponent of the eigenvalue spectrum depends on time binning. The exponent $\mu$ increases with $\delta$. Difference in $\mu$ is particularly marked between $\delta = 1T_s$ (no time binning) and $\delta = 2 T_s = 3.3$ ms. Dashed lines represent the fit for $\delta = 1T_s = 1.7$ ms and $\delta = 5T_s = 8.3$ ms.  All quantities are calculated for an individual subject and threshold $3$ SD. Results are consistent across subjects.}
\end{figure*}

\clearpage


\clearpage

\clearpage

\begin{figure*}[ht]
    \centering
    \includegraphics[width=0.5\linewidth]{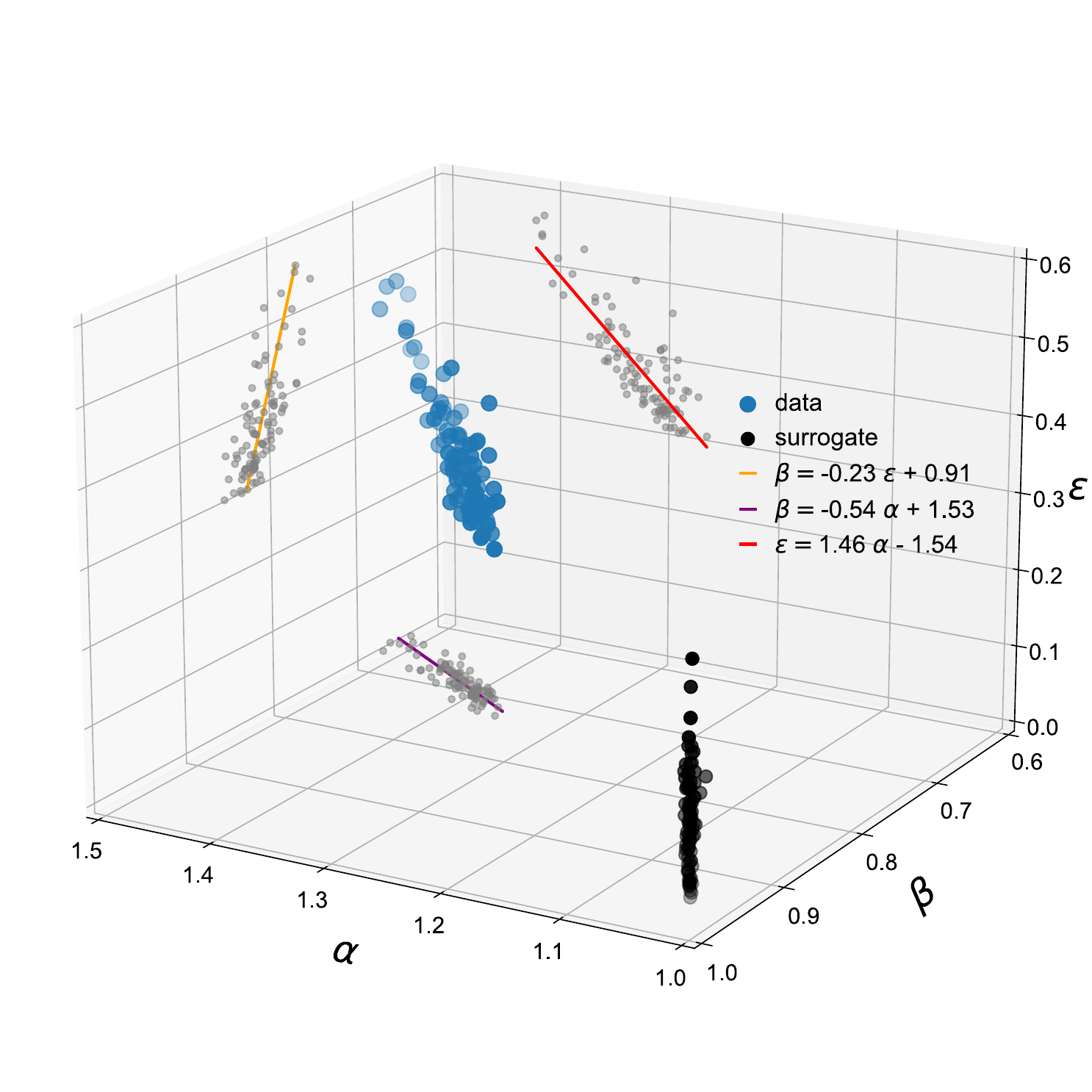}
    \caption{\textbf{Relationships between scaling exponents, $\alpha$, $\beta$, and $\epsilon$ in original and surrogate MEG data.} Exponent relationships for original and phase-randomized data (black dots).}
\end{figure*}

\clearpage

\begin{figure*}[ht]
    \centering
    \includegraphics[width=\linewidth]{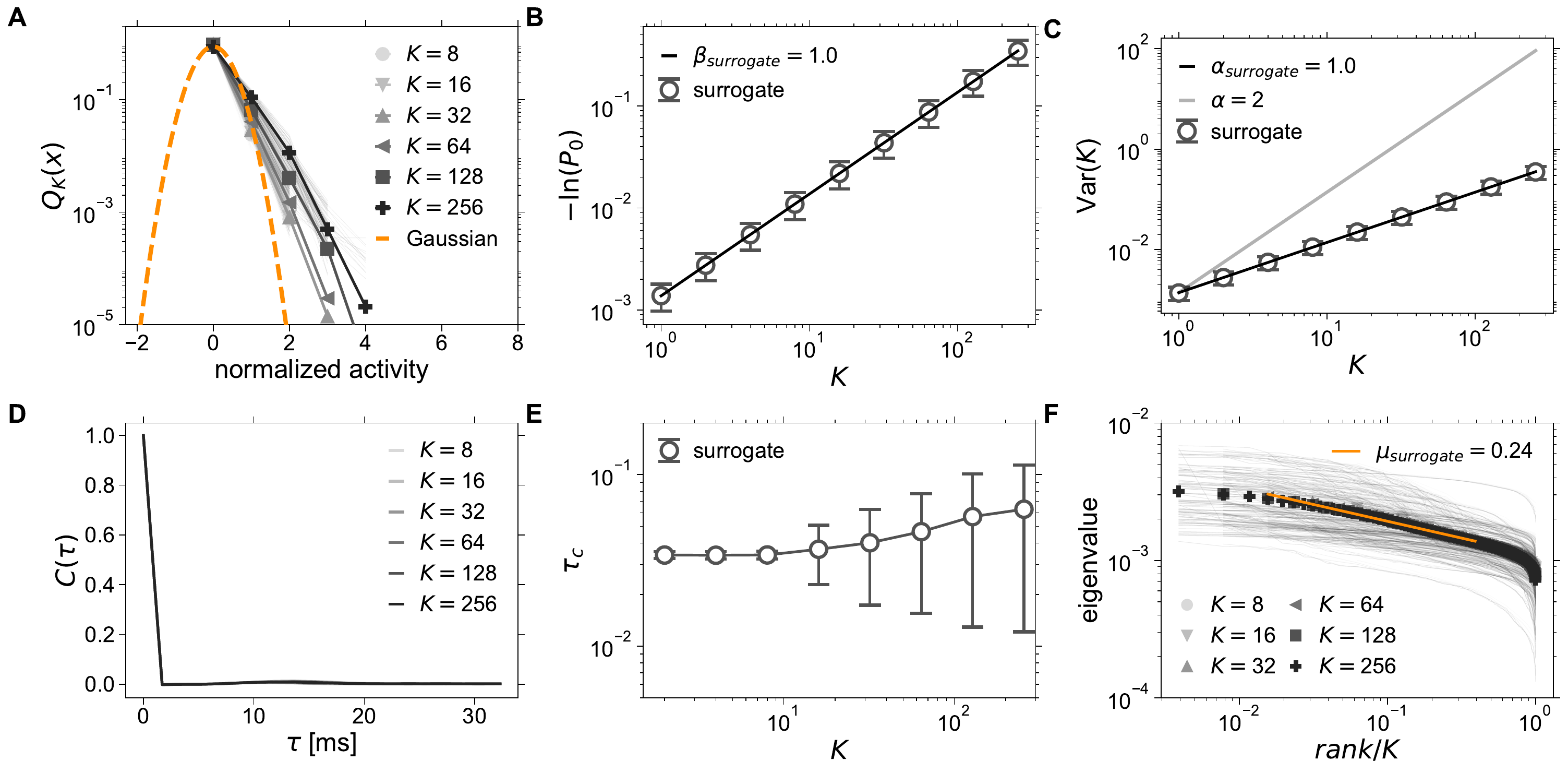}
    \caption{\textcolor{black}{\textbf{Renormalization group analysis of phase randomized MEG signals.} \textbf{A}. Probability distributions for normalized non-zero activity ($K \geq 8$) approach a Gaussian in phase-randomized data (gray scale). Darker color = Increasing cluster size.   \textbf{B}. Probability of silence in coarse-grained variables ($-ln \ P_0$) as a function of the cluster size,  $K$. Probability of silence decays as  $P_0(K) = e^{-bK^\beta}$ with $\beta = 0.9982 \pm 0.0005$ (fit $\pm$ error on the fit), significantly different from original data ($1.0003 \pm 0.0001$, mean $\pm$ SEM). 
     \textbf{C.} Variance (Var) of the coarse-grained variables  as a function of the cluster size, $K$. The solid  lines correspond to  linear least-square fits,  $log \ Var(K) = \alpha \cdot logK + b$. For phase randomized data $\alpha = 0.9994 \pm 0.0015$ (mean $\pm$ SEM; solid black line), significantly different from original data ($0.9995 \pm 0.0002$, mean $\pm$ SEM). Gray thick line: $Var(K) \sim K^2$.  \textbf{D}. Average correlation functions for different cluster sizes, $K$. Autocorrelation is independent of $K$ and shows a very fast decay to zero.
    \textbf{E}. 
    The autocorrelation time, $\tau_c$, is independent of $K$ for the phase randomized data (black circles).
    \textbf{F.} Eigenvalues of covariance matrices within clusters for different  cluster sizes. For randomized data (gray scale: lines = individual subject, squares = average), we observe high variability across samples and disrupted scaling with an average exponent $\mu = 0.2448 \pm 0.0030$ (fit $\pm$ error on the fit) significantly different from the one observed in original data ($0.2277 \pm 0.0130$, mean $\pm$ SEM). Least squares fit to $\log  \lambda = \log  b (rank/K)^{-\mu}$ is performed for $K = 128$. \textcolor{black}{The fitting range is $2/128 - 50/128$.} Results are averaged over $n=100$ subjects. Error bars represent SD.  Where not shown, error bars are smaller than the marker size.}
    }
\end{figure*}

\clearpage

\begin{figure*}[ht]
    \centering
    \includegraphics[width=\linewidth]{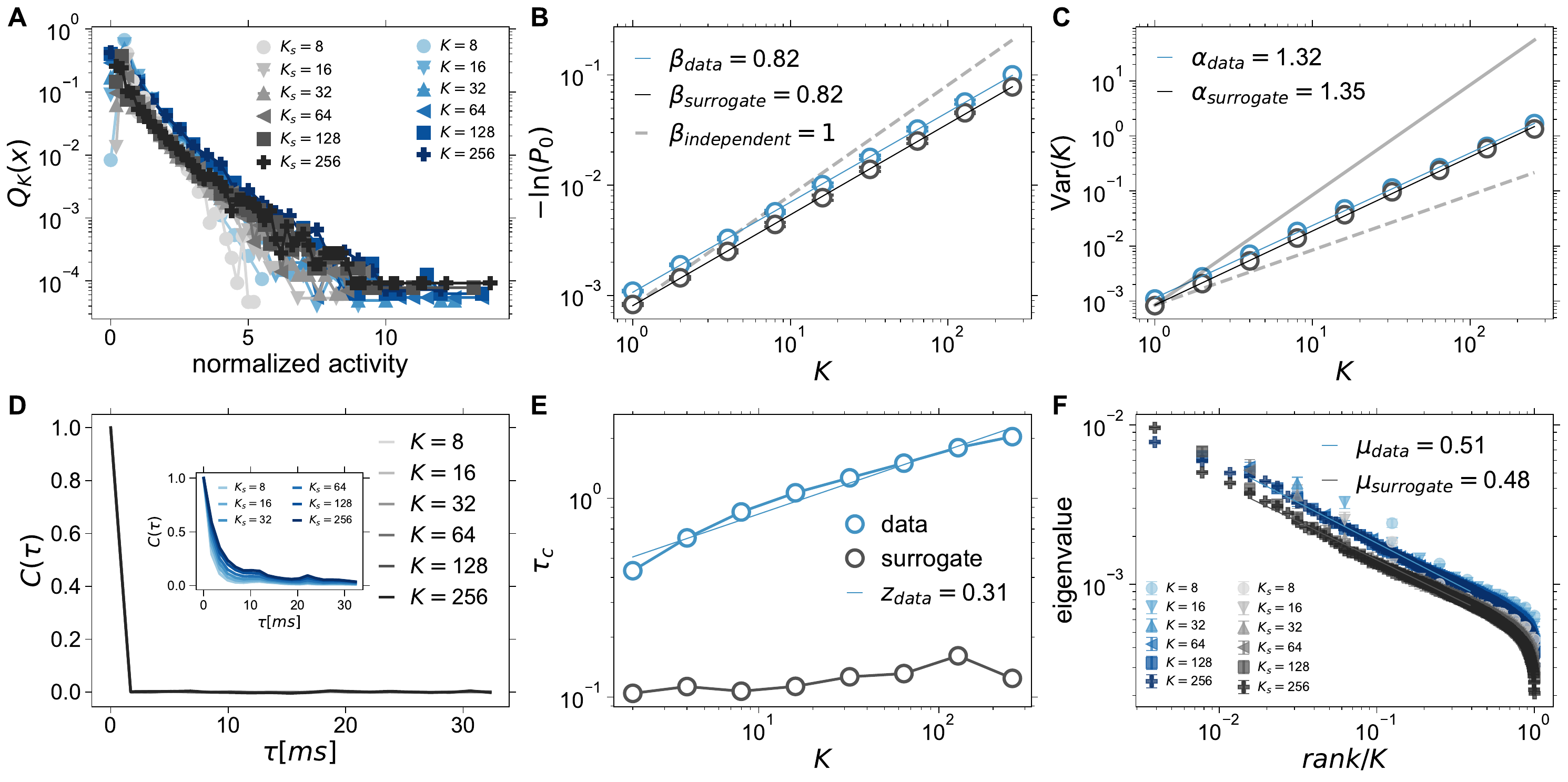}
    \caption{\textbf{Renormalization group analysis of trace randomized MEG signals.}  Trace randomized data are obtained by random shuffling of signal amplitudes for all MEG sensors (Methods). All signal timeseries are shuffled using the same random  permutation of time indices. Unlike in the phase randomized data, the distributions of normalized activity approaches a non-Gaussian form for increasing $K$ (\textbf{A}, black-gray scale). The probability of silence scales with $K$ as for the original data, i.e. $P_0(K) = e^{-bK^\beta}$ with $\beta = 0.82$ (\textbf{B}). Similarly, the variance of coarse-grained variables as a function of $K$ scales as $Var(K) \sim K^\alpha$ , with $\alpha = 1.32$ as for the original data (\textbf{C}). On the other hand, 
    the autocorrelation time, $\tau_c$, is very short (\textbf{D}, main panel) and is  independent of $K$ as for trace  randomized  data (\textbf{E}).  For the original data  $\tau_c \propto K^z$, with $z = 0.31$, while $z \approx 0$ for phase randomized data.
    For trace randomized data (gray scale), the exponent $\mu = 0.48$ of the eigenvalue spectra is as in the original data (\textbf{F}). In all panels, original data are shown in blue and randomized data in black. Larger clusters correspond to darker shades of colors.
    Solid lines in \textbf{B}, \textbf{C}, and \textbf{E} are linear least-square fit of the form $\log y = a\cdot \log x + b$, where $x = K$ and $y = -\ln P_0$, $Var(K)$ and $\tau_c$, respectively.
    Solid blue and black line in \textbf{F}: least squares fit to $\log  \lambda = \log  b (rank/K)^{-\mu}$ performed for $K = 128$. The fitting range is $2/128 - 50/128$. Error bars represent SEM.
    \textcolor{black}{Across 100 subjects, the scaling exponents $\beta$, $\alpha$, and $\mu$ do not show  statistically significant differences between the original and trace-randomized data (two-sample $t$-tests, $p > 0.9$ for all three exponents).  
    }}
\end{figure*}

\clearpage

\begin{figure*}[ht]
    \centering
    \includegraphics[width=\linewidth]{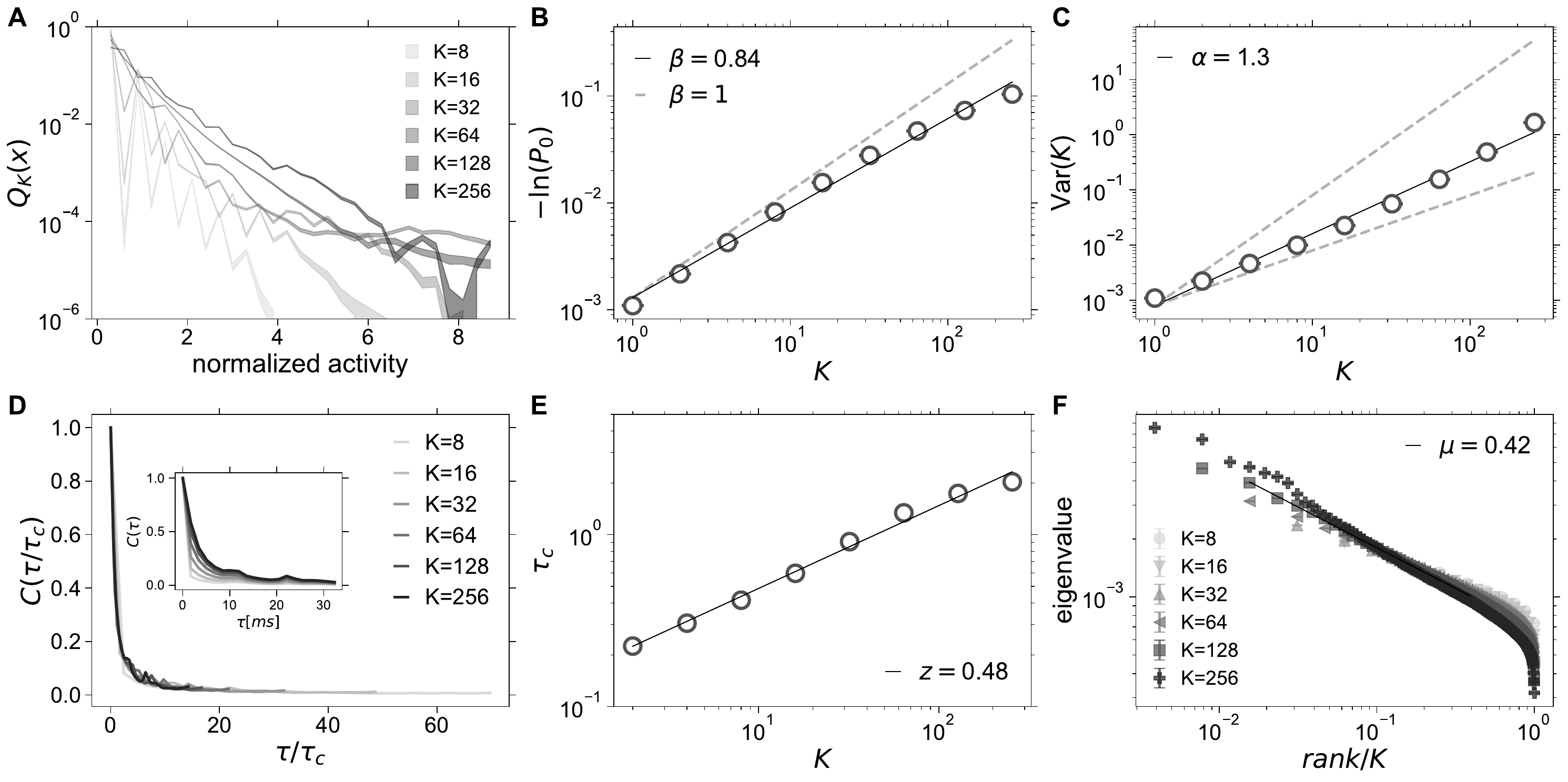}
    \caption{\textbf{Renormalization group analysis with random pairing of MEG sensors.} In random pairing variables are not combined according to their correlations. Each variable is uniquely paired with another variable that is selected randomly (Methods). \textbf{A.} Distributions of normalized activity for $K \geq 8$. \textbf{B.} Probability of silence, $P_0$, as a function of $K$. \textbf{C.} Variance of non-normalized coarse-grained variables as a function of $K$. \textbf{D.} Average autocorrelation for clusters of size $K \geq 8$. \textbf{E.} Autocorrelation time $\tau_c$ as a function of $K$. \textbf{F.} Eigenvalue spectra for $K \geq 8$.  Despite random pairing many of the scaling features observed in original data seem to be preserved (see B, C, E), with slight changes in the exponents. Scaling is also observed in the eigenvalue spectra, although the exponent appear to depend on $K$. The exponent for $K = 128$ is $\mu = 0.42$ (fitting range: 2-50). The distribution $Q_K(x)$ tends to a non-Gaussian behavior for large $K$. However, such behavior is not consistent across $K$'s, as observed instead for original data. Solid lines in \textbf{B}, \textbf{C}, and \textbf{E} are linear least-square fit of the form $\log y = a\cdot \log x + b$, where $x = K$ and $y = -\ln P_0$, $Var(K)$ and $\tau_c$, respectively. Solid line in \textbf{F} is a least squares fit to $\log  \lambda = \log  b (rank/K)^{-\mu}$ performed for $K = 128$.  Results are averaged over 100 realizations of the random pairing. For each quantity, fits are performed on the average.  Error bars and shaded areas represent SEM calculated of 100 realizations of the random coarse-graining. Where not shown (the variance and $P_0$), error bars are smaller than the symbol size. }
\end{figure*}

\clearpage

\begin{figure*}[ht]
    \centering
    \includegraphics[width=\linewidth]{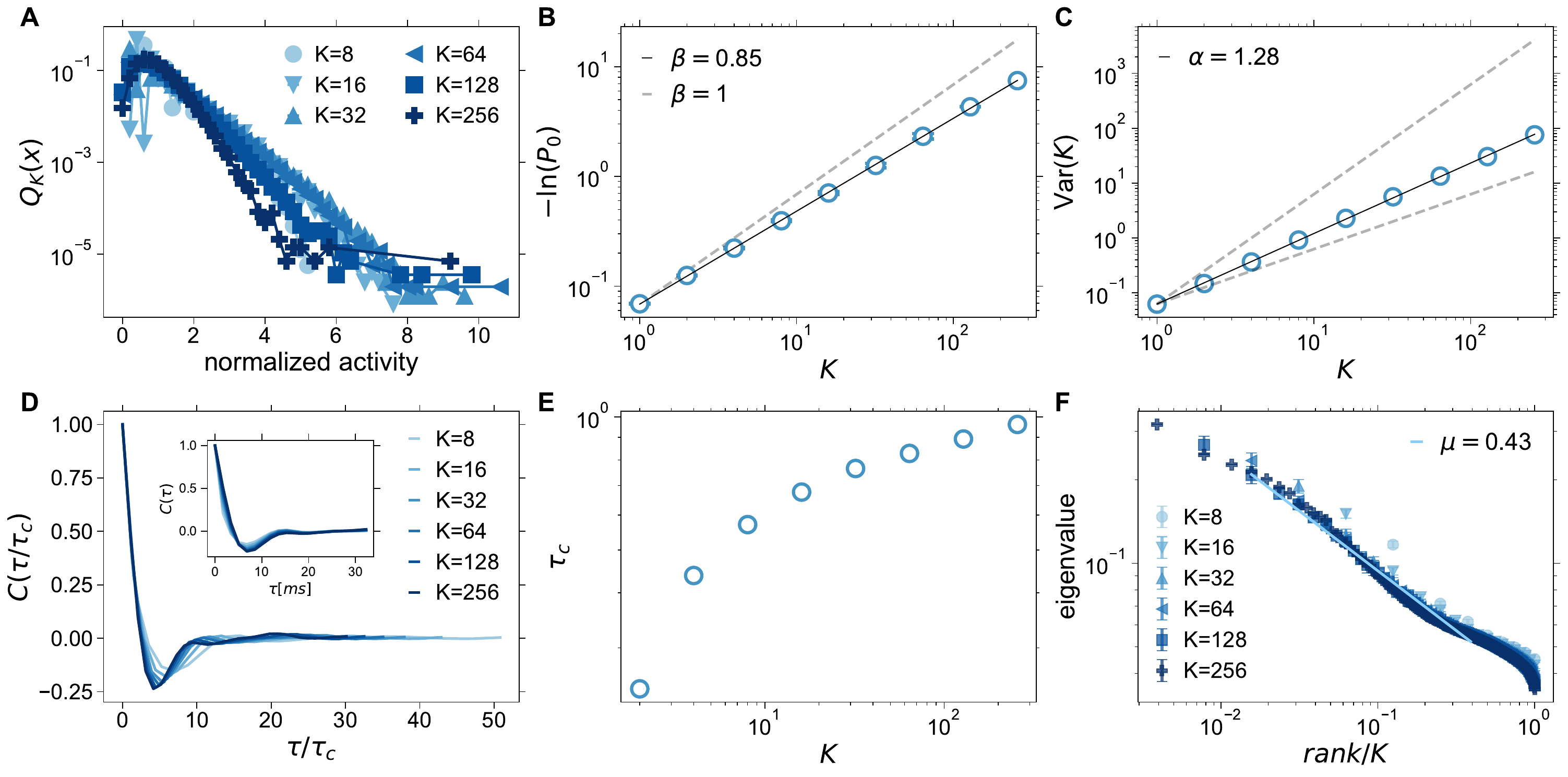}
    \caption{\textbf{PRG analysis performed using a binarization threshold $e=0.5$SD.} The distribution of cluster activity tends to a Gaussian for increasing cluster sizes \textbf{(A)}, while the scaling exponent $\beta$ slightly increases \textbf{(B)} and  $\alpha$ slightly decreases \textbf{(C)} compared to the values obtained for $e = 3$ SD. The autocorrelation instead shows no scaling across $K$ \textbf{D-E}. Scaling and data collapse of the eigenspectra of the cluster covariance matrices are also altered compared to higher binarization thresholds \textbf{(F)}. Error bars (SEM) are smaller than symbol sizes (B and F).}
\end{figure*}

\clearpage

\begin{figure*}[ht]
    \centering
    \includegraphics[width=\linewidth]{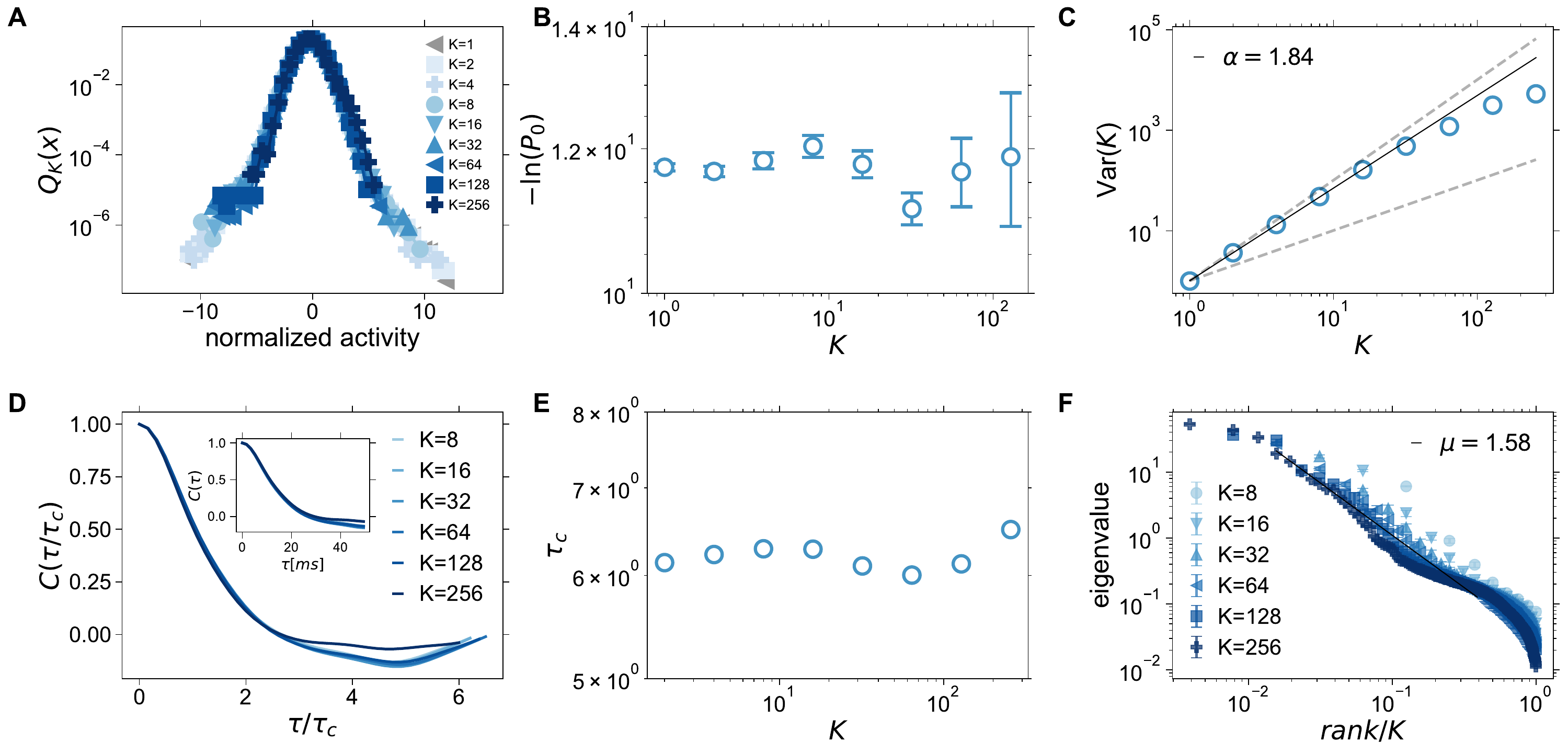}
    \caption{\textbf{PRG analysis performed on continuous MEG data (individual subject).}  The probability of activity \textbf{(A)}, which initially  deviates from a Gaussian for large activity values, progressively approaches a Gaussian form ($K = 256$) that corresponds to the Gaussian core of the original signal amplitude distribution ($K=1$, gray triangles). The probability of silence is independent of $K$ \textbf{(B)}, while the variance  \textbf{(C)} scales with an exponent ($\alpha = 1.84$; linear least-square fit of the form $\log Var(K) = a\cdot \log K + b$) closer to 2 (fully correlated variables, upper dashed line). The autocorrelation of cluster activity is nearly independent of the cluster size $K$ (\textbf{D-E}).  Eigenspectra of the covariance matrix do not show a clear scaling regime and data collapse \textbf{(F)}.  The solid black line is a least squares fit to $\log  \lambda = \log  b (rank/K)^{-\mu}$ performed for 
    $K = 128$ (fitting range is $2/128 - 50/128$) and is   shown only as a reference and for comparison with Fig. 1 (main text). Values of original MEG signals and coarse-grained variables lower than $10^{-5}$ were set equal to zero to ensure that $P_0 \neq 0$ throughout all coarse-graining steps \textcolor{black}{$P_0=0$ in $K=256$}. This does not affect results for the other analyzed quantities, which show the same behavior for original data and coarse-grained variables with no zeros. Scaling of $P_0$ and all other quantities remain stable on a broad range of zero-defining threshold ($< 0.1$). Error bars (SEM) are smaller than symbol size (B and F). \textcolor{black}{Z-normalization is applied to the summed continuous signals at each PRG step. Variance is calculated on the non-normalized variables.} }
\end{figure*}

\clearpage

\begin{figure*}[ht]
    \centering
    \includegraphics[width=\linewidth]{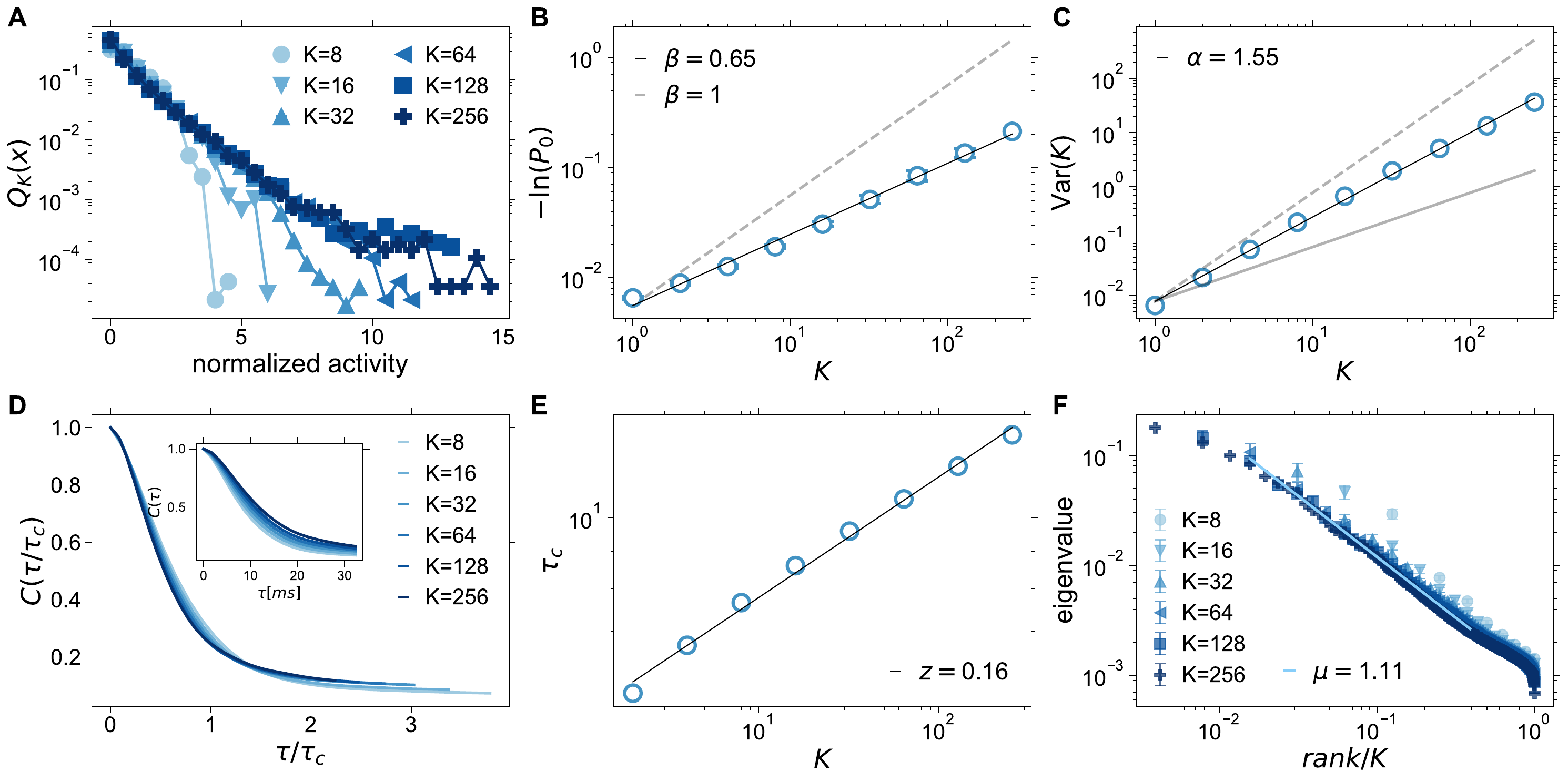}
    \caption{\textbf{PRG analysis performed considering all points of signal excursions above and below thresholds for an individual subject.} As for the case in which one considers only the tip of the excursions (Appendix D), the distribution of activity approaches a universal non-Gaussian form for increasing $K$ (\textbf{A}; only data in the range $(0,15]$ are shown.). Scaling of silence and variance remains non-trivial (\textbf{B-C}).  $\beta$ decreases and $\alpha$ increases compared to the values measured when considering only the extremes of the excursions (see Table II). The decay of the variable autocorrelation is  slower for increasing $K$, and all autocorrelation $C(\tau)$ collapse on the same curve when $\tau$ is rescaled by the autocorrelation time $\tau_c (K)$ (\textbf{D}). $\tau_c$ scales  as $K^z$, with $z = 0.16 \pm 0.04$  (mean $\pm$ SD; $n = 100$ subjects) (\textbf{E}). Eigenvalues of cluster covariance matrices decay as a power-law of their scaled ranks ($rank/K$), with an exponent $\mu  = 1.08 \pm 0.08$ ($n = 100$ subjects; Table II).  The solid line is a least squares fit to $log \ \lambda = log \ b (rank/K)^{-\mu}$ performed for $K=128$. The fitting range is $2/128 - 50/128$. Error bars (SEM) are smaller than symbol size (B and F).}. 
\end{figure*}

\clearpage

\begin{figure*}[ht]
    \centering
    \includegraphics[width=0.65\linewidth]{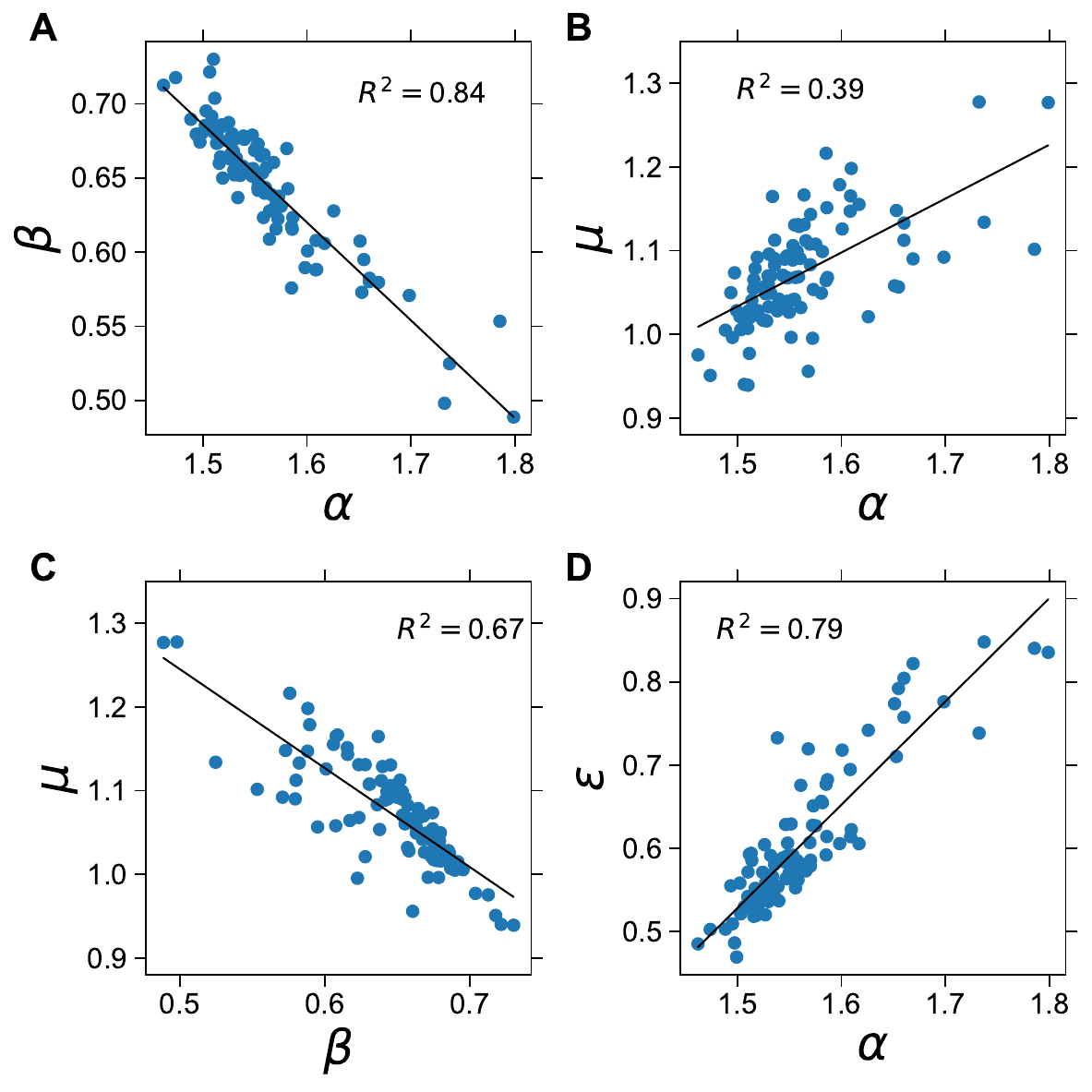}
    \caption{\textbf{Relationships between scaling exponents in the MEG of the resting state considering all points of above and below-threshold  excursions.}  \textbf{A.} The scaling exponent of the cluster variance is inversely proportional to the exponent $\beta$ that characterizes the the probability of silence $P_0 (K)$, i.e. $\beta = -0.66\alpha + 1.68$ (linear least square fit; $R^2 = 0.84$).  \textbf{B.} The exponent $\mu$ characterizing the eigenvalue spectrum of the covariance matrix increases linearly with $\alpha$: $\mu = 0.64\alpha + 0.06$ (linear least square fit; $R^2 = 0.39$).  \textbf{C.} As expected from A and  B, the exponent $\mu$ decreases for increasing values of $\beta$. The relationship between $\mu$ and $\beta$ is well described by a linear equation, $\mu = -1.18\beta + 1.83$ (linear least square fit; $R^2 = 0.67$). \textbf{D.} The scaling exponent of the largest eigenvalue of the covariance matrix, $\epsilon$, increases for increasing values of $\alpha$, as shown by the linear relationship  
     $\epsilon = 1.24\alpha - 1.33$ (linear least square fit; $R^2 = 0.79$).  
    }
\end{figure*}

\clearpage

\begin{figure*}[ht]
    \centering
    \includegraphics[width=\linewidth]{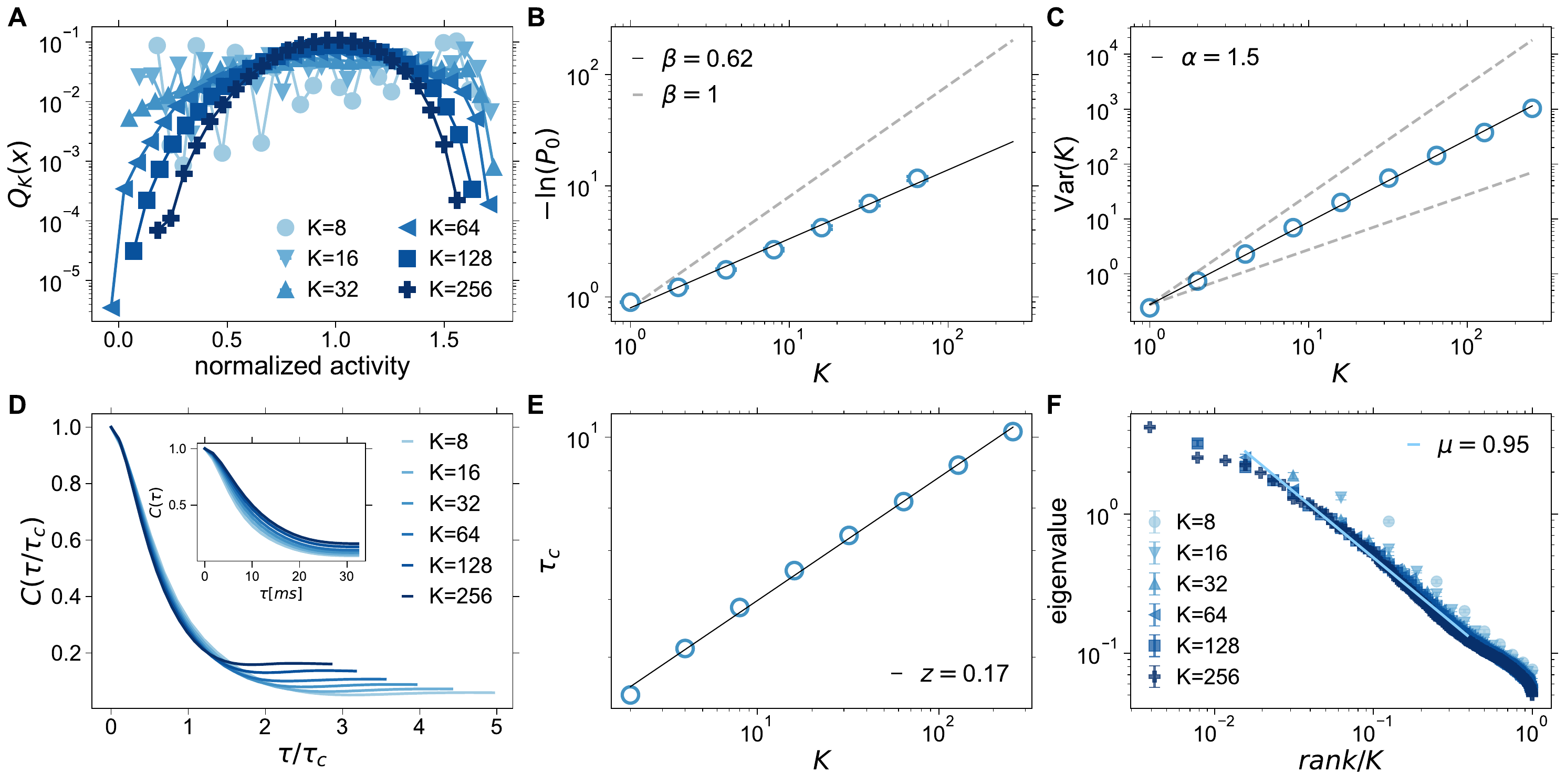}
    \caption{\textbf{PRG analysis performed using a binarization threshold $e = 0.5$ SD and considering all points belonging to  beyond-threshold excursions.} Scaling  is comparable to the one observed for $e = 3$ SD for all quantities (Fig.~S12).  Error bars (SEM) are smaller than symbol size (B and F).}
\end{figure*}

\clearpage

\begin{figure*}[ht]
    \centering
    \includegraphics[width=\linewidth]{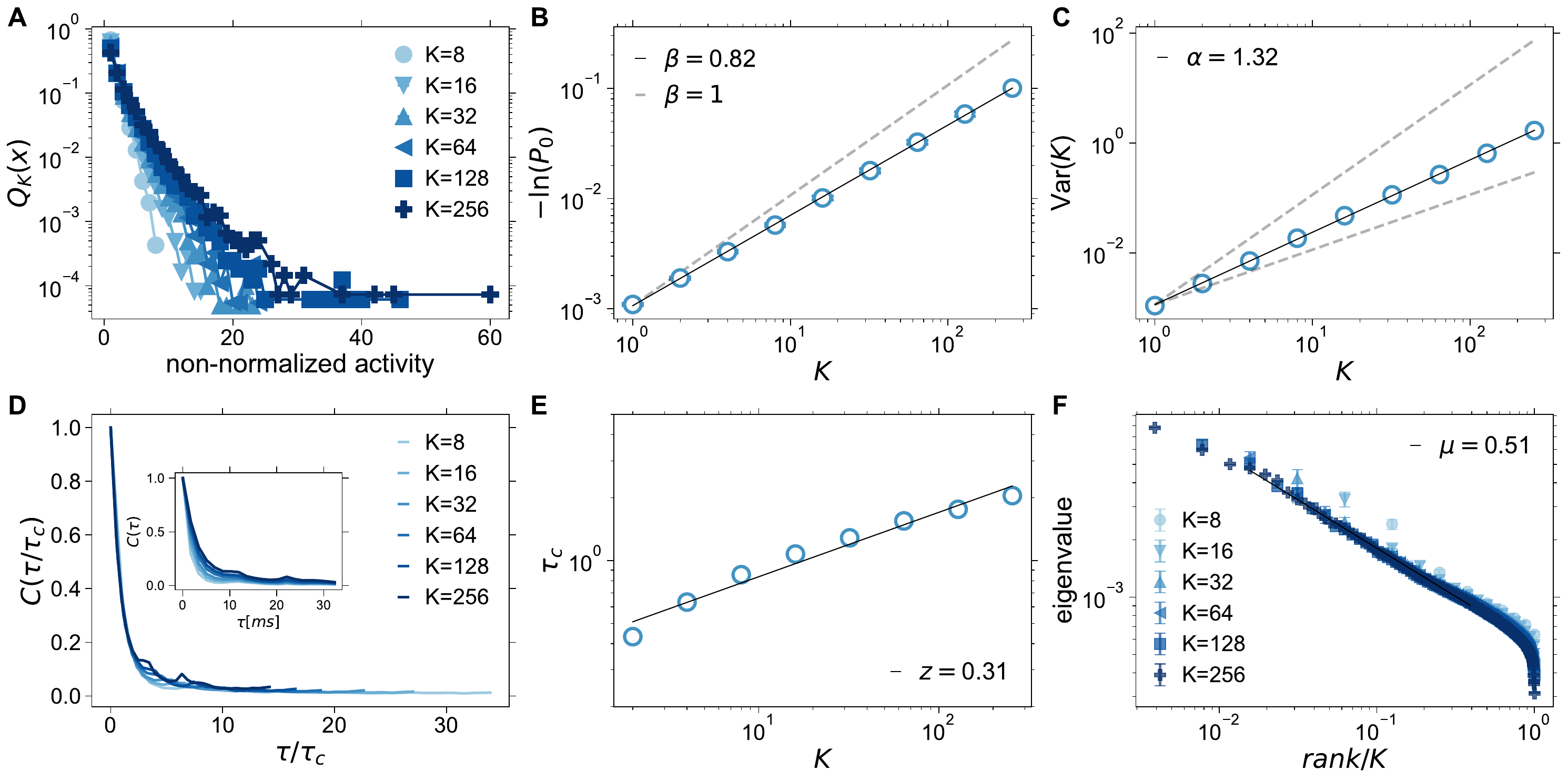}
    \caption{\textbf{PRG analysis performed on non-normalized variables for an individual subject.} For all analyzed quantities, results of the PRG analysis performed on non-normalized variables are in line with those obtained from normalized variables (Fig.~1, main text, individual subject; Fig.~S1, average over subjects). Error bars (SEM) are smaller than symbol size (B and F).
    }. 
\end{figure*}

\clearpage


\clearpage

\begin{figure*}[ht]
    \centering
    \includegraphics[width=0.6\linewidth]{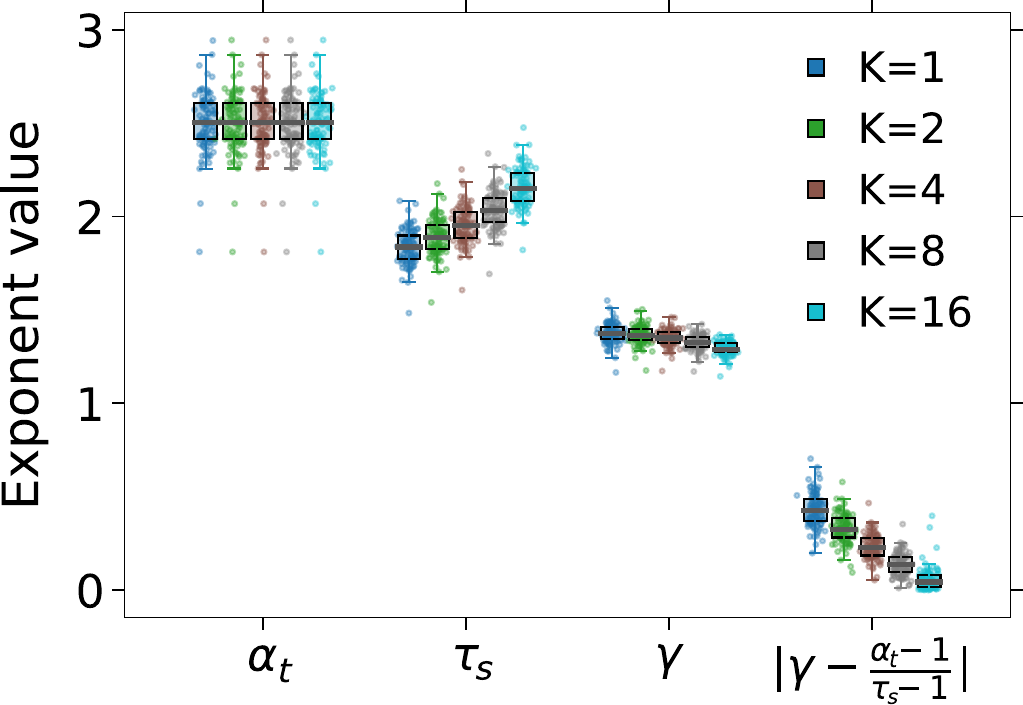}
    \caption{{\bf Exponents $\tau_s$, $\alpha_t$, $\gamma$ and corresponding  scaling relationship  for neuronal avalanches during coarse-graining.} 
    The exponent $\tau_s$ of the avalanche size distribution slightly increased for increasing cluster sizes, $K = 1, 2, 4, 8, 16$:  $1.83 \pm 0.09$, $1.89 \pm 0.10$, $1.95 \pm 0.10$, $2.03 \pm 0.10$, $2.16 \pm 0.10$ ($mean \pm SD$). The exponent $\alpha_t$ of the avalanche duration distribution remained unchanged across $K$: $\alpha_t = 2.50 \pm 0.16$ ($mean \pm SD$). The exponent $\gamma$ slightly decreases for increasing  $K = 1, 2 , 4, 8, 16$: $1.37 \pm 0.05$, $1.36 \pm 0.05$, $1.35 \pm 0.04$,  $1.32 \pm 0.04$, $1.29 \pm 0.04$ (mean $\pm$ SD). The difference $\big | \frac{\alpha_t -1}{\tau_s - 1} - \gamma \big|$ tends to zero for increasing $K$, suggesting that the system is getting closer to criticality. From $K =1$ to $K = 16$, the difference is: $0.43 \pm 0.09$, $0.33 \pm 0.08$, $0.23 \pm 0.07$, $0.14 \pm 0.06$, $0.06 \pm 0.06$ (mean $\pm$ SD).
    Maximum likelihood estimates were performed following \cite{alstott2014powerlaw} using the range  between  $T_{min} = S_{min} = 1$ and $T_{max} = max(T)$, $S_{max} = 1.5 N_k$. Black box and line: interquartile range (IQR), $Q_3 - Q_1$. and median. Whiskers: $1.5$ IQR.
    }
\end{figure*}


\clearpage

\begin{figure*}[ht]
    \centering
    \includegraphics[width=0.85\linewidth]{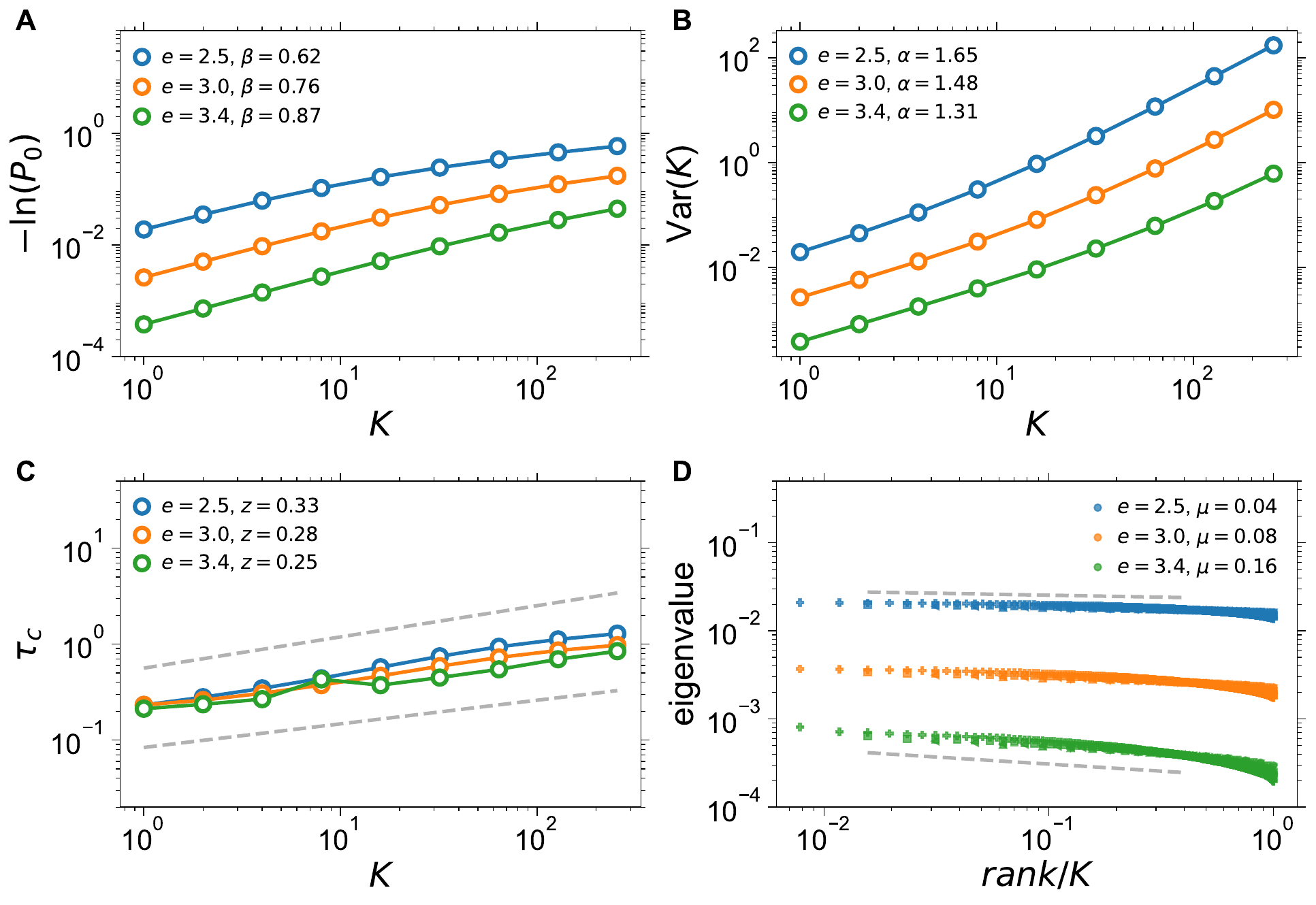}
    \caption{\textbf{Robustness of scaling behaviors with respect to  binarization threshold in model simulations.} \textbf{A.} Distributions of silence robustly follow the functional form  $P_0 = e^{-aK^\beta}$ for a range of binarization threshold,  $2.5 \leq e \leq 3.4$ SD.  This is expected since more and more points are removed from the timeseries as one increases the threshold. \textbf{B.} The variance scaling $Var(K)$ remains robust across binarization thresholds. For $e > 3$ SD the exponent tends to 1, the prediction for independent variables. \textbf{C.} Autocorrelation time, $\tau_c$, as a function of $K$. Scaling exponent decreases with increasing $e$, and show little variability in a range around $e = 3$ SD, the value used for empirical and model data analysis in Figs.~2, 5 and 6. Fit is performed between $K = 1$ and $K = 256$. Dashed lines represent the fit for $e = 2.5$ and $e = 3.4$ SD. \textbf{D.} Dependence of eigenvalue spectra on the threshold $e$.  Dashed lines represent the fit for $e = 2.5$ and $e = 3.4$ SD. The analysis is performed using $\delta = 3dt$.
    }
\end{figure*}

\begin{figure*}[ht]
    \centering
    \includegraphics[width=0.8\linewidth]{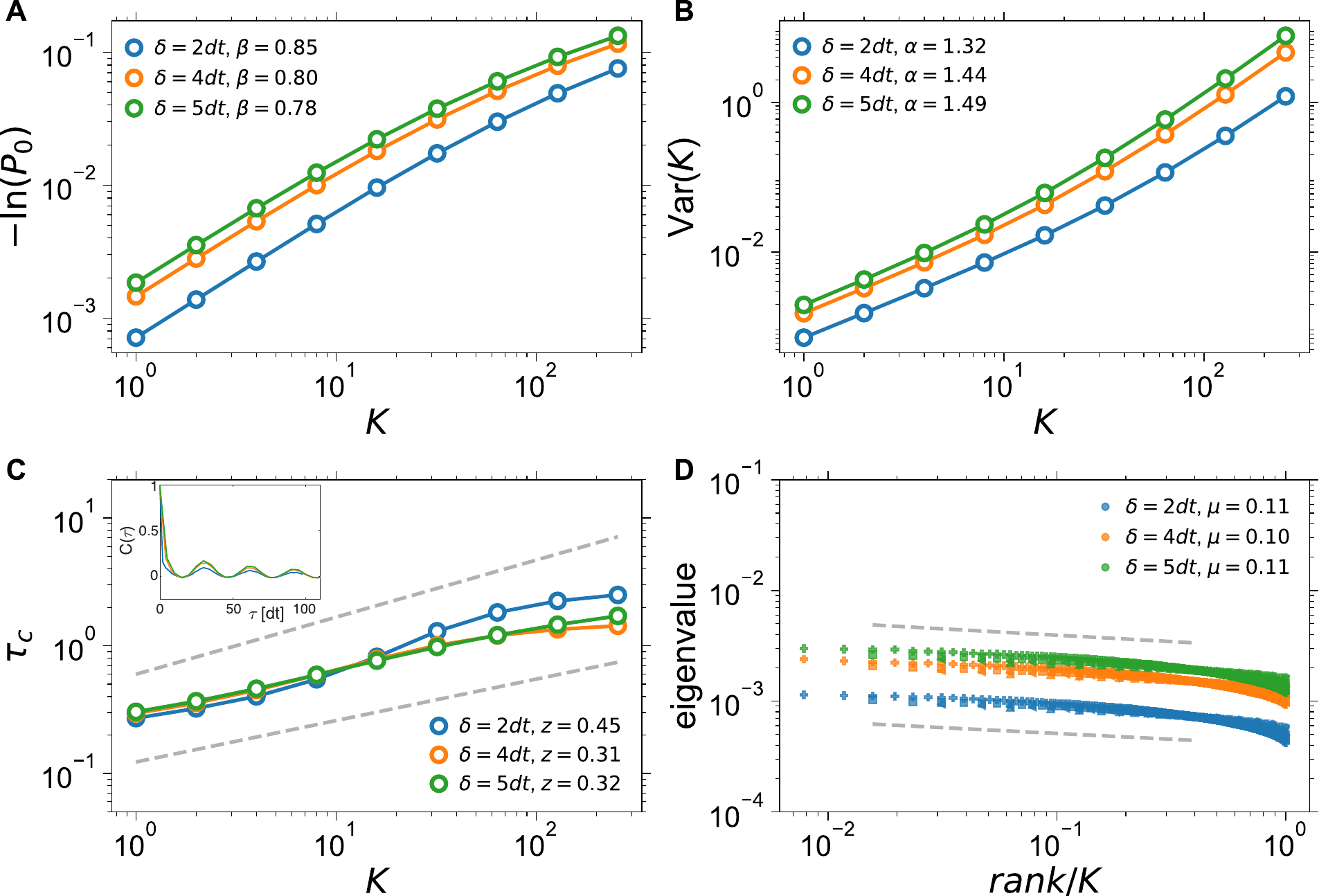}
    \caption{\textbf{Dependence of scaling behaviors of the model on temporal binning.} \textbf{A.} Distributions of silence follow the equation $P_0 = e^{-aK^\beta}$ in the range of time bin sizes, from $2dt$  to $5dt$, where $dt$ is the simulation timestep corresponding to one system sweep. The exponent $\beta$ tends to decrease for increasing bin sizes, and goes from $0.85$ at $\delta = 2dt$ to $\beta = 0.78$ at $\delta = 5dt$. \textbf{B.} The scaling exponent of the variance, $\alpha$, increases with the bin size $\delta$. Although rather limited, this increase indicates that binning increases correlations among variables.  \textbf{C.} Autocorrelation time, $\tau_c$, as a function of $K$. 
    The approximate data collapse of curves  corresponding to different $\delta$ shows that the relationship between   $\tau_c$ and $K$ is weakly dependent on time binning, except for the largest clusters.  Inset: Representative $C(\tau)$ for different $\delta$ and $K = 8$.  
    \textbf{D.} The scaling behavior and the corresponding exponent of the eigenvalue spectrum show little  dependence on the choice of time binning.  Dashed lines represent the fit for $\delta = 2dt$ and $\delta = 5dt$.  All quantities are calculated for a simulation of a network with $N = 273000$ and $M = 273$ subsystems made of $n_{sub} = 1000$ neurons. The threshold used is $e = 3.2$ SD.}
\end{figure*}

\clearpage

\begin{figure*}[ht]
    \centering
    \includegraphics[width=0.65\linewidth]{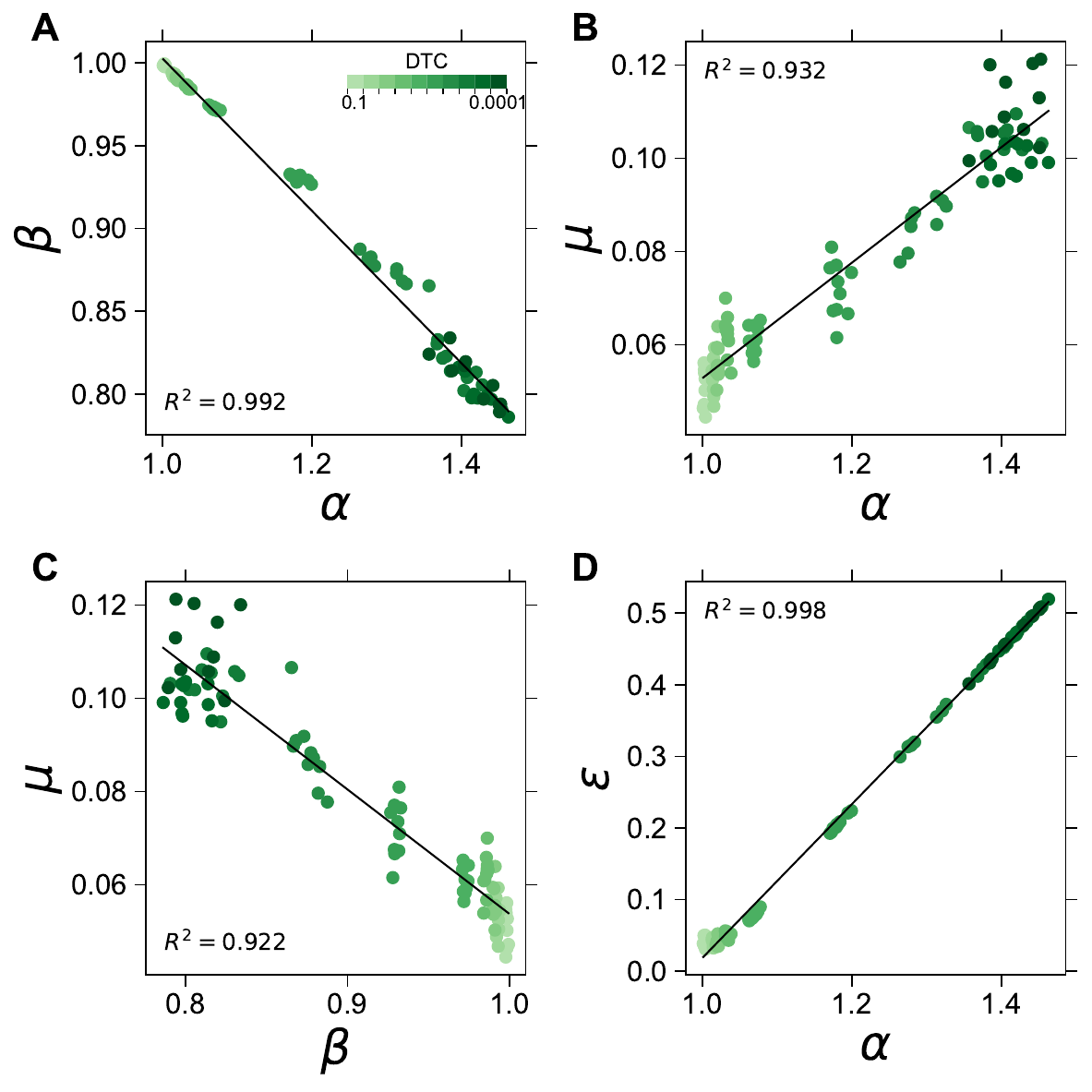}
    \caption{\textcolor{black}{\textbf{Relationships between scaling exponents in the model.} $\beta = -0.46\alpha + 1.46$ (linear least square fit: $R^2 = 0.99$), $\mu = 0.12\alpha - 0.07$ (linear least square fit: $R^2 = 0.93$), $\mu = -0.26\beta + 0.32$ (linear least square fit: $R^2 = 0.92$), $\epsilon = 1.07\alpha - 1.05$ (linear least square fit: $R^2 = 0.99$). Simulations (10 for each DTC value) were performed on a network with $N = 273000$ neurons. The network was parcelized in $M = 273$ subsystems, each containing $n_{sub} = 1000$ neurons. Feedback strength was set to $c = 0.01$ for all simulations.}}
\end{figure*}

\clearpage

\begin{figure*}[t]
    \centering
    \includegraphics[width=1\linewidth]{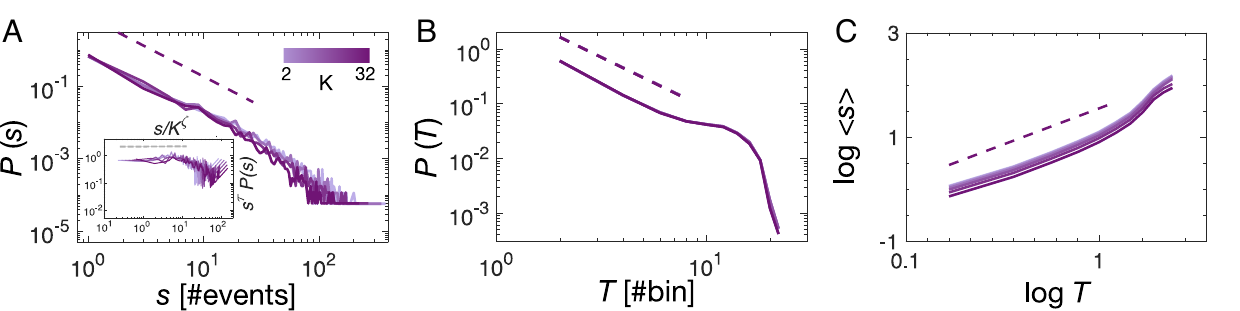}
    \caption{\textcolor{black}{\textbf{Invariant avalanche scaling in the neural network model under PRG.} 
    The distribution of avalanche sizes (A), $P(S)$, and avalanche durations (B), $P(T)$, do not change over consecutive steps of the coarse-graining procedure. Dashed lines: $y=x^\tau$ and $y = x^\alpha$, with $\tau = 1.6$. and $\alpha = 1.9$. The scaling relation between $s$ and $T$ is also invariant under coarse-graining (C), i.e. $\langle s \rangle \propto T^\gamma$, with $\gamma = 1.5$. Inset: the plot of $s^\tau P(s)$ versus $s/K^\zeta$ shows that the distributions of avalanche sizes for different cluster sizes, $K$, collapse onto a single curve  for $\tau = 1.5$ and $\zeta = 0.2$, indicating that the scaling ansatz, $P(s) = s^{-\tau}f(s/K^{\zeta})$ is valid also for avalanches in the neural network model. 
    The analysis is performed using $e = \pm 3.2$SD and $\delta = 1$  (Methods).}
    }
\end{figure*}

\clearpage

\begin{figure*}[ht]
    \centering
    \includegraphics[width=\linewidth]{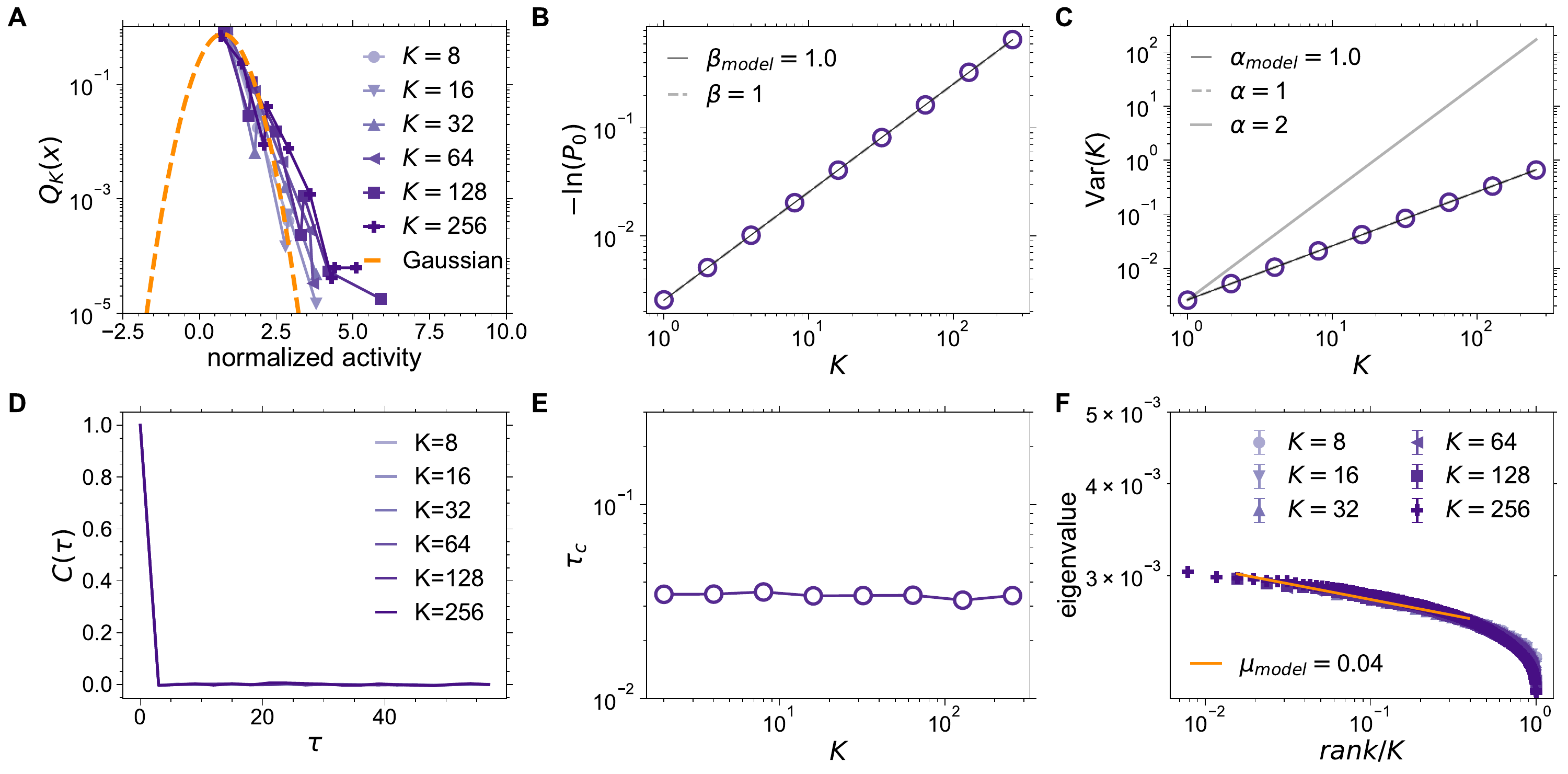}
    \caption{\textbf{PRG in the  adaptive Ising model with no synaptic interactions.}  \textbf{A.} Probability distribution of normalized non-zero  activity for cluster sizes $K \ge 8$. The distributions are Gaussian, as expected for independent variables. \textbf{B.} The log-probability of silence (-$\ln P_0$) as a function of cluster size ($K$). The solid red line represents a linear least-squares fit to $log(-\ln P_0) = logK^{\beta_{model}} + b$, where $\beta_{model} = 1$, the  behavior expected for independent variables.
    \emph{$\beta_{model} = 1$}, fully consistent with the independent-variable prediction. \textbf{C.} Variance of coarse-grained variables as a function of cluster size ($K$). A linear least-squares fit of $log(Var)$ vs $logK$ gives 
      \emph{$\alpha_{model} = 1$}, as expected for independent variables. \textbf{D.} Average autocorrelation functions for different cluster sizes ($K$) show absence of correlations. All curves are identical across cluster sizes. \textbf{E.} Autocorrelation time ($\tau_c$) as a function of cluster size. Since all clusters have the same autocorrelation function, no scaling in {$\tau_c$} is observed. \textbf{F.} Eigenvalues of covariance matrices within clusters for different cluster sizes. The eigenspectra show no scaling. Error bars (SEM) are smaller than the symbol size.
Simulations were performed near the critical regime ($\beta=0.99$, $c=0.0$),  on a network with with $J_{ij} = 0$ and $N = 273000$ neurons. The number of sensors was $M = 273$, each including $n_{sub} = 1000$ neurons ($e = 3$SD; $\delta = 1dt$).}
\end{figure*}

\clearpage

\begin{figure*}[ht]
    \centering
    \includegraphics[width=0.8\linewidth]{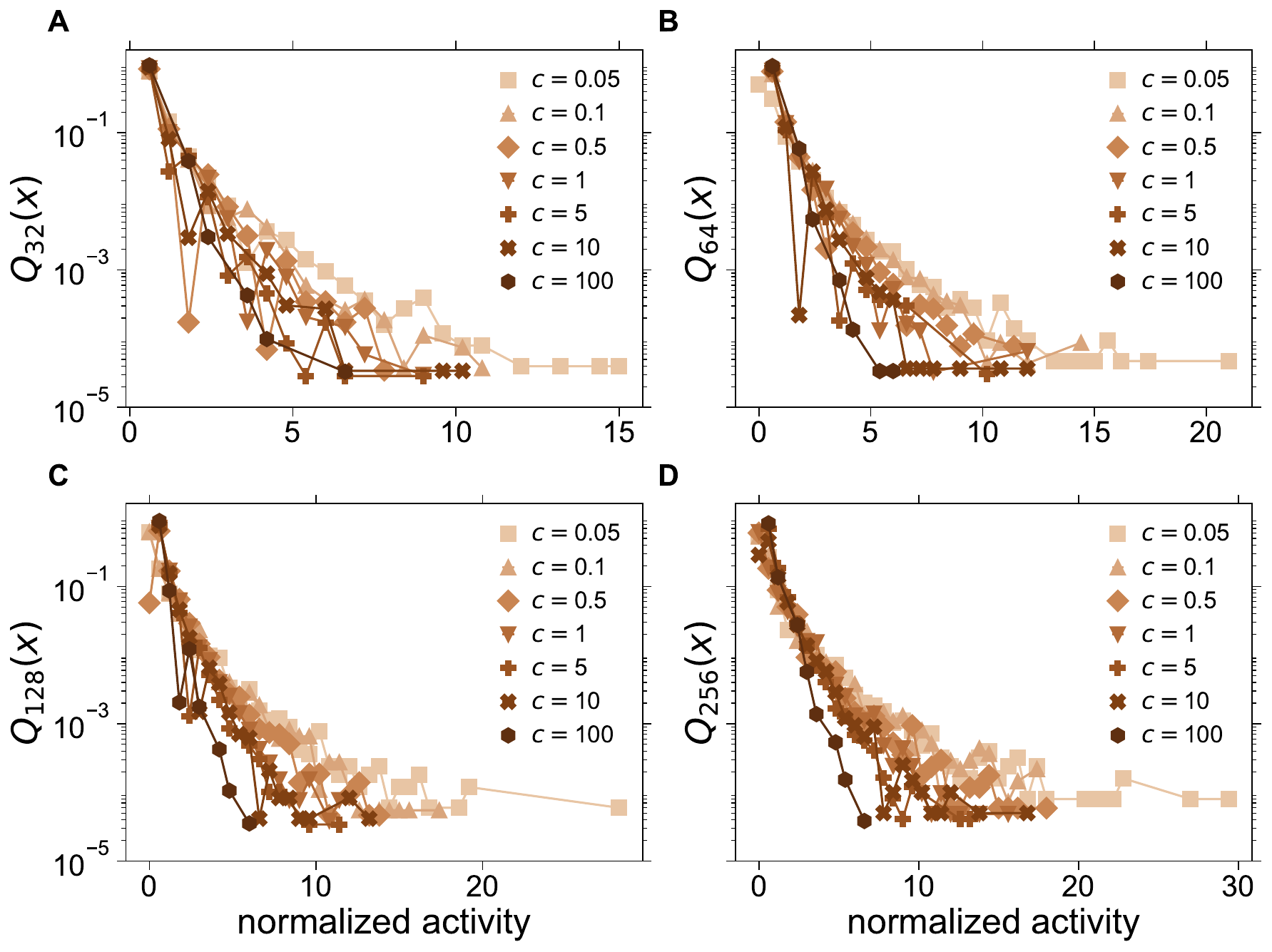}
    \textcolor{black}{\caption{\textbf{Dependence of activity distributions on the parameter $c$.} Probability distributions of normalized non-zero activity, $Q_K(x)$, for cluster sizes $K=32$ (\textbf{A}), $K=64$ (\textbf{B}), $K=128$ (\textbf{C}), and $K=256$ (\textbf{D}). As $c$ increases, the distributions become shorter-tailed and tend to a Gaussian-like behavior. For each value of $c$, the threshold, $e$, was identified by approximately matching the empirical  activity rate, using a time bin $\delta=3dt$. Each curve corresponds to a single simulation for the respective value of $c$. Simulations were performed near the critical regime ($\beta=0.9999$),  on a network with with $J_{ij} = 1$ and $N = 273000$ neurons. The number of sensors was $M = 273$, each including $n_{sub} = 1000$ neurons.} }
\end{figure*}

\clearpage

\begin{figure*}[ht]
    \centering
    \includegraphics[width=0.8\linewidth]{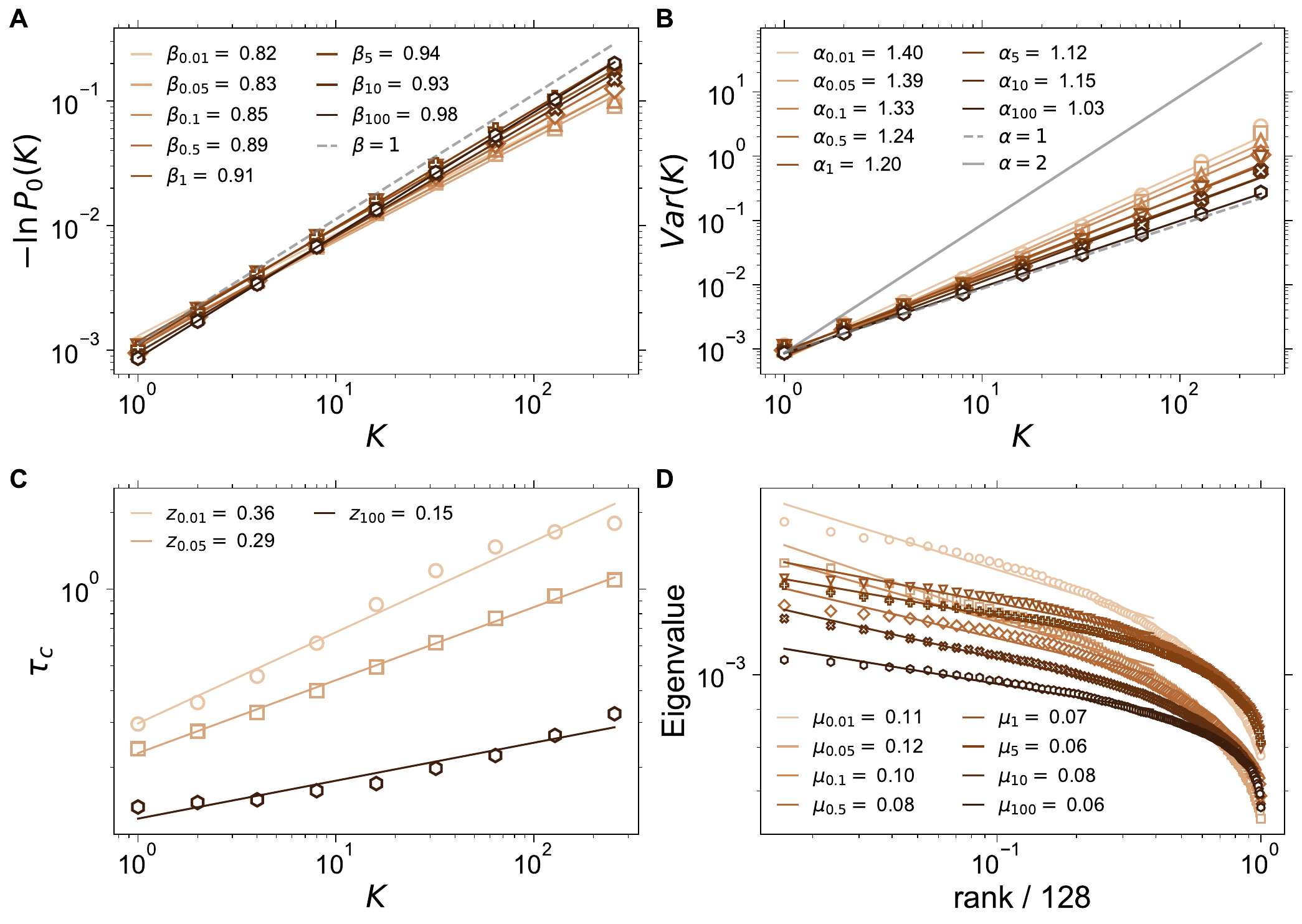}
    \textcolor{black}{\caption{\textbf{Dependence of scaling exponents on the parameter $c$.} As $c$ increases,  the exponent $\beta$ of the  log-probability of silence  (\textbf{A}), and $\alpha$ of variance of non-normalized variables (\textbf{B})  approach $1$,  characteristic of a collection of independent variables. A similar behavior is observed for the the exponent $z$ characterizing the scaling of the autocorrelation time,  $\tau_c$ (\textbf{C}) and for the scaling exponent $\mu$ of the eigenspectra of the covariance matrix for $K=128$ (\textbf{D}; $\mu \to 0$ for increasing $c$ values). 
    Simulations were performed near the critical regime ($\beta=0.9999$),  on a network with with $J_{ij} = 1$ and $N = 273000$ neurons. The number of sensors was $M = 273$, each including $n_{sub} = 1000$ neurons. For each value of $c$, the threshold, $e$, was identified by approximately matching the empirical  activity rate, using a time bin $\delta=3dt$.}}
\end{figure*}

\clearpage

\begin{figure*}[ht]
    \centering
    \includegraphics[width=0.8\linewidth]{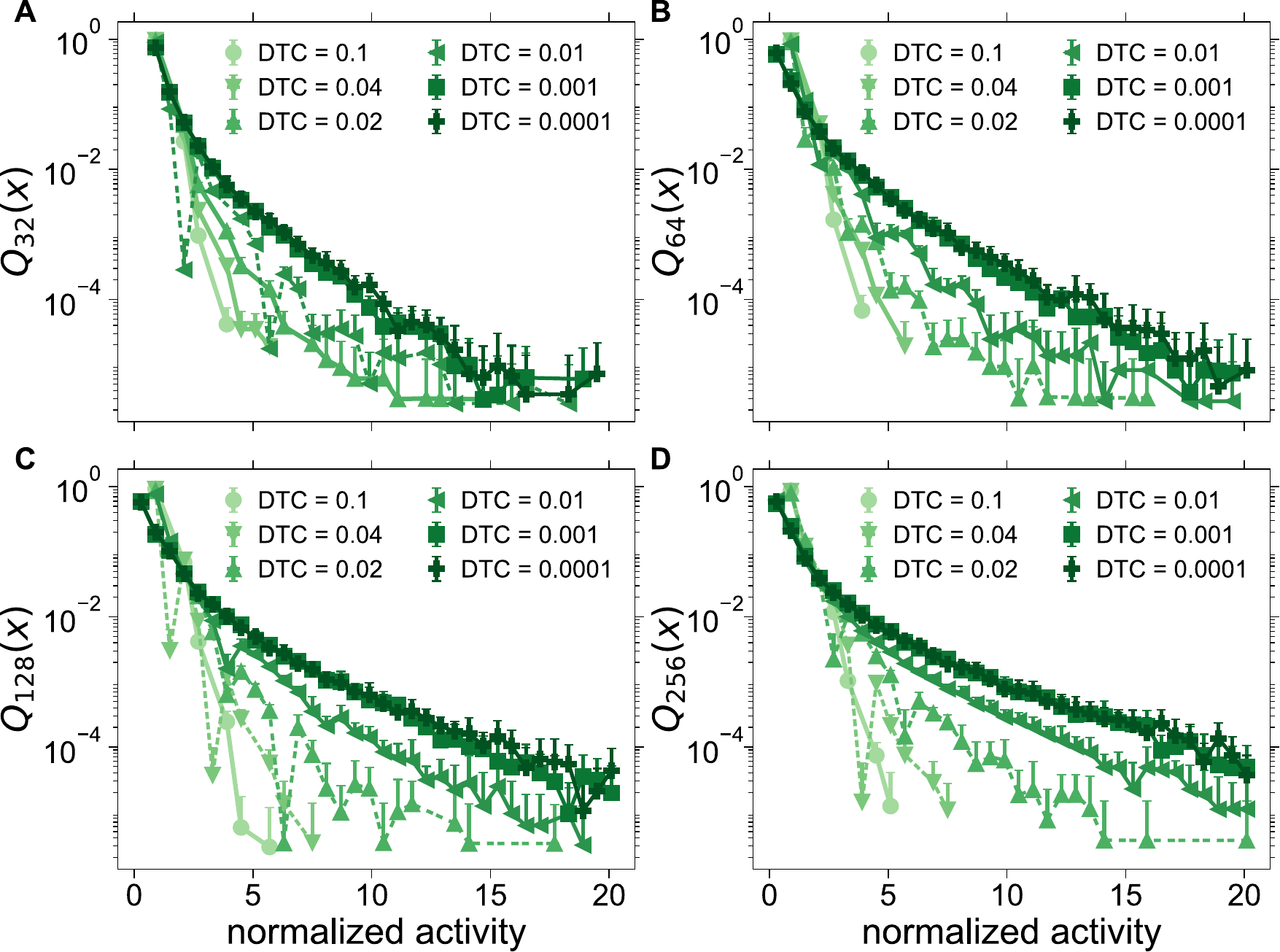}
    \caption{\textbf{Effect of the distance from criticality DTC $ = \beta_c - \beta_T$  on the distribution of activity for $K = 32, 64, 128, 256$.}  For $DTC > 0.05$ ($\beta_T < 0.95$) the distributions are short-tailed and  approximately Gaussian for all $K$. As $\beta_T$ approaches the critical point the distributions become distinctly non-Gaussian.  Simulation of a network with $N = 273000$ and $M = 273$ subsystems made of $n_{sub} = 1000$ neurons. For each DTC, the threshold for the analysis is determined by (approximately) matching the average activity rate to the empirical one (Methods).The bin size is $\delta = 3dt$.}
\end{figure*}

\begin{figure*}[ht]
    \centering
    \includegraphics[width=\linewidth]{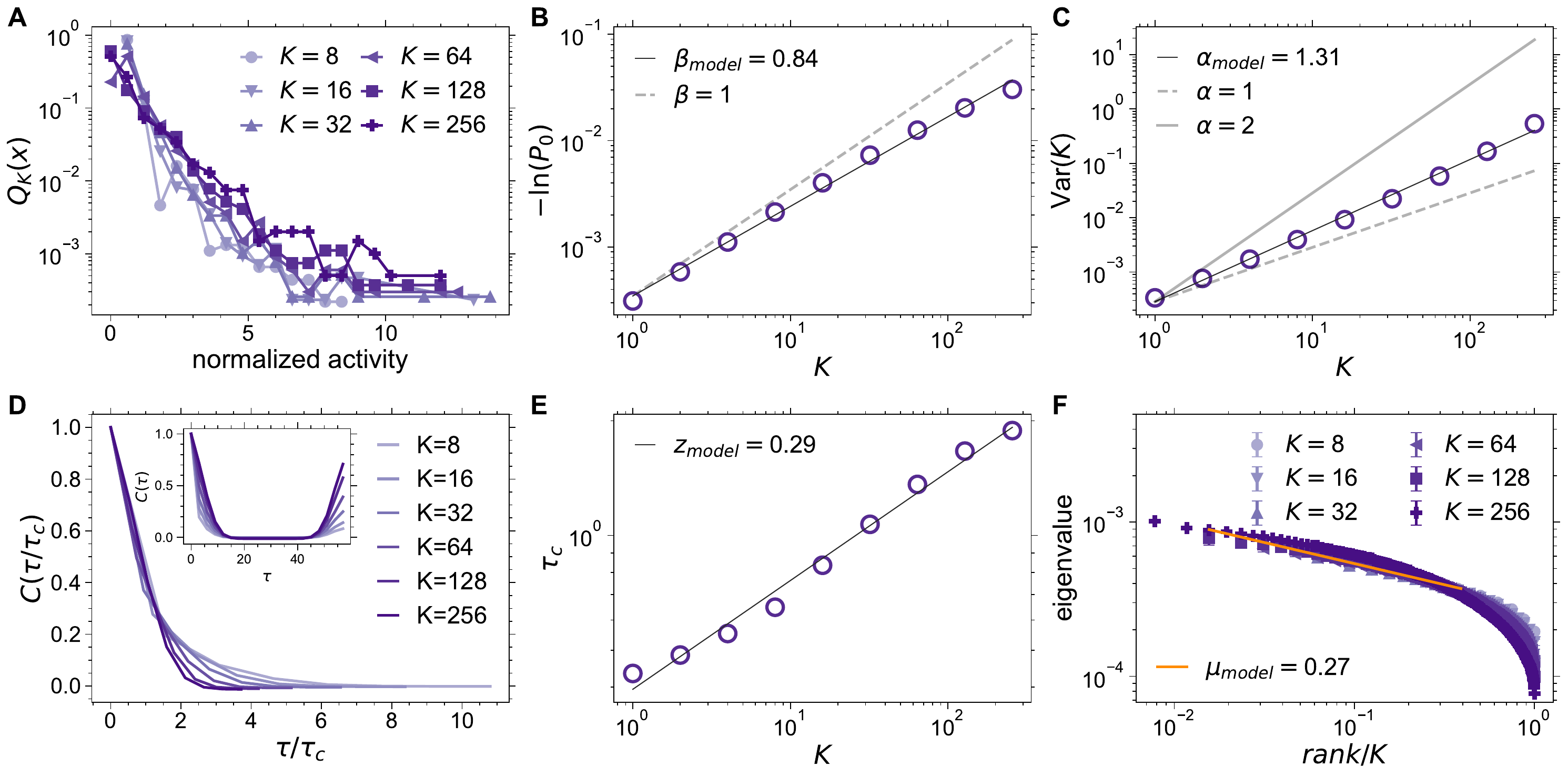}
    \caption{\textcolor{black}{{\bf PRG analysis of the adaptive Ising model on a small-world network.} \textbf{A.} Probability distribution of normalized non-zero activity for cluster sizes $K \geq 8$. Darker shades of purple correspond to increasing cluster sizes. Only data in the range $(0,20]$ are shown. \textbf{B.} The log-probability of silence in coarse-grained variables ($-\ln P_0$) as a function of cluster size ($K$). The solid red line represents a linear least-squares fit to $log(-\ln P_0) = log \ bK^{\beta_{model}}$. \textbf{C.} Variance (Var) of the coarse-grained variables as a function of cluster size, $K$. The solid line corresponds to the linear least-squares fit, $log \ Var(K) = \alpha \cdot logK + b$, where $\alpha_{model} = 1.31$. \textbf{D.} Rescaled average autocorrelation functions for different cluster sizes ($K$) collapse in the range of the initial decay exponential decay.    Inset: Non-rescaled $C(\tau)$. \textbf{E.} The autocorrelation time, $\tau_c$, scales as a power-law of $K$, with an exponent $z_{model} = 0.29$.  \textbf{F.} Eigenvalues of cluster covariance matrices  for different cluster sizes. Larger clusters correspond to darker colors. The solid line is a least squares fit to $log \ \lambda= log \ b(rank/K)^{-\mu}$ performed for $K = 128$. The fitting range is $[2/128, 50/128]$. Error bars represent SEM and are smaller than the symbol size. Simulations were performed near the critical point of the model on a small-world network with $N = 5120 × 5120 = 26.2 \cdot 10^6$, 
    $p=0.279$, $\beta = 0.329$ and $c=0.01$ ($J=1$; see Methods). 
    The number of sensors was $M = 256$, each including $n_{sub} = 10^5$ neurons (\textcolor{black}{$e = 2.05$ SD; $\delta = 3dt$)}.    }
    }
\end{figure*}


\clearpage



\clearpage

\clearpage

\begin{figure*}[ht]
    \centering
    \includegraphics[width=\linewidth]{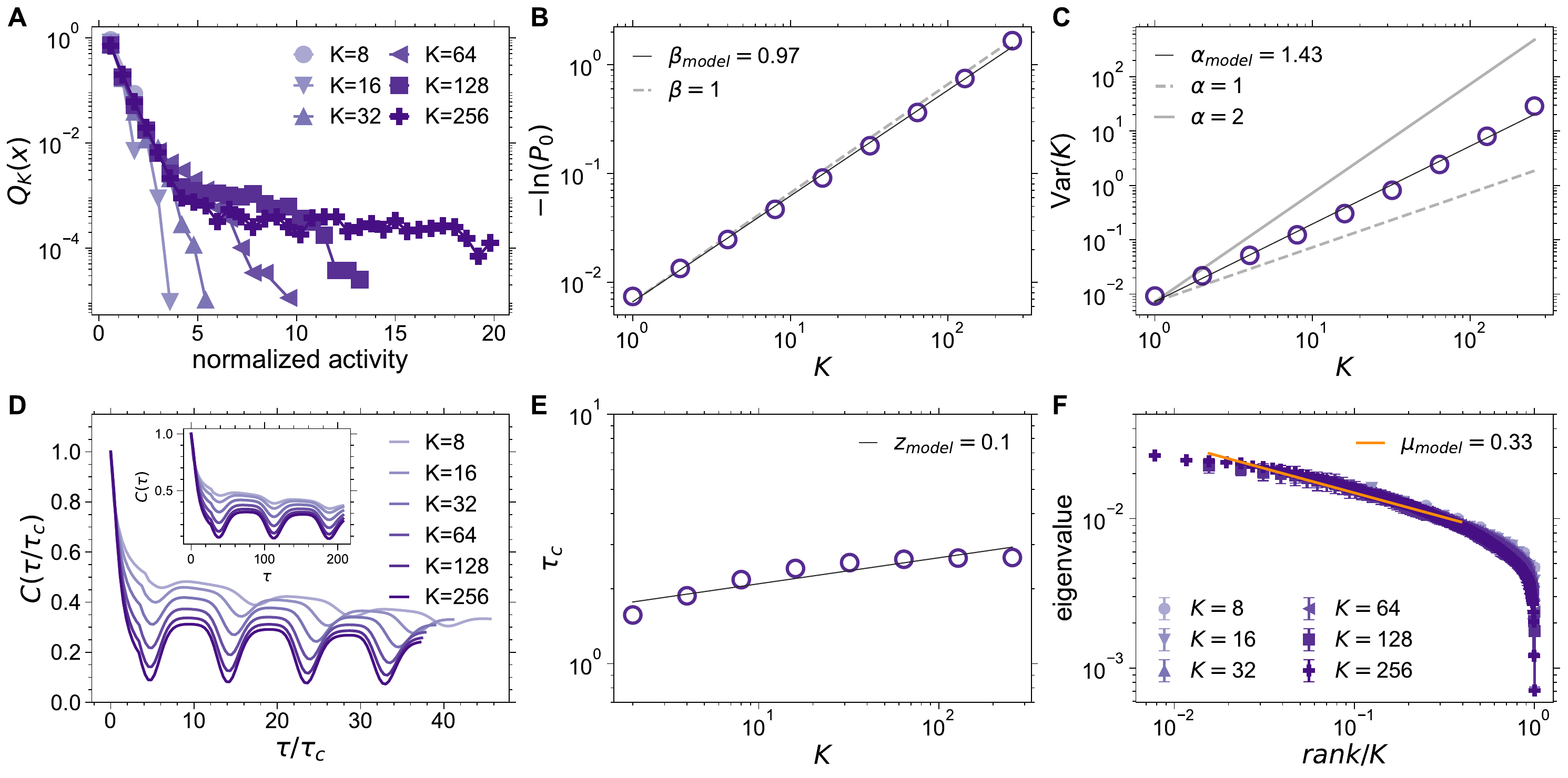}
    \caption{{\bf PRG analysis of the adaptive Ising model on a 2D network.} \textbf{A.} Probability distribution of normalized non-zero activity for cluster sizes $K \geq 8$. Darker shades of purple correspond to increasing cluster sizes. Only data in the range $(0,20]$ are shown. \textbf{B.} The log-probability of silence in coarse-grained variables ($-\ln P_0$) as a function of cluster size ($K$). The solid red line represents a linear least-squares fit to $log(-\ln P_0) = log \ bK^{\beta_{model}}$, where $\beta_{model} = 0.97$, very close to the  behavior expected for independent variables ($\beta = 1$). \textbf{C.} Variance (Var) of the coarse-grained variables as a function of cluster size, $K$. The solid red line corresponds to the linear least-squares fit, $log \ Var(K) = \alpha \cdot logK + b$, where $\alpha_{model} = 1.43$. \textbf{D.} Rescaled average autocorrelation functions for different cluster sizes ($K$) collapse in the range of the initial exponential decay.   The $C(\tau)$ are smoothed with a $20-$points moving average (approximate period of the oscillatory trend). $\tau_c$ is estimated via non-linear least square fit of the form $y= A\cdot e^{-\tau/\tau_c} + B\cdot e^{-\tau/\tau_1}$.    Inset: Non-rescaled $C(\tau)$. \textbf{E.} The autocorrelation time, $\tau_c$, scales as a power-law of $K$, with an exponent $z_{model} = 0.12$.  \textbf{F.} Eigenvalues of cluster covariance matrices  for different cluster sizes. Larger clusters correspond to darker colors. The solid line is a least squares fit to $log \ \lambda= log \ b(rank/K)^{-\mu}$ performed for $K = 128$. The fitting range is $[2/128, 50/128]$. Error bars represent SEM and are smaller than the symbol size. Simulations were performed near the critical regime of the 2D model ($\beta = 0.43$, $c=0.01$, $J=1$) on a network with $N = 256 \cdot 10^5$ neurons. The number of sensors was $M = 256$, each including $n_{sub} = 10^5$ neurons ($e = 2.5$ SD; $\delta = 3dt$).    }
\end{figure*}

\clearpage

\centering
{\Large {\bf Supplementary Tables}}

\vspace{1cm}

\begin{table}[H]
\begin{center}
\begin{tabular}{|c| c |c |} 
 \hline
 \multirow{3}{2cm}{\footnotesize{Scaling exponent}} & \multirow{3}{3cm}{\footnotesize{Extreme point of positive excursions}} & \multirow{3}{3cm}{\footnotesize{Extreme point of negative excursions}} \\ [1cm]
 \hline
 $\beta$ & $0.83 \pm 0.03$  & $0.84 \pm 0.02$  \\ 
 \hline
 $\alpha$ & $1.31 \pm 0.04$ & $1.29 \pm 0.04$\\
 \hline
 $z$ & $0.31 \pm 0.05$ & $0.30 \pm 0.04$ \\
 \hline
 $\mu$ & $0.47 \pm 0.04$ & $0.48 \pm 0.05$ \\
 \hline
 \hline
\end{tabular}
 \caption{\textbf{Scaling exponents  when considering only positive or only negative extremes of above-threshold excursions.} The  threshold is $e = 3.0$SD. Errors are standard deviations over  $100$ subjects.}
\label{exponent_table_1}
\end{center}
\end{table}

\vspace{-2.5cm}

\begin{table}[H]
\begin{center}
\begin{tabular}{|c| c |c |} 
 \hline
 \multirow{3}{2cm}{\footnotesize{Scaling exponent}} & \multirow{3}{3cm}{\footnotesize{All points in the excursion}} & \multirow{3}{3cm}{\footnotesize{Extreme point of excursion  only}} \\ [1cm]
 \hline
 $\beta$ & $0.64 \pm 0.07$  & $0.82 \pm 0.03$  \\ 
 \hline
 $\alpha$ & $1.56 \pm 0.07$ & $1.32 \pm 0.04$\\
 \hline
 $z$ & $0.16 \pm 0.04$ & $0.33 \pm 0.04$ \\
 \hline
 $\mu$ & $1.08 \pm 0.08$ & $0.46 \pm 0.04$ \\
 \hline
 \hline
\end{tabular}
 \caption{\textbf{Scaling exponents for the two different event detection approaches.} The threshold is $e = 3.0$SD. Errors are standard deviations over $100$ subjects.}
\label{exponent_table_2}
\end{center}
\end{table}

\begin{table}[H]
\begin{center}
\begin{tabular}{|c| c |c |} 
 \hline
 \multirow{3}{2cm}{\footnotesize{Scaling exponent}} & \multirow{3}{3cm}{\footnotesize{All points of positive excursions}} & \multirow{3}{3cm}{\footnotesize{All points of negative excursions}} \\ [1cm]
 \hline
 $\beta$ & $0.66 \pm 0.07$  & $0.67 \pm 0.06$  \\ 
 \hline
 $\alpha$ & $1.53 \pm 0.08$ & $1.51 \pm 0.07$\\
 \hline
 $z$ & $0.14 \pm 0.04$ & $0.14 \pm 0.04$ \\
 \hline
 $\mu$ & $1.08 \pm 0.09$ & $1.08 \pm 0.08$ \\
 \hline
 \hline
\end{tabular}
 \caption{\textbf{Scaling exponents  when considering only positive or only negative  above-threshold excursions (all points in the excursions).} The threshold is $e = 3.0$SD. Errors are standard deviations over $100$ subjects.}
\label{exponent_table_3}
\end{center}
\end{table}


\end{document}